\documentclass[twocolumn]{aastex63}

\usepackage{rotating}
\usepackage{graphicx}
\usepackage{amssymb}
\usepackage{amsmath}
\usepackage{multirow}
\usepackage{hyperref}
\usepackage{lineno}
\usepackage{lipsum}
\usepackage{gensymb}
\usepackage{todonotes}

\accepted{June 19, 2025}

\def\g-rays{{$\gamma$ rays}}

\def\aap{A\&A}

\def\apjl{ApJL}
\def\apjs{ApJS}

\newcommand{\RRR}[1]{{{#1}}}
\newcommand{\RR}[1]{#1}
\newcommand{\R}[1]{#1}
\begin{document}
\title{The Vela pulsar and its pulsar wind nebula Vela--X using 13 years of {\it Fermi}--LAT Observations}

\author[0000-0003-3540-2870]{Alexander Lange}
\affiliation{Department of Physics, The George Washington University, 725 21st Street NW, Washington, DC 20052, USA.}

\author[0000-0001-9633-3165]{J. Eagle}
\affiliation{NASA Goddard Space Flight Center, Greenbelt, MD 20771, USA}

\author[0000-0002-6447-4251]{O. Kargaltsev}
\affiliation{Department of Physics, The George Washington University, 725 21st Street NW, Washington, DC 20052, USA.}
\author[0000-0002-7889-6586]{Lucien Kuiper}
\affiliation{SRON-Netherlands Institute for Space Research, Leiden, 2333 CA, The Netherlands}
\author[0000-0002-8548-482X]{Jeremy Hare}
\affiliation{X-ray Astrophysics Laboratory, Code 662 NASA Goddard Space Flight Center, Greenbelt Rd., MD 20771, USA}
\affiliation{University of Maryland, Baltimore County, Baltimore, MD 21250, USA}
\affiliation{The Catholic University of America, 620 Michigan Ave., N.E. Washington, DC 20064, USA}

\begin{abstract}
We present results of more than 13\,years of {\it Fermi}--LAT data analysis for the Vela pulsar from 60\,MeV to 100\,GeV and its pulsar wind nebula (PWN), Vela--X, for $E>1$\,GeV in the off-pulse phases. We find the Vela--X PWN can be best characterized using two extended components: a large radial Gaussian accompanied by an off-set, compact radial disk, both with a similar spectral index, $\Gamma \sim 2.3$. The common spectral properties support a common PWN origin, but a supernova remnant component is plausible for the compact radial disk. With an updated Vela--X model, the phase resolved spectral properties of the Vela pulsar are explored through a phase-resolved analysis. The phase-resolved spectral properties of the pulsar are presented, such as the SED peak energy $E_p$, the width of the SED at its peak, $d_p$, and the asymptotic (low-energy) spectral index, $\Gamma_{0}$, are presented. 
The best-fit spectral models for each LAT pulse peak (Peak 1 and Peak 2) are extrapolated to UV energies and compared
to archival, phase-resolved spectra at UV, X-ray, soft $\gamma-$ray and TeV energies. We also discuss the physical implications of our modeling and the data comparisons. 
\end{abstract}

\section{Introduction}

One of the closest and brightest  pulsars, the Vela pulsar (PSR J0835--4510), has been the subject of extensive multi-wavelength studies ranging from radio waves to $\gamma-$rays. 
Its relatively young age,
$\sim$11,000 years, and 
proximity to Earth, 
$\sim$290 pc \citep{dodson_vela_2003}, make Vela a prime candidate to study pulsar and pulsar wind nebula (PWN) emission, guiding the theoretical understanding of pulsar magnetospheres and particle acceleration mechanisms. 
The Vela pulsar is 
the brightest 
persistent source in GeV $\gamma-$rays \citep{kanbach_egret_1994}. The 89-ms pulsations were promptly 
detected by the {\it Fermi} Gamma-Ray Observatory's Large Area Telescope \citep[LAT,][]{abdo_fermi_2009,abdo_psr_2010} with the first year of observations. The early Fermi--LAT observations measured the Vela pulsar's spin period $P = 0.089\,\mathrm{s}$, period derivative $\dot{P} = 1.24\times10^{-13}\, \mathrm{s \,s}^{-1}$, and other physical properties, finding good agreement with multi-wavelength observations \citep[e.g.,][]{manchester_australia_2005}.

\citet{abdo_psr_2010} 
also reported 
energy-dependent pulse profiles with structure resolved down to $\sim$0.3\,ms and 
phase-resolved spectral properties for 0.1--300\,GeV energies.
\RR{An analysis of the off-pulse emission in the 0.2--20\,GeV energy range} revealed 
spatially-extended emission 
associated with the PWN of the
Vela pulsar \citep{abdo_pwn_2010}. 

Multi-wavelength pulse profiles  \citep{kuiper_soft_2015,rudak_modeling_2017,harding_multi-tev_2018,aharonian_discovery_2023} 
show two distinct peaks  (labeled Peak 1 and Peak 2), as well as a third peak or bridge structure, in the optical, UV, X-ray and $\gamma-$ray, none of which precisely match the single peak observed in the radio \citep[see, e.g., Figure 4 in][]{kuiper_soft_2015}. 
In the GeV band, Peak 1 (below $\sim20$\,GeV) and Peak 2 occur at phases $\phi_{PK1} = 0.133$ and $\phi_{PK2} = 0.564$ \citep{3pc}, respectively, with a peak phase separation $\Delta \phi \approx 0.431$ where $\phi = 0$ corresponds to the radio peak \citep{abdo_psr_2010}.

The theoretical models of the pulsar magnetosphere ``Outer Gap'' (OG) \citep{romani_gamma-ray_1996} or the ``Slot Gap'' model \citep{harding_high-altitude_2008}, attribute pulsed $\gamma-$ ray emission to relativistic electrons/positrons in the open field line region of the pulsar magnetosphere . 
In the OG model, the majority of the observed pulsed $\gamma-$ray emission isdue to synchrotron/curvature radiation (SR/CR) at MeV--GeV energies and Inverse Compton (IC) scattering off thermal photons from the hot neutron star (NS) surface and off non-thermal photons produced from pair-production near the boundary of the OG region at GeV--TeV energies. The non-thermal X-ray emission can be attributed to SR \citep{cheng_multicomponent_1999}. CR in the OG region can explain the observed cut-off energy in the Fermi--LAT band, but it cannot explain the observed inter-pulse structure nor the winged structures just before the first peak and following the second peak \citep{abdo_psr_2010}.

More recent theoretical models \citep[for a review of models, see][]{harding_pulsar_2021} suggest that the GeV emission results from particles accelerated in an induced electric field (by a dissipating magnetic field) outside the light cylinder which radiate CR, called the force-free inside dissipative outside (FIDO) model \citep{kalapotharakos_gamma-ray_2014,brambilla_testing_2015}. The GeV emission can also be explained by synchrotron cooling of particles accelerated by magnetic reconnection inside a current sheet extending past the light-cylinder \citep{
philippov_ab-initio_2018}. 

The Vela-X PWN is powered by the pulsar through the continuous injection of relativistic electrons and positrons. In fact, most of the pulsar's rotational energy is converted into a highly magnetized \RR{relativistic particle wind} \citep{slane2006}. These particles radiate away their energy in the form of synchrotron emission from interactions with the PWN magnetic fields and through IC emission from interactions with the ambient photon fields. The Vela--X PWN is brightly detected from radio to $\gamma-$ray bands, exhibiting varying morphology throughout the multi-wavelength spectrum \citep{dodson_radio_2003,hales_vela_2004,abdo_pwn_2010,grondin2013,slane_investigating_2018,tibaldo2018}. A bright 2--3$^\circ$ radio nebula surrounds the pulsar, though displaced southward, with a relatively flat radio spectral index of $\alpha\approx-0.08$ \citep[$S_\nu \propto \nu^\alpha$, e.g.,][]{weiler1980}. An off-set, compact counterpart is detected in the X-ray \citep[the ``cocoon'',][]{frail1997,slane2018}. The radio and X-ray PWN emission are  encompassed by bright diffuse $\gamma-$ray emission \citep{abdo_pwn_2010,abdalla_hess_2019}. It has long been suspected that the PWN has two different particle components contributing to the broadband emission \citep{dejager2008,hinton2011}, indicated by the off-set positions and sizes of the radio nebula and X-ray cocoon and their varying spectral shapes. Such a distinction between particle populations in PWNe is commonly observed in the later stages of evolution, where the morphologies of evolved PWNe can be explained by an asymmetric interaction with the supernova remnant (SNR) reverse shock, which crushes and displaces the PWN from the continuous injection of high-energy particles by the central pulsar \citep[e.g.,][]{dejager2008,hinton2011,temim2015}. As a result, a younger, compact nebula is observed near the pulsar, surrounded by a diffuse, older nebula. 

We present new Fermi--LAT analysis results of both the Vela pulsar and Vela--X PWN using more than 13\,years of data. In Section \ref{sec:fermi},
we describe the data reduction and analysis. In Sections \ref{sec:pwn} and \ref{sec:pulsar},  we provide the spatial and spectral results characterizing the Vela--X PWN from off-pulse data followed by the phase-resolved analysis results for the Vela pulsar. In Section~\ref{sec:seds}, we compare archival multi-wavelength data of the Vela pulsar to the extrapolated Fermi--LAT best-fit spectral models of each $\gamma-$ray peak. In Section \ref{sec:discuss}, we present the physical implications of our results including the comparison to previous results and conclude in 
Section \ref{sec:conclude}.

\section{Fermi--LAT Data Analysis}\label{sec:fermi}
The LAT instrument onboard the Fermi Gamma-ray Space Telescope  is sensitive to $\gamma-$rays with energies ranging from 50\,MeV to $> 300$\,GeV \citep{atwood2009}. 
We analyze $\sim$13\, years (from 2008 August to 2021 October) of Pass~8 \texttt{SOURCE} class data \citep{atwood2013,pass82018} between 60\,MeV and 2\,TeV. Photons detected at zenith angles larger than 90\,$^\circ$ are excluded to limit the contamination from $\gamma-$rays generated by cosmic ray interactions in the upper layers of Earth's atmosphere. We perform a binned likelihood analysis using the latest \texttt{FermiTools} package\footnote{\url{https://fermi.gsfc.nasa.gov/ssc/data/analysis/software/}} (v.2.2.11) and \texttt{FermiPy} Python~3 package \citep[v.1.2.0,][]{fermipy2017}. We perform the likelihood analysis on a square 
$10\times 10\,^\circ$ region of interest (ROI) in equatorial coordinates using a pixel bin size $0.1\,^\circ$ and 8 bins per decade in energy (36 total bins). The $\gamma-$ray sky for the 15$^\mathrm{o}$ around Vela is modeled from the Fermi--LAT source catalog based on 12 years of data, 4FGL--Data Release 3 \citep[DR3,][]{4fgldr3} for point and extended sources\footnote{{\url{https://fermi.gsfc.nasa.gov/ssc/data/access/lat/12yr_catalog/}.}}, as well as the latest Galactic diffuse and isotropic diffuse templates (\texttt{gll\_iem\_v07.fits} and \texttt{iso\_P8R3\_SOURCE\_V3\_v1.txt}, respectively)\footnote{LAT background models and appropriate instrument response functions can be found at: \url{https://fermi.gsfc.nasa.gov/ssc/data/access/lat/BackgroundModels.html}.}.

The maximum likelihood technique is used to find the best model to the Fermi--LAT data, achieved when the Test Statistic (TS) of the source model is maximized. The TS value is defined to be twice the natural logarithm of the ratio of the likelihood of one hypothesis, $\mathcal{L}_{1}$, (e.g. presence of one additional source) to the likelihood for the null hypothesis, $\mathcal{L}_{0}$, (e.g. absence of a source):
\begin{equation}
\label{eq:TS}
  TS = 2 \times \log\left({\frac{{\mathcal{L}_{1}}}{{\mathcal{L}_{0}}}}\right)
\end{equation}
For sources characterized as a power law and have a freed position, TS values $>25$ correspond to a detection significance $> 4 \sigma$ for 4 degrees of freedom \citep[DOF,][]{mattox1996}. 

We apply an updated ephemeris solution\footnote{Provided courtesy of M. Kerr and the National Radio \\ Astronomy Observatory (NRAO).
}, derived from Fermi--LAT photon arrival times, using {\it Tempo2} software \citep{hobbs_tempo2_2006} and perform a spatial and spectral study of the Vela--X PWN, 4FGL~J0834.3--4542e, in the off-pulse phases. We then perform a phase-resolved analysis in the on-pulse phases of the Vela pulsar, 4FGL J0835.3--4510. \RR{The updated ephemeris has a} phase shift of $\approx0.031$ from previous studies \RR{that use a radio ephemeris}. 

\begin{figure}
\centering
\includegraphics[width=1.0\linewidth]{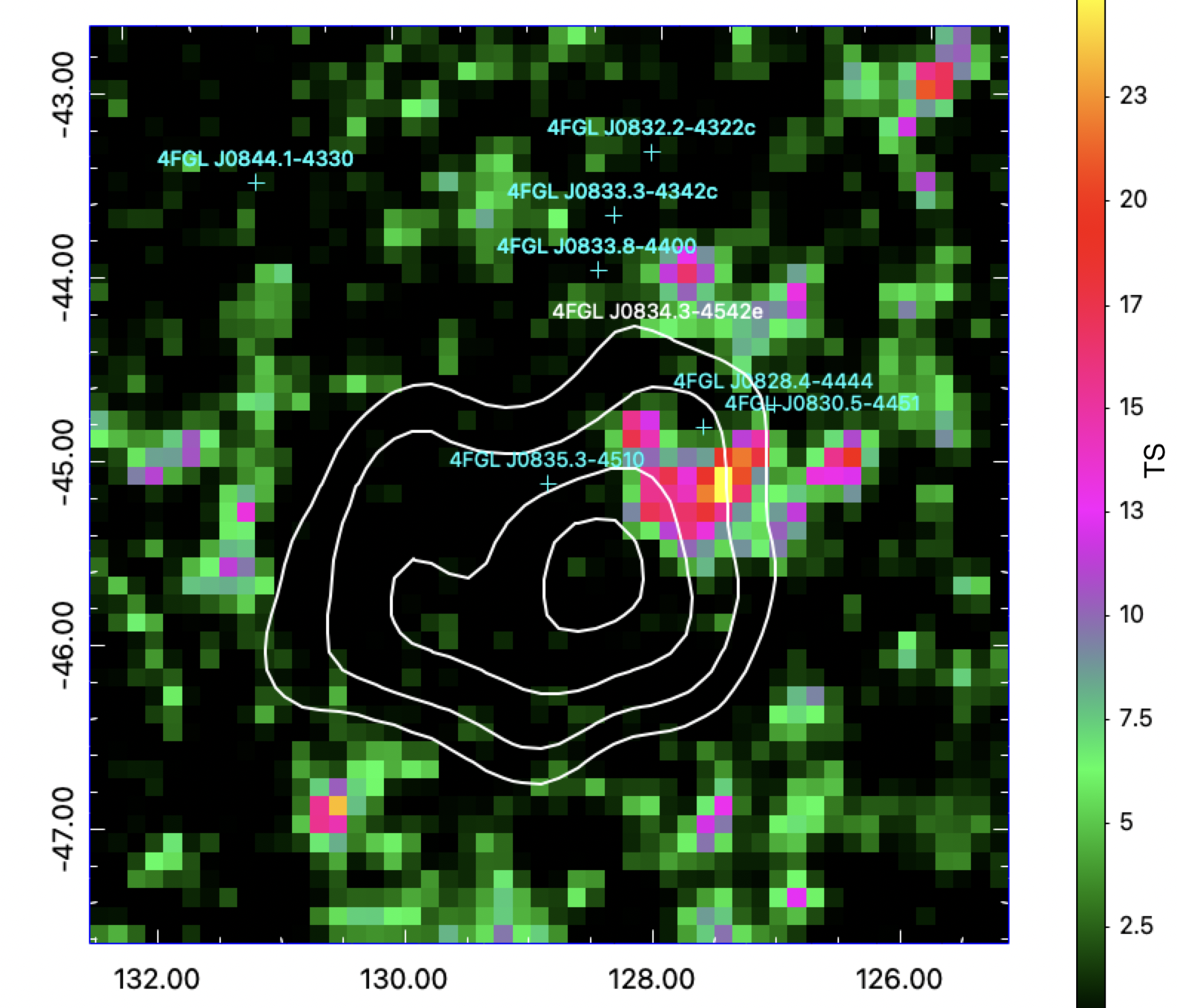}
\caption{$5\,\degree \times 5\,\degree$ TS map \RR{in J2000 equatorial degrees} of the 4FGL--DR3 source model for $E>1\,$GeV in off-pulse phases as defined in the text. There are TS residuals not accounted for in the global model, concentrating near the northwestern region of the PWN. The 4FGL--DR3 source model using the 330\,MHz radio template to characterize PWN emission is shown with the white contours. Other 4FGL--DR3 sources are labeled in cyan.}\label{fig:4fgl_tsmap}
\end{figure} 

\subsection{Vela--X PWN}\label{sec:pwn}
To analyze the PWN emission from Vela--X, we make selection cuts to the data such that we only consider photon events from the off-pulse window of the $\gamma-$ray pulsations. We choose an off-pulse interval of $\phi = [0,0.08]$ and $\phi = [0.8,1.0]$, which corresponds to \RR{28\% of the total exposure time}, motivated by minimizing associated pulsed emission as much as possible. We estimate the \RR{pulsar contribution} in the off-pulse phases by deriving the \RR{off-pulse} integrated energy flux from the pulsar and comparing it with the flux found for the Vela--X PWN, 4FGL~J0834.3--4542e. The level of \RR{pulsar} emission in the off-pulse window is $\sim 1\% $ of the PWN flux in the 60\,MeV--2\,TeV energy band. For the analysis of the PWN, we only consider energies 1\,GeV $< E <$ 2\,TeV to avoid the higher background below this energy. \RR{Since only 28\% of the total dataset is considered, we scale the normalization and flux values by a factor $\frac{1}{0.28} = 3.57$} to account for the removal of on-pulse phase events. We include the 4FGL--DR3 source that corresponds to the Vela pulsar, 4FGL~J0835.3--4510, in the source model and allow only the normalization to vary. Sources with TS $\geq 25$ and with distances from the ROI center/PWN position $\leq3.0$\,$\degree$ are free to vary in flux normalization along with the diffuse and isotropic background components.

In the 4FGL--DR3 catalog \citep{4fgldr3}, J0834.3--4542e assumes a custom spatial template, which is derived from the 330\,MHz radio observation of the PWN.
Displayed in Figure~\ref{fig:4fgl_tsmap} is a TS map ($E > 1\,$GeV) of the region using the 4FGL--DR3 source model (Model 1 in Table~\ref{tab:gamma_1gev}), where significant residual emission is seen in the northwest section of the PWN. We perform a spatial analysis for energies above 1\,GeV, to re-characterize the PWN and the residual excess.
\begingroup
\begin{table*}[!htp]
\centering
\begin{tabular}{|c|c|ccccc|}
\hline
\ Model & Sources Tested & Spatial Model & Power Law Index & $\log{L}$ & $G_E$ ($\times 10^{-4}$ MeV cm$^{-2}$ s$^{-1}$) & TS \
\\
\hline
\ 1 & 4FGL~J0834.3--4542e & 330MHz radio template & $2.37 \pm 0.03$ & --49635.5 &  $0.86 \pm 0.05$ & 1798 \\
\hline
\ 2 & RG & 1.16\,$\degree$ Radial Gaussian & $2.32 \pm 0.03$ & --49634.6 & $1.0 \pm 0.06$ & 1790 \\
\hline
\ 3 & RG & 1.16\,$\degree$ Radial Gaussian & $2.34 \pm 0.04$ & --49575.7 & $0.87 \pm 0.06$ & 1332 \\
\ & RD & 0.37\,$\degree$ Radial Disk & $2.33 \pm 0.08$ & & $0.19 \pm 0.03$ & 225 \\
\hline
\ 4 & 4FGL~J0834.3--4542e & 330MHz radio template & $2.39 \pm 0.04$ & --49630.9 &  $0.84 \pm 0.05$ & 1782 \\
\ & RD & 0.63\,$\degree$ Radial Disk & $0.58 \pm 0.73$ & & $0.29 \pm 0.18$ & 9.5 \\
\hline
\end{tabular}
\caption{Summary of the best-fit parameters and the associated 68\% C.L. statistics for all models tested for E$> 1\,$GeV. $G_E$ is the integrated energy flux for energies 1\,GeV--2\,TeV and is multiplied by 3.57 to account for the phase cut selection. Model 4 corresponds to \citet{tibaldo2018}.}
\label{tab:gamma_1gev}
\end{table*}
\endgroup
\begin{figure*}[!htp]
\begin{minipage}[b]{.33\textwidth}
\centering
\includegraphics[width=0.95\linewidth]{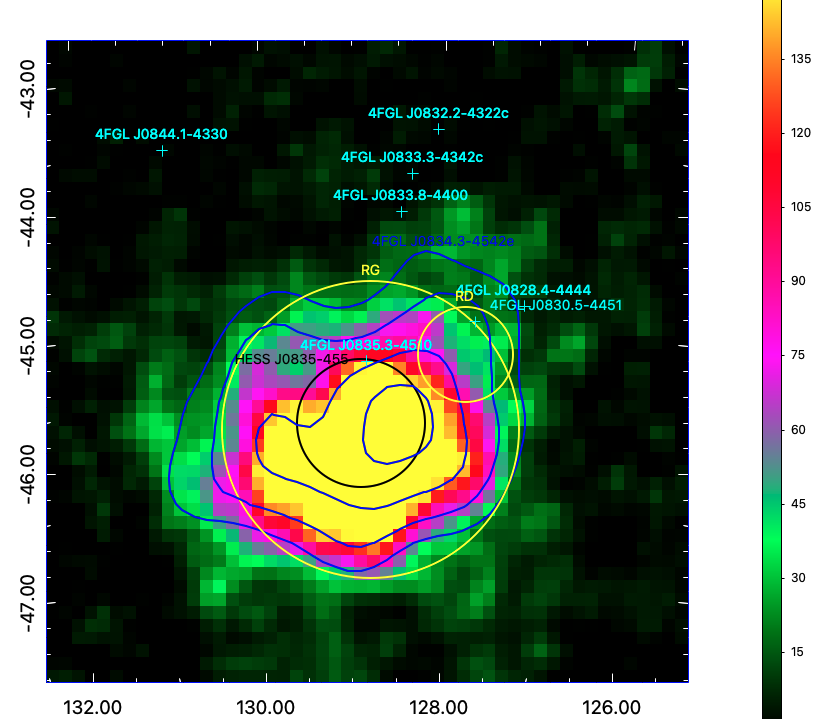}
\end{minipage}
\begin{minipage}[b]{.33\textwidth}
\centering
\includegraphics[width=0.95\linewidth]{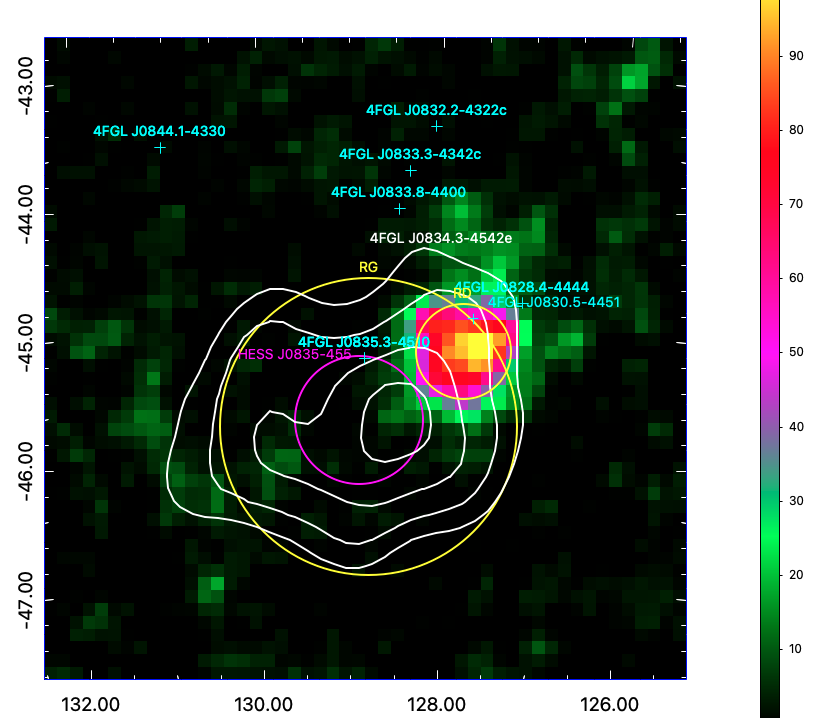}
\end{minipage}
\begin{minipage}[b]{.33\textwidth}
\centering
\includegraphics[width=0.99\linewidth]{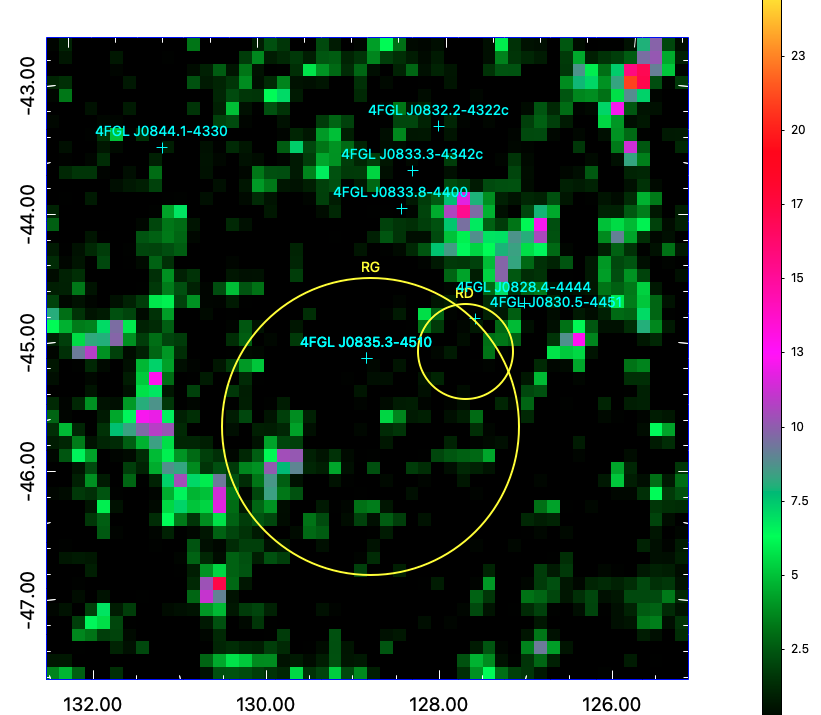}
\end{minipage}
\caption{{\it Left:} A $5\,\degree \times 5\,\degree$ TS map for $E>1$\,GeV of the large radial Gaussian component in Model~3 (the yellow circle labeled ``RG''), $r = 1.16\,\degree$. A possible second extended component in Model~3 is fit using a radial disk, $r = 0.37\,\degree$, and is the smaller yellow circle (see middle panel). The radio template used in the Fermi--LAT 4FGL--DR3 catalog is shown as the blue contours. The HESS TeV position and size is indicated in black. {\it Middle:} The same as the left but now showing the TS of the smaller radial disk (smaller yellow circle labeled ``RD'') that is found in this work. The radio template used in the Fermi--LAT 4FGL--DR3 catalog is shown as the white contours. The HESS TeV position and size is indicated in magenta. {\it Right:} A $5\,\degree \times 5\,\degree$ TS map for $E > 1$\,GeV demonstrating the residual TS signal is sufficiently modeled by the two extended sources in Model 3. \RR{All maps are shown in J2000 equatorial degrees.} }\label{fig:tsmaps}
\end{figure*} 
We replace J0834.3--4542e with a point source at the center position of the 330\,MHz radio PWN emission (RA, Dec) = (128.814, --45.596)\,$\degree$ in J2000 and assume a simple power-law spectrum, 
\begin{equation}
  \frac{dN}{dE} = N_{0} \left(\frac{E}{E_0}\right)^{-\Gamma}
\end{equation}
where $E_0$ is fixed to 1000\,MeV. We free the position of the point source and vary the normalization and spectral index. We then run extension tests starting from the two spatial templates supported in FermiPy, the radial disk and radial Gaussian templates.
Each template assumes a symmetric 2D shape with respective width parameters radius and sigma. We measure the significance for extension using {$\text{TS}_{\text{ext}} = 2 \log\big({\frac{\mathcal{L}_{\text{ext}}}{\mathcal{L}_{\text{ps}}}}\big) > 15$ which corresponds to a $\sim 4\,\sigma$ significance}.   We find that a radial Gaussian source at (RA, Dec) = (128.856, --45.783)\,$\degree$ provides the best-fit to the extension, $\text{TS}_\text{ext} = 874$, and is Model 2 in Table~\ref{tab:gamma_1gev}. The Gaussian size $r_{68} = 1.16\,\degree$ is similar to the radio size $r \sim 1.6\,\degree$ and provides a comparable statistical fit to the radio template (Table~\ref{tab:gamma_1gev}).

There remains significant ($\text{TS} > 25$) residuals that persist \RR{below 10\,GeV} in the Northern quadrant of the Vela--X PWN region  as in the 4FGL source model, see Figure~\ref{fig:4fgl_tsmap}. We model the persisting residuals by placing another point source with a power-law spectrum at the peak position and find an improvement to the fit ($\text{TS} = 53$). We free the point source location and allow the spectral index and normalization to vary. We run extension tests on this source, too, and find significant evidence for extension, $\text{TS}_\text{ext} = 44$ using the radial disk template. 

The final best-fit source model comprises the new $1.16\,\degree$ Radial Gaussian source in addition to a Radial Disk source at (RA, Dec)  = (127.743, --45.123)$\degree$ with radius $r_{68} = 0.37\,\degree$ (Model 3 in Table~\ref{tab:gamma_1gev}). The left panel in Figure~\ref{fig:tsmaps} shows a TS map for $E > 1\,$GeV showing the $\gamma-$ray significance of the larger Gaussian component. Similarly, the middle panel shows a TS map for $E > 1\,$GeV for emission modeled by the small radial disk component. The right panel displays the TS of the residuals after modeling both the larger Radial Gaussian and small Radial Disk source in the same energy range.

\begin{figure}
\centering
\includegraphics[width=1.0\linewidth]{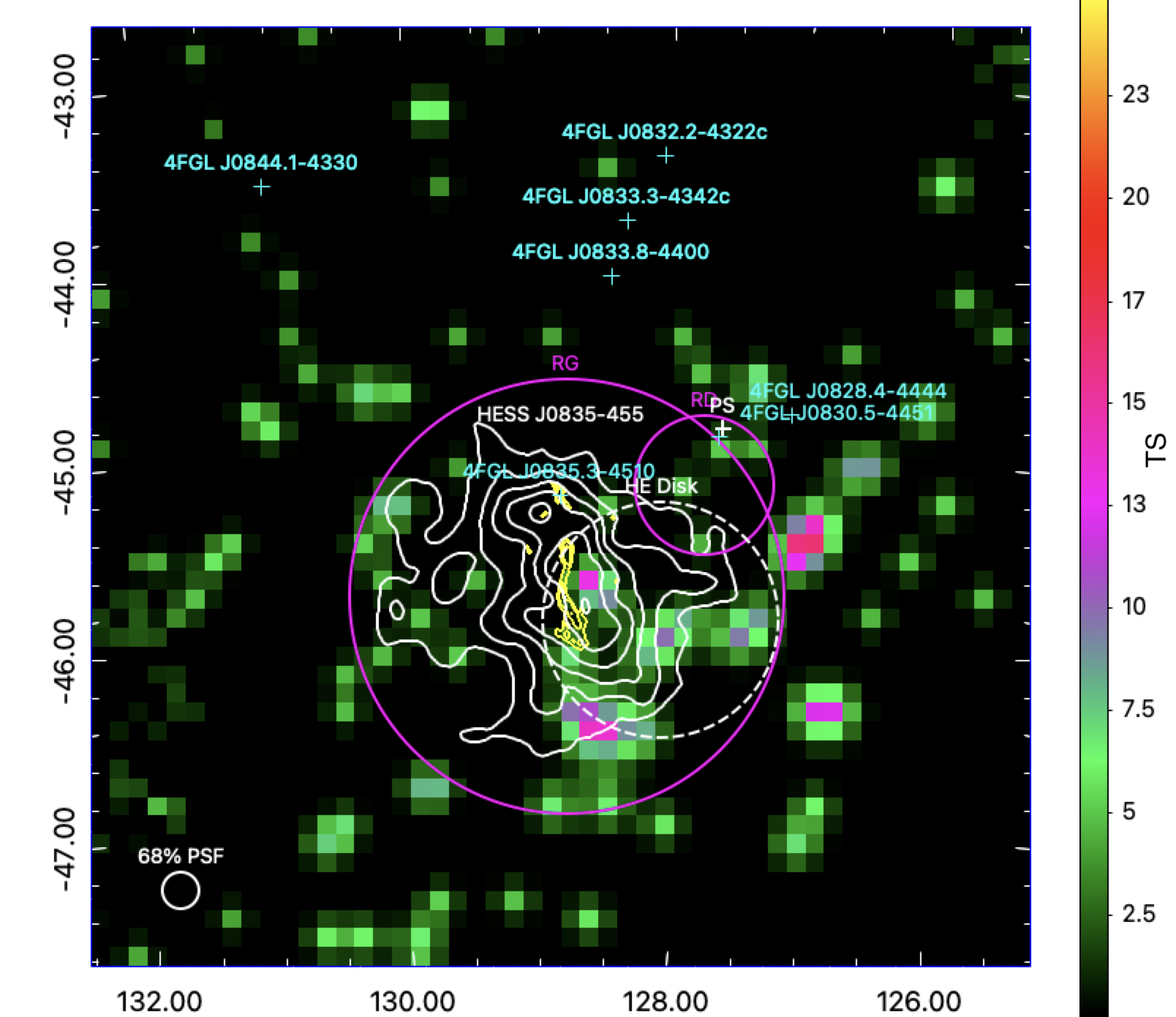}
\caption{\RR{A $5\,\degree \times 5\,\degree$ TS map \RR{in J2000 equatorial degrees} $E > 40$\,GeV, corresponding to a PSF $\sim 0.1\,\degree$, showing the \RR{total} TS \RR{signal from} both extended sources associated to the Vela--X PWN. The HESS TeV PWN significance contours starting from 3\,$\sigma$ and ending at 13\,$\sigma$ are shown as the white contours. The 843\,MHz radio PWN is shown as the yellow contours. The Fermi--LAT PWN emission is characterized as a radial Gaussian (the magenta circle labeled ``RG''), $r = 1.16\,\degree$, with a possible second extended component fit using a radial disk, $r = 0.37\,\degree$, and is the smaller magenta circle labeled as ``RD''. The high-energy radial disk (white dashed circle labeled ``HE Disk'') and the PS (white cross) are from \citet{tibaldo2018}.}}\label{fig:special_tsmap}
\end{figure}

\begin{figure*}
\begin{minipage}[b]{.5\textwidth}
\centering
\includegraphics[width=1.0\linewidth]{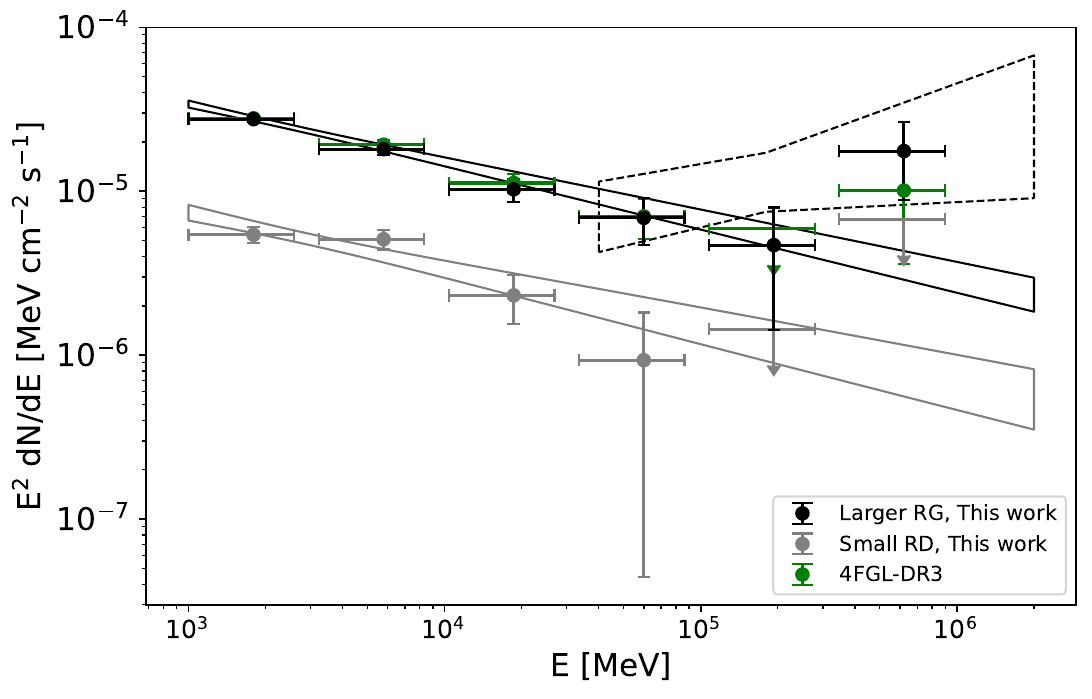}
\end{minipage}
\begin{minipage}[b]{.5\textwidth}
\centering
\includegraphics[width=1.0\linewidth]{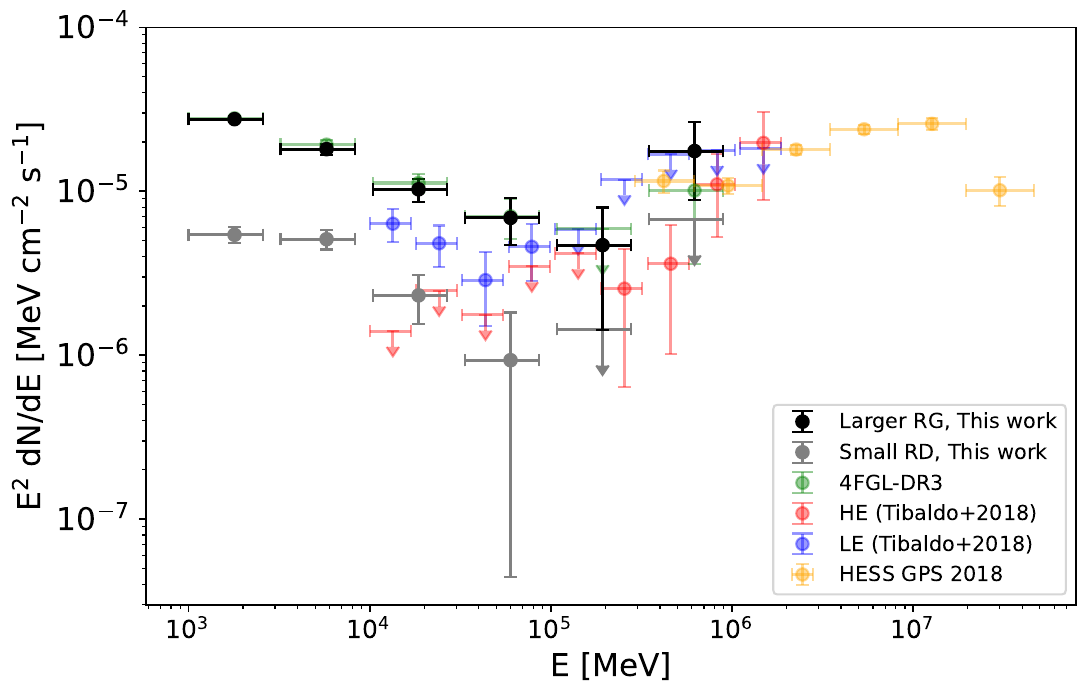}
\end{minipage}
\caption{{\it Left:} 
The 1\,GeV--2\,TeV $\gamma-$ray spectral energy distribution for the Vela--X PWN modeled as 4FGL~J0834.3--4542e (green) or Model 3 (black and grey). \RR{The best-fit $E>1\,$GeV spectral models are shown as the solid regions for both sources. The dashed black region is the best-fit model of the larger RG above 40\,GeV} (see text for details).
{\it Right:} The same flux data as the left panel compared to the low- and high-energy components measured from \citet{tibaldo2018} in addition to the HESS TeV flux data from \citet{hessgps2018}.}\label{fig:gamma_sed}
\end{figure*} 

\RR{Lastly, we test the spatial model described in \citet{tibaldo2018}, which reports a best-fit above 10\,GeV using three sources: an extended source modeled using the radio template, a second extended source modeled as a 0.63\,$\degree$ radial disk, and a point source. Using this model, we} do not find an improvement to the fit (Model 4 in Table~\ref{tab:gamma_1gev}). \RR{We note that} 4FGL~J0830.5--4451 coincides with the point source reported in \citet{tibaldo2018}.  \RR{We have checked that 4FGL~J0830.5--4451 is required and is therefore included in all source models.} No additional point source is required in our $E > 1\,$GeV analysis, so only the radio template and the 0.63\,$\degree$ radial disk are reported in Table~\ref{tab:gamma_1gev}. \RR{The best-fit parameters for each source model tested are provided in Table~\ref{tab:gamma_1gev}.} The best-fit source model found here, Model 3 in Table~\ref{tab:gamma_1gev}, gives an improvement over the 4FGL--DR3 source model (Model 1) $\text{TS} = 119.6$\RR{, $\text{TS} = 117.8$ over Model 2, and $\text{TS} = 110.4$ over Model 4.} 

\RR{We plot a TS map displaying the residuals of both extended components above $E >40\,$GeV in Figure~\ref{fig:special_tsmap}, which shows the high-energy contribution, mostly from the Radial Gaussian source. The bright, diffuse component (the Gaussian) and the fainter, more compact component (the disk) are both contributing predominantly at energies below 10\,GeV, see also the $\gamma-$ray spectral energy distribution (SED) in Figure~\ref{fig:gamma_sed}. Both of the extended sources have similar spectral index $\Gamma_\gamma \sim 2.3$ for energies above 1\,GeV (Table~\ref{tab:gamma_1gev}), but the Gaussian source has a notable high-energy component appearing above 40\,GeV and is demonstrated in both Figures~\ref{fig:special_tsmap} and \ref{fig:gamma_sed}. Therefore, we measure the spectral index of both sources above and below 40\,GeV to quantify the degree of spectral hardening. We find that the larger Gaussian source hardens above 40\,GeV to $\Gamma_\gamma = 1.68 \pm 0.25$ and the smaller disk source is not detected above 40\,GeV. The $E>1\,$GeV and $E > 40\,$GeV spectral models for the larger Gaussian source are included in the left panel of Figure~\ref{fig:gamma_sed}, shown as the solid and dashed regions.}

While we have not recovered the exact result in \citet{tibaldo2018}, there is general agreement between the presented source models. As in \citet{tibaldo2018}, we find that two extended sources are needed to characterize the Vela--X PWN emission. \citet{tibaldo2018} find that there are two distinct components: a high-energy component (the 0.63\,$\degree$ radial disk) that has a power-law spectral index $\Gamma = 0.9 \pm 0.3$ and a low-energy component (the radio template) that has a power-law spectral index $\Gamma = 2.2 \pm 0.2$.  The diffuse component corresponds to our Gaussian source, which connects to the flux data of \citet{tibaldo2018} and to the HESS TeV data nicely. 
\subsection{Vela pulsar}\label{sec:pulsar}

\begin{figure}[!h]
\centering
\includegraphics[width=0.9\linewidth]{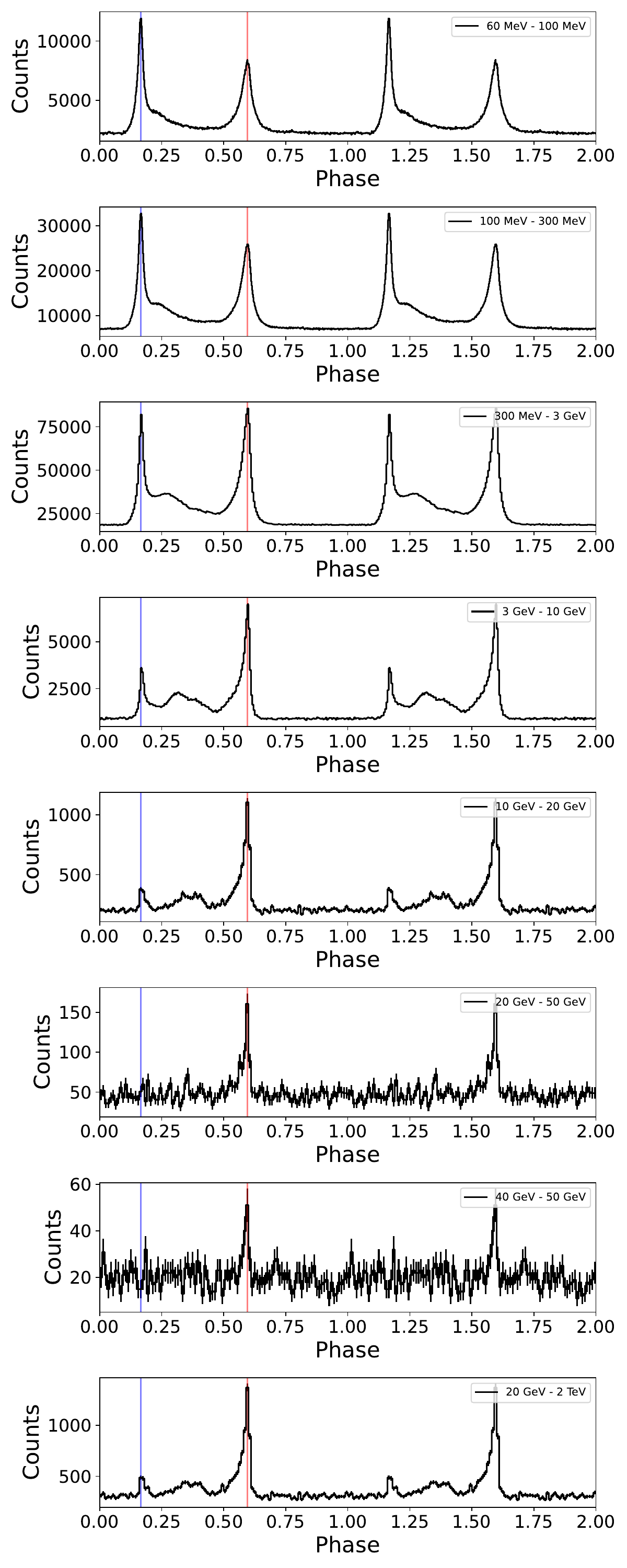}
\caption{Pulse profile for the Vela pulsar. From top to bottom: 60\,MeV--100\,MeV, 100\,MeV--300\,MeV, 300\,MeV--3\,GeV, 3--10\,GeV, 10--20\,GeV, 20--50\,GeV, 40--50\,GeV and 20\,GeV--2\,TeV. Blue and red vertical lines indicate the peak $\gamma-$ray emission, Peak 1 ($\phi_{\mathrm{PK1}}=0.165$) and Peak 2 ($\phi_{\mathrm{PK2}}=0.595$), respectively. The top 2 panels are binned with 500 bins of equal width. The remaining panels are binned with 200 bins of equal width. Two cycles are shown for clarity.}\label{fig:pulse_profiles}
\end{figure}

While we assigned pulse phases for all photons within $15^\circ$ of Vela, to investigate the 
dependence of the pulse profile on energy, we \RR{o}nly use the photons within $1^\circ$ of the Vela pulsar to create the energy-resolved pulse profiles shown in Figure~\ref{fig:pulse_profiles}. The resolution (bin size) of the pulse profiles varies from 0.178\,ms (500 bins per spin period) at lower energies to 0.445\,ms (200 bins per spin period) at higher energies to maintain small and comparable statistical uncertainties per bin. As in previous studies \citep[e.g.,][]{hess_collaboration_first_2018}, we find that the first peak, Peak 1 ($\phi_{\rm P1} = 0.165$), dominates the softer $\gamma-$rays E$<10$\,GeV. The second peak, Peak 2 ($\phi_{\rm P2} = 0.595$), persists until $40$--$50$ \,GeV. 
The inter-pulse structure (between phase $\phi \sim 0.25$ and $\sim0.45$) is prominent between $300$\,MeV--$20$\,GeV and shifts from $\phi\approx0.3$ to $\phi\approx0.4$ with increasing energies. This is referred to as a ``bridge'', \RR{labeled} as P3 in \citet{abdo_psr_2010} (alternatively called ``bridge emission'' elsewhere).  The shape of the inter-pulse structure evolves from a broad peak to a flat-top (or, possibly, two smaller peaks) as the energy increases. 
The structure is not detectable above 20\,GeV. We do not see any significant phase shifts for Peak 1 or Peak 2. The phase separation between both peaks remains constant $\Delta \phi \approx 0.4$ across all energy selections. 

For the phase-integrated and phase-resolved analyses, we consider the energy range 
$ 60$\,MeV $< E < 100$\,GeV and consider all phase-assigned photons within $ 15^\circ$ of Vela.
We first
perform a phase-integrated analysis with no phase selection applied. The phase-integrated fit determines the best-fit values within the 10$\,\degree$ ROI and is considered to be the ``baseline'' source model used for the phase-resolved analysis. All 4FGL--DR3 sources within the baseline model (including our new Vela-X model) are allowed to vary only in normalization. The Vela pulsar, 4FGL~J0835.3--4510, has all its spectral parameters freed (with the exception of the scale, E$_0$). Sources with a TS value $< 25$ are removed from the model. 
We model the pulsar spectrum as a power-law with a super exponential cutoff (PLEC4)\footnote{\url{https://fermi.gsfc.nasa.gov/ssc/data/analysis/scitools/source_models.html}}. 
This model, introduced for pulsars in the 4FGL--DR3 catalog \citep{4fgldr3}, defines the differential photon flux as 
\begin{equation}\label{eq:PLEC4}
\frac{d N}{d E} =N_0\left(\frac{E}{E_0}\right)^{-\Gamma+\frac{d}{b}} \exp \left[\frac{d}{b^2}\left(1-\left(\frac{E}{E_0}\right)^b\right)\right]
\end{equation}
where $E_0$ is the reference energy and fixed to the 4FGL--DR3 value of 1914.6\,MeV,
$b$ is the spectral asymmetry parameter (its value is anti-correlated to the degree of symmetry of the spectrum and dictates the strength of the cutoff at high energies), and $d$ is the SED curvature at $E_0$
\citep[$d$ is anti-correlated to the width of the peak of the spectrum,][]{3pc}. The best-fit spectral values from the phase-integrated fit are $N_0 = (4.357 \pm 0.001)\times10^{-10}$\,cm$^{-2}$\,s$^{-1}$\,MeV$^{-1}$, $\Gamma = 2.271 \pm 0.001$, $d = 0.579 \pm 0.002$ and $b = 0.394 \pm 0.005$. In comparison, the 4FGL--DR3 catalog \citep{4fgldr3} reports $N_{0,\mathrm{DR3}} = (4.422 \pm 0.014)\times10^{-10}$\,cm$^{-2}$\,s$^{-1}$\,MeV$^{-1}$, $\Gamma_{\mathrm{DR3}} = 2.243 \pm 0.004$, $d_{\mathrm{DR3}} = 0.572 \pm 0.004$ and $b_{\,\mathrm{DR3}} = 0.493 \pm 0.011$.   We find our best-fit values are close to the values reported in the 4FGL--DR3 catalog, with the exception of our $b$ value, which may be explained by the use of likelihood weights in the 4FGL--DR3. 
\citet{4fgldr3} introduced the PLEC4 model to provide a better characterization of pulsar spectral properties by minimizing the correlation between the spectral parameters, a problem that exists when using the PLEC2 spectral model (Equation~\ref{appendix:PLEC2}) used in \citet{abdo_psr_2010}.
PLEC4 parameters can be converted to PLEC2 parameters (the cutoff energy of the SED, $E_c$, spectral index, $\Gamma_0$, and the normalization, $N$) 
using the equations provided in the Appendix, Section~\ref{appendix:plec conversion} \citep[see also][]{3pc}. 

\begin{figure*}[!hp]
\includegraphics[scale=.5]{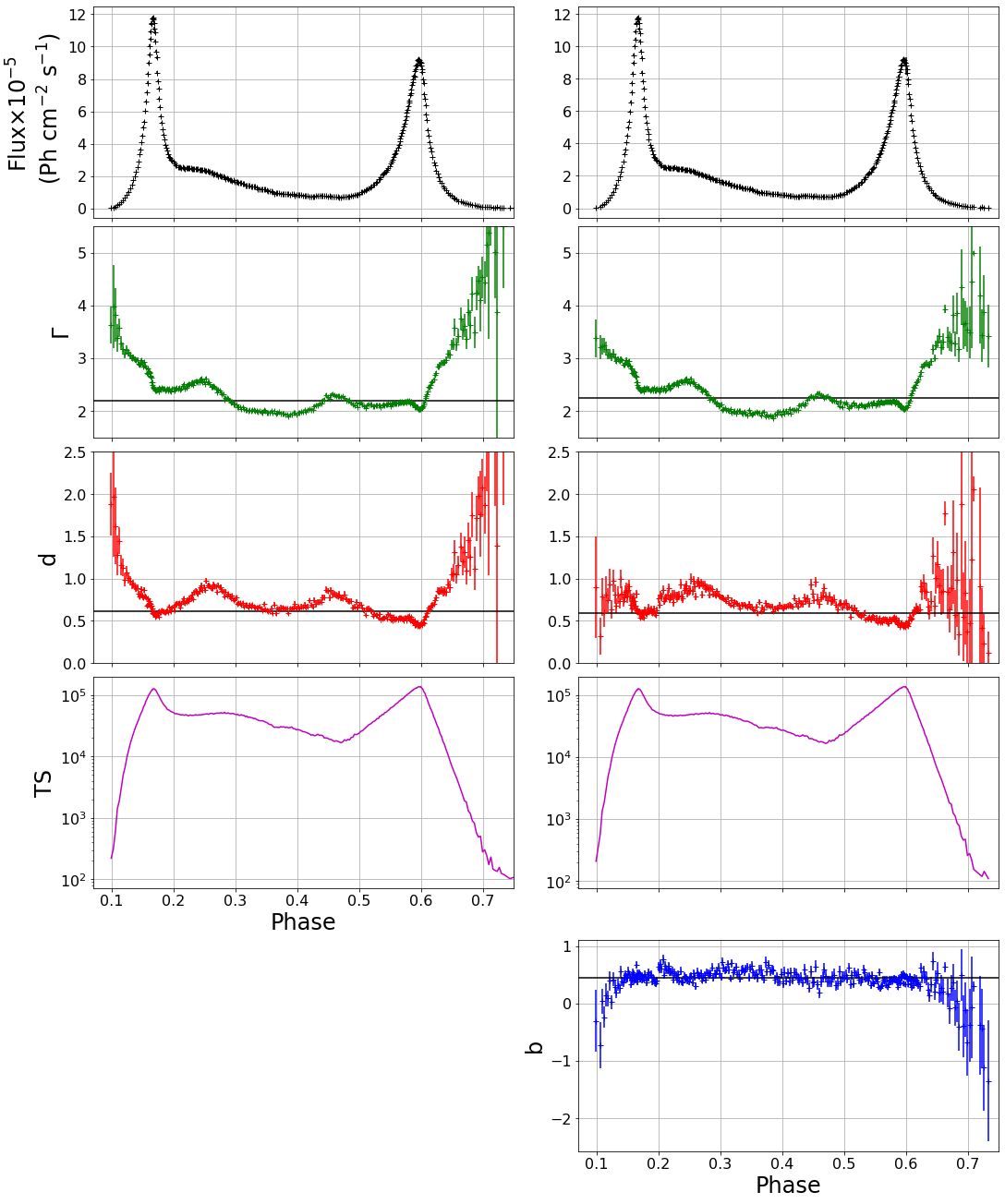}
\begin{minipage}{0.5\textwidth}
\vspace{-6cm}
\caption{Spectral evolution with phase as folded on the spin-period $\sim$ 89\,ms using the PLEC4 model. We
explore two cases: where $b$ is frozen to our best-fit phase-integrated value of 0.394 (left) and where $b$ is allowed to vary (right). From top to bottom: the photon flux, spectral index $\Gamma$, spectral curvature $d$, and the TS are plotted. The asymmetry parameter $b$ is shown in the last panel on the right for the case where $b$ varies. Lastly, the best-fit value from the phase-integrated is shown as a black horizontal line for $\Gamma$, $d$, and $b$. 
}\label{fig:Fixed_b}
\end{minipage}  
\end{figure*}

For the phase-resolved analysis, we fix all sources in the ROI to our ``baseline'' model except the Vela pulsar to investigate the evolution in the phase range of $0.08 \leq \phi \leq 0.80$. We bin all events into phase bins with 30,000 counts \citep[$\sim$ 20 times more counts than][]{abdo_psr_2010} and apply an exposure correction corresponding to the phase bin width (\texttt{expscale = 1/width}). The phase evolution of the photon flux, and PLEC4 parameters $\Gamma$ and $d$, and the TS is shown in Figure~\ref{fig:Fixed_b} for two models: ``fixed-b'', where $b=0.394$ (left), and ``free-b'', where $b$ is a free parameter (right). We show the phase-resolved results for the bins where the pulsar has a TS $>$ 100, corresponding to a phase range of $\phi \approx$ [0.09,0.73]. We observe $b$ to vary between 0.29 and 0.59, generally decreasing with increasing phase. The highest asymmetry in the SED (large values of $b$) occurs near Peak 1, while the more symmetric SEDs (small values of $b$) occur during the inter-pulse phases and near Peak 2. The SEDs for phases $\phi$ = [0.08, 0.77] (rebinned into 26 bins for a more condensed view of the SED of the pulsar at Peak 1, Peak 2 and the bridge emission) can be found in the Appendix Figures~\ref{app:count_fixedb}, \ref{app:e2_fixedb}, \ref{app:count_freeb} and \ref{app:e2_freeb}.

Whether $b$ is fixed or allowed to vary, we find similar behavior between the ``fixed-b'' and ``free-b'' fits.
In the ``free-b'' case, $\Gamma=2.444\pm0.020$for Peak 1 and $\Gamma=2.060\pm0.014$ for Peak 2 ($\phi_{\rm P2} = 0.595$)\RR{softening} to $\sim 2.0$ during the inter-pulse phases. Spectral width $d$ appears to be less variable during the inter-pulse phases with values varying between 0.5 and 0.9. 
For both the ``fixed-b'' and ``free-b'' fits, we observe $d$ to grow rapidly to the left of Peak 1 and to the right of Peak 2. In general, the phase-resolved values of $\Gamma$ and $d$ are in agreement with those obtained from the phase-integrated values.   

\begin{figure*}[!hpt]
\begin{minipage}[b]{.5\linewidth}
\centering
\includegraphics[width=1.0\linewidth]{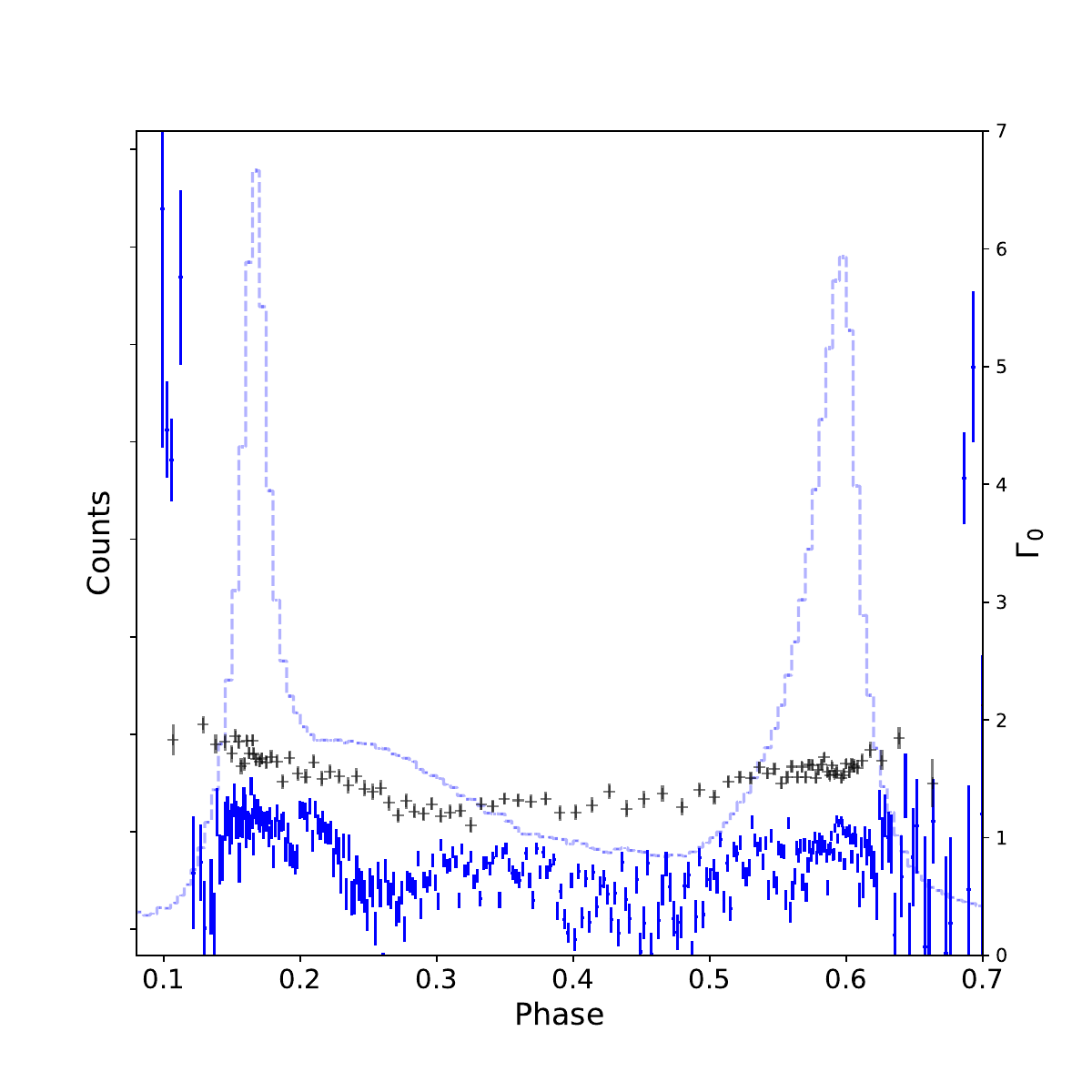}
\end{minipage}
\begin{minipage}[b]{.5\linewidth}
\centering
\includegraphics[width=1.0\linewidth]{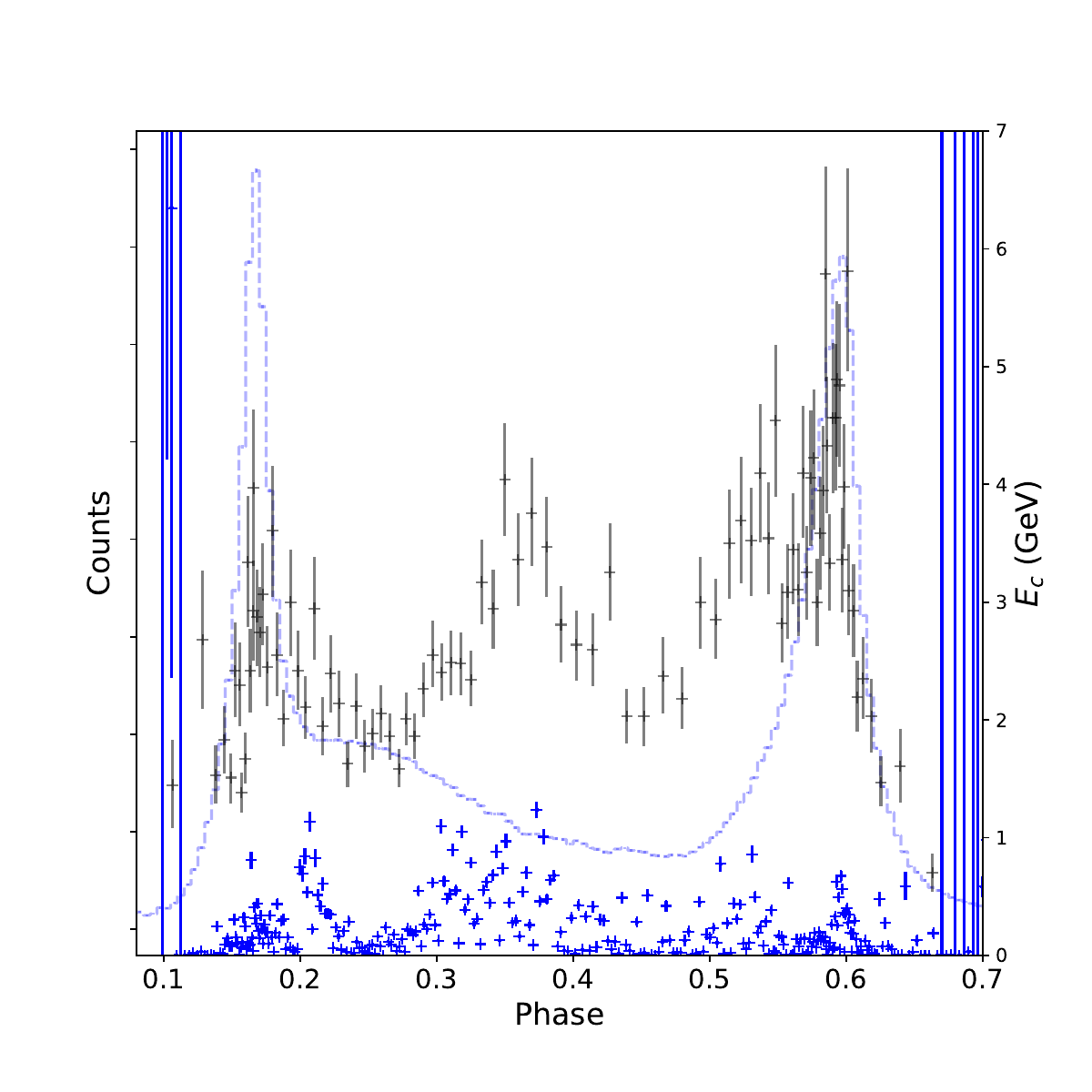}
\end{minipage}
    \vspace{-2pt} 

\begin{minipage}[b]{.5\linewidth}
\centering
\includegraphics[width=1.0\linewidth]{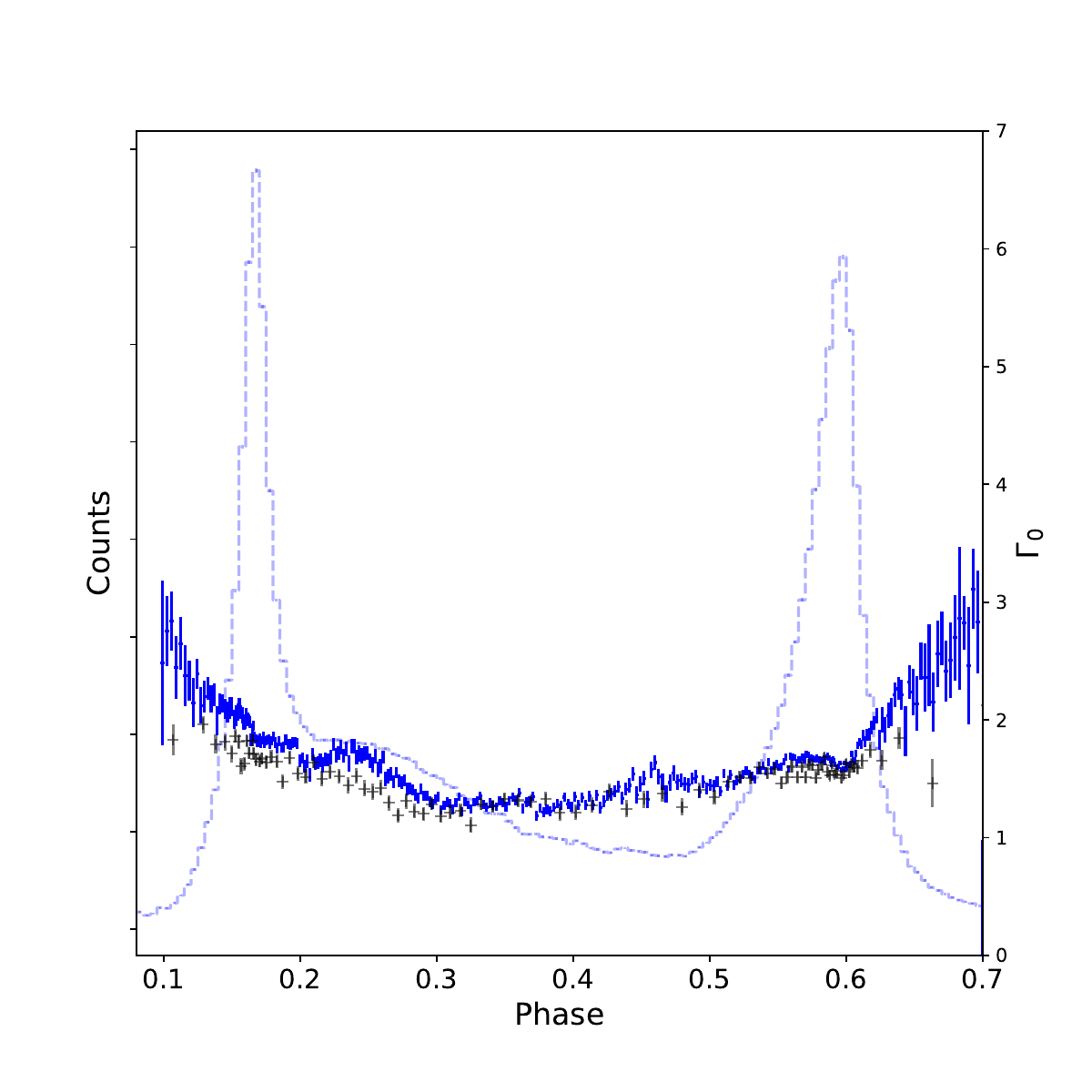}
\end{minipage}
\begin{minipage}[b]{.5\linewidth}
\centering
\includegraphics[width=1.0\linewidth]{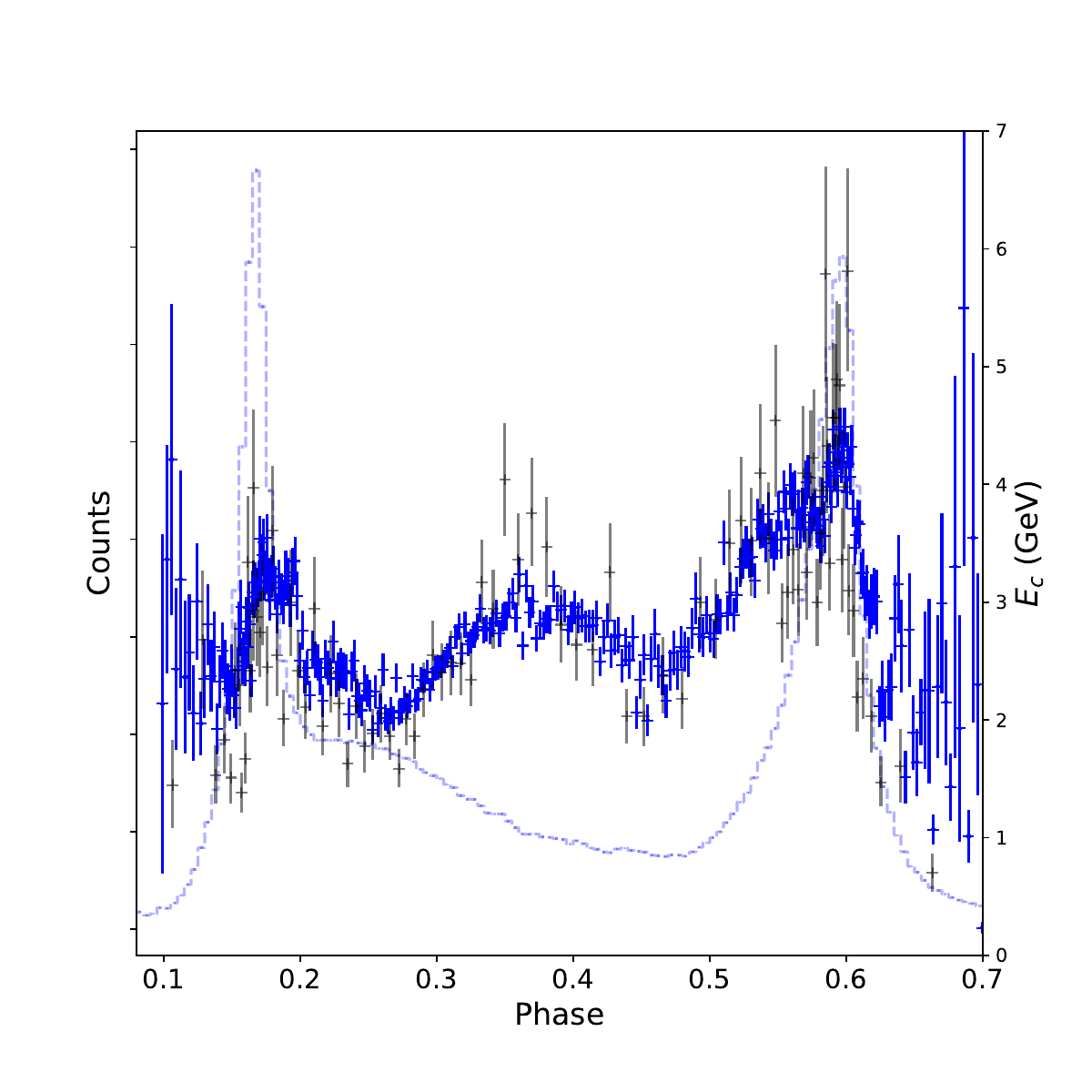}
\end{minipage}
\caption{Spectral evolution between best-fit values of the PLEC4 model converted to PLEC2 values (blue data points) using Equation~\ref{appendix:plec conversion} compared to \citet{abdo_psr_2010} PLEC2 values (grey data points). Overlaid on each figure is the pulse profile in blue. 
\textbf{Top:} {\it Left}: $\Gamma_0$ as a function of phase from the ``free-b'' model. 
{\it Right}: Cutoff energy (GeV) as a function of phase from the ``fixed-b'' model.  
\textbf{Bottom:} {\it Left}: $\Gamma_0$ as a function of phase when $b$ is fixed to 1. {\it Right}: Cutoff energy (GeV) as a function of phase when $b$ is fixed to 1. 
}
\label{fig:comparison}
\end{figure*}

We directly compare our PLEC2 results converted from the PLEC4 model to those of
\citet{abdo_psr_2010}, shown in Figure~\ref{fig:comparison}. The $\Gamma_0$ and $E_c$ values differ to those of \citet{abdo_psr_2010} due to the treatment of the $b$ parameter, which is free in our model, see the top panels of Figure~\ref{fig:comparison}. 
Assuming $b = 1$, \citet{abdo_psr_2010} report phase-resolved $E_c\sim 3-4$\,GeV between phases 0.1 and 0.7, while our own $E_c$ values, calculated according to \ref{appendix:plec conversion} for the fit with free $b$ range between a few hundred\,MeV. Fixing $b=1$, we recover the results found in \citet{abdo_psr_2010}, see the bottom panels of Figure~\ref{fig:comparison} \citep[see also Figures 8 and 9 in][]{abdo_psr_2010}. We find the same soft spectra at the peaks and harder spectra in the inter-pulse phases with sharp increases of the cutoff energy $E_c$ at Peak 1, Peak 2, and the inter-pulse phases around $\phi\approx0.4$.

{The ``fixed-b'' and ``free-b'' models provide similar \R{spectral fit} results within $\phi=[0.15, 0.65]$, but the ``free-b'' model provides a better description to the data, so we \R{choose}} the ``free-b'' model for the remainder of this work. Finally, we note that because the Vela pulsar is extremely bright and we are using 13 years of Fermi--LAT data as compared to 1 year in \citet{abdo_psr_2010}, we are able to better constrain the spectral parameters.

\section{Multi-wavelength Analysis}
\label{sec:seds}
\subsection{SEDs of Peak 1 and Peak 2}
We produce multi-wavelength SEDs for Peak 1 and Peak 2 using archival far-UV (FUV), X-ray, soft $\gamma-$ray, and TeV flux measurements based on previously published results, comparing them to the Fermi--LAT PLEC4 ``free-b'' spectral model and data.
For these fits, we choose the phase ranges that match those of recent the Nuclear Spectroscopic Telescope Array (NuSTAR) observations \citep{karg2023}.
The best-fit PLEC4 models for Peak 1 ($0.16 < \phi_{\mathrm{PK1}} < 0.20$) and Peak 2 ($0.50 < \phi_{\mathrm{PK2}}< 0.66$) and their uncertainties are extrapolated  to cover the energy range between 1\,eV to 100\,TeV. Far-UV data is from the Hubble Space Telescope \citep[HST,][]{romani_vela_2005}, hard X-ray data from NuSTAR \citep{karg2023} and the Rossi X-ray Timing Explorer \citep[RXTE,][]{harding_rxte_1999}, soft $\gamma-$ray data from the Oriented Scintillation Spectrometer Experiment \citep[OSSE,][]{strickman_osse_1996} and the 
Compton Telescope \citep[COMPTEL,][]{hermsen_comptel_1993}, as well as TeV data from the High Energy Stereoscopic System \citep[H.E.S.S.,][]{hess_collaboration_first_2018,aharonian_discovery_2023}. A discussion on the physical implications are discussed in Section~\ref{sec:discuss_psr}.

\begin{figure*}[!hpt]
\begin{minipage}[b]{.5\linewidth}
\centering
\includegraphics[width=1.0\linewidth]{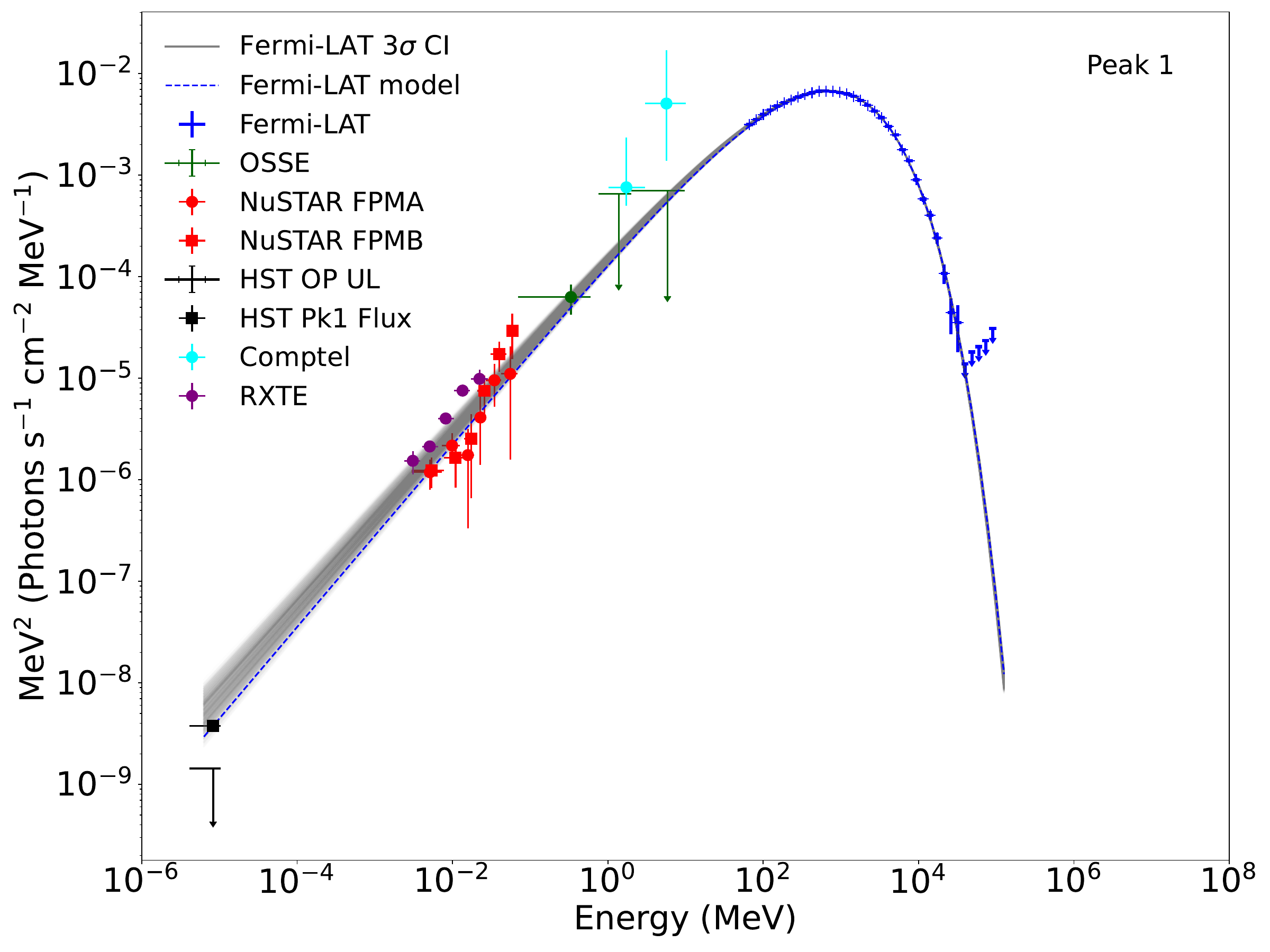}
\end{minipage}
\begin{minipage}[b]{.5\linewidth}
\centering
\includegraphics[width=1.0\linewidth]{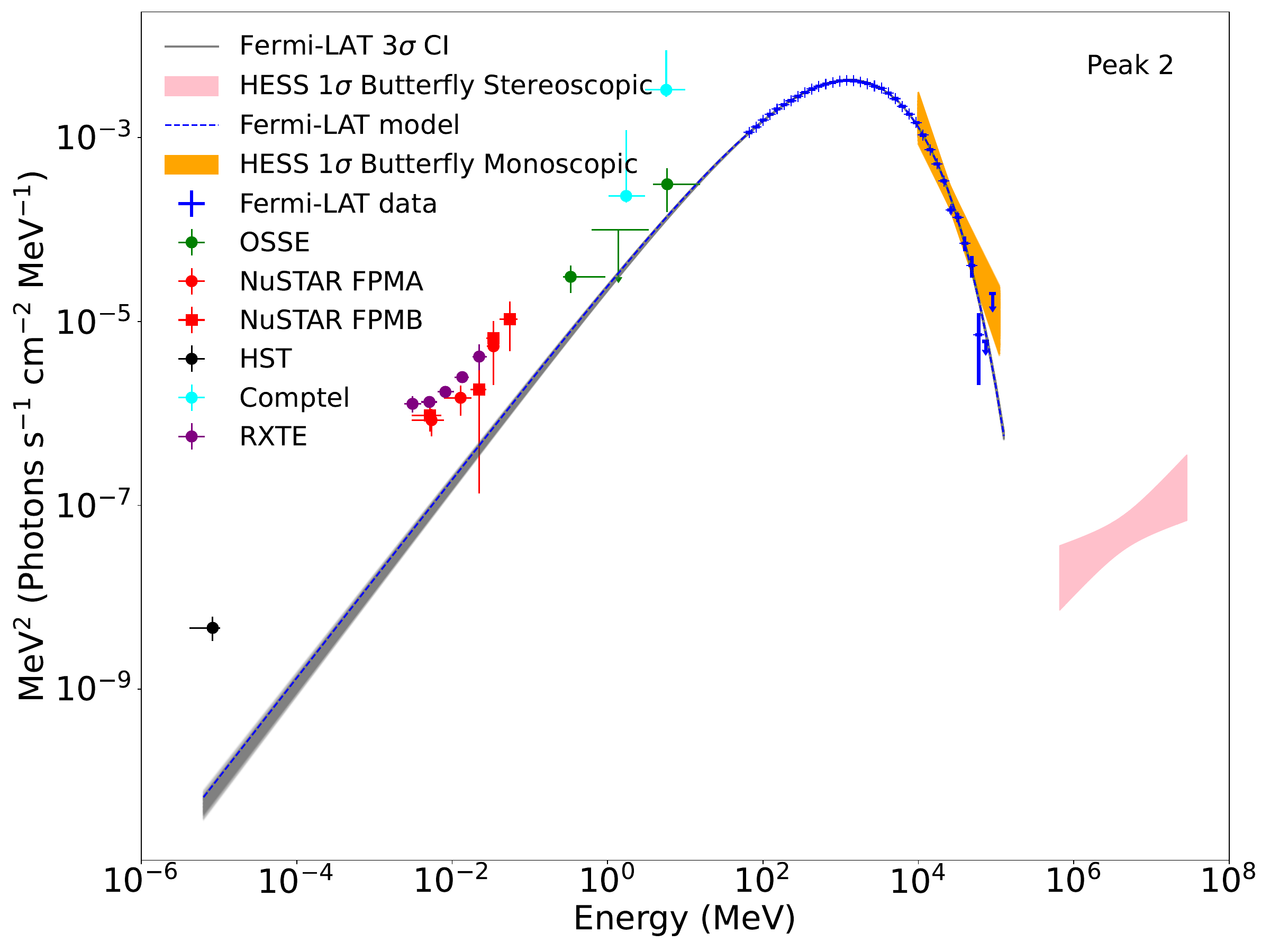}
\end{minipage}
\caption{Broadband SEDs of Peaks 1 and 2. Upper-limits are denoted by one-sided error bars. Fermi--LAT data along with the best-fit spectral model is shown in blue with the 95\% confidence interval in grey. {\bf Left}: Broadband SED of Peak 1 ($0.15<\phi_{\mathrm{PK1}}<0.19$). HST (including the FUV Peak 1 and the upper-limit from the UV off-pulse (OP, black)), NuSTAR (red squares and diamonds; FPMA and FPMB), RXTE (purple), OSSE (green) and COMPTEL (cyan) fluxes and upper-limits are plotted. {\bf Right:} Broadband SED of Peak 2 ($0.58<\phi_{\mathrm{PK2}}<0.62$). H.E.S.S. monoscopic (HE) and 
stereoscopic (VHE) spectral models with $1 \sigma$ uncertainty are also shown as orange and pink shaded bands. }

\label{fig:Pk_sed}
\end{figure*}

An important caveat here is that the fluxes that we find in the literature for the MeV, keV, and far-UV energies are not always measured in the same phase range as the Fermi--LAT spectra for Peak 1 and Peak 2. Therefore, we correct the flux measurements to match the phase width of the NuSTAR observations (with the exception of the H.E.S.S. fluxes) and \R{describe each telescope's observations and data reduction in detail below}. Figure~\ref{fig:Pk_sed} shows the resulting multi-wavelength SEDs for Peak 1 and 2.

{\sl HST:}  
A flux measurement and an upper-limit are taken from \cite{romani_vela_2005} for the \R{FUV} (central frequency $\nu = 2\times10^{15}$ Hz) band.
Since Peak 2 is greater in flux than Peak 1 in the FUV and less offset from the MeV--GeV Peak 2 phase interval, we use the flux measurement for Peak 2 and an upper limit for Peak 1. The Peak 1 FUV flux is measured from a phase range of $0.21 < \phi_{\mathrm{PK1}} < 0.35$ (``PK1'', though this does not correspond to the $\gamma-$ray peak) and Peak 2 from a range $0.46 < \phi_{\mathrm{PK2}} < 0.57$ in HST data. We also include the Peak 1 flux upper-limit extracted from the $\gamma-$ray ``off-pulse'' (or ``OP'') from \citet{romani_vela_2005}, measured from a joint fit of the phase ranges: $0.06<\phi_{OP1}<0.174$, $0.91<\phi_{OP2}<1.04$ and $0.57<\phi_{OP3}<0.86$.

{\sl NuSTAR:}
Recent results of NuSTAR 3--79\,keV observations are reported in \citet{karg2023},
providing fluxes for Peaks 1 and 2 in the phase ranges $0.16 < \phi_{\mathrm{PK1}} < 0.20$ and $0.50 < \phi_{\mathrm{PK2}} < 0.66$, respectively. 

{\sl RXTE}: Archival 2--60\,keV RXTE  \citep{strickman_rossi_1999,harding_rxte_1999,harding_multi-component_2002} flux measurements\footnote{These phase-resolved spectral points were created ``using \\ the straightforward `on-pulse' minus `off-pulse' technique, \\ then time-normalized to the average over the entire light \\ curve'' \citep{strickman_osse_1996}.} from Peak 1 and Peak 2 are defined between $0.100 < \phi_{\mathrm{PK1}} < 0.135$ and $0.495 < \phi_{\mathrm{PK2}} < 0.605$, respectively. To correct for these ``time-normalized'' fluxes, we divide by the reported phase widths of 0.035 and 0.11 for Peak 1 and Peak 2, returning similar values as the NuSTAR fluxes reported in \citet{karg2023}.

{\sl OSSE}:
Three soft $\gamma-$ray bands are analyzed in \citet{strickman_osse_1996}: 0.07--0.6\,MeV, 0.76--2.0\,MeV and 2.0--9.7\,MeV. 
Peak 1 data are measured within the phase range $0.03 < \phi_{\mathrm{PK1}} < 0.19$ and Peak 2 data are measured within $0.33 < \phi_{\mathrm{PK2}} < 0.77$. 
Only an upper limit is measured in the 0.76--2.0\,MeV and 2.0--9.7\,MeV bands for Peak 1. An upper limit is used for Peak 2 in the 0.76--2.0\,MeV band. The reported OSSE fluxes are normalized similar to RXTE (see the above footnote for RXTE data), so we divide by the reported phase widths in \citet{strickman_osse_1996} for Peak 1 and Peak 2.

{\sl COMPTEL:}  
Phase-resolved flux measurements for the energy range 0.\RR{7}--30\,MeV are extracted
from Figure 5 in \citet{hermsen_comptel_1993}, corresponding to the following phase ranges: $0.95 < \phi_{\mathrm{PK1}} < 0.025$ and $0.375 <\phi_{\mathrm{PK2}} < 0.475$. \citet{hermsen_comptel_1993} use the ``time-averaged'' OSSE spectrum but do not explicitly state if the COMPTEL spectrum is normalized in a similar fashion. Assuming this is the case, we apply a similar normalization method to the COMPTEL fluxes as we do to the RXTE and OSSE fluxes, dividing by the reported phase ranges.  

{\sl H.E.S.S.:}  
Flux measurements from high-energy \R{(HE, }10--110\,GeV) and very high-energy \R{(VHE, }260\,GeV--28.5\,TeV) ranges were reported by \citet{hess_collaboration_first_2018} and \citet{aharonian_discovery_2023}. Peak 1 is not \R{detected} by H.E.S.S., therefore only Peak 2 is analyzed. The phase interval for the high-energy (HE) Peak 2 data ranges between $0.565< \phi_{\mathrm{PK2}} < 0.571$, while the very-high energy (VHE) phase range is $0.55 < \phi_{\mathrm{PK2}} <  0.60$. Here, we do not correct normalizations due to the difference in phase ranges following \citet{aharonian_discovery_2023}. 
\subsection{Multi-wavelength Timing Analysis} 
\label{sec:MW Timing}
In addition to Fermi--LAT events, we use Neutron Star Interior Composition Explorer Mission \citep[NICER,][]{Gendreau2016}, NuSTAR and Fermi\R{ Gamma-ray Burst Monitor (GBM)} to probe the soft and hard X-ray/soft $\gamma$-ray regime in Section~\ref{sec:MW pulse profiles} \R{and investigate the energy dependence of the pulse profile with this new timing data} \R{The reduction of these events are described below}. 

{\sl NICER:}
Low-energy X-ray (0.2--15\,keV) observations from  of Vela have been taken since its launch in 2017 for a total of 715\,ks (cleaned exposure) covering the time period July 7, 2017 01:23:59.00 (UTC) -- April 26, 2024 04:12:59.00 (UTC) (ObsIDs: 0020180102-7020180411). Event files are cleaned using the standard \texttt{NICERDASv012} after an additional particle screening based on the 12--15\,keV event rate (where events are ignored if the rate exceeds 0.5\,cts/s). The resulting events are then barycentered and binned into 120 equal width phase bins. We create two different pulse profiles  between 0.3--0.8\,keV and 0.8--2.0\,keV to demonstrate the phase shift between peaks due to thermal and non-thermal emission (as seen in \citet{kuiper_soft_2015}).

{\sl NuSTAR:}
\RR{A} (200\,ks) \RR{NuSTAR} observation from 2019  of the Vela pulsar and its PWN as described in \citealt{karg2023} \RR{was used to update the 20--50\,keV X-ray pulse profile}. X-ray events are binned into 30 equal width phase bins folded onto the ephemeris used in Section~\ref{sec:fermi}.

{\sl Fermi--GBM:}
Fermi--GBM's NaI detectors, contemporaneous with Fermi--LAT, provide soft $\gamma$-ray coverage from 10\,keV--2\,MeV. About 5.5 years of data have been collected on Vela from 2019 January to 2024 June. We use the same screening procedure outlined in Sections~2 and 3 of \citet{2018MNRAS.475.1238K}, to filter and barycenter the events. With time-tagged events, we generate two pulse profiles with 36 equal-width phase bins \R{between} 50--150\,keV and 150--750\,keV.
\section{Discussion}\label{sec:discuss}

\subsection{Vela-X PWN}\label{sec:discuss_pwn}

We have newly characterized the extended emission associated to the Vela--X PWN using two extended components: a diffuse Gaussian source and a compact disk source (Model 3 in Table~\ref{tab:gamma_1gev}). 
The similar spectra of both components may suggest a single (PWN) origin, but it is also possible that the smaller disk source is related to the Vela SNR shell or originates from an unknown source that is positionally coincident. Overall, the best-fit source model we find here is in agreement with previous Fermi--LAT analysis of the PWN. In the 4FGL--DR3 catalog where 50\,MeV--1\,TeV data with no phase-cut selection is used, one large extended source (4FGL~J0834.3--4542e) characterizes the Vela--X PWN, corresponding to the 330\,MHz radio template, and has a power-law spectral index $\Gamma_\gamma = 2.18 \pm 0.03$. In the $E>1\,$GeV off-pulse analysis in Section~\ref{sec:pwn}, 4FGL~J0834.3--4542e has a spectral index $\Gamma_\gamma = 2.37 \pm 0.03$ (Model 1 in Table~\ref{tab:gamma_1gev}). \RR{The larger Gaussian source for the best-fit model (Model 3 in Table~\ref{tab:gamma_1gev}) has a power-law spectral index $\Gamma_\gamma = 2.34 \pm 0.04$ and the smaller disk source has $\Gamma_\gamma = 2.33 \pm 0.08$ for $E > 1\,$GeV in the off-pulse data.} The two extended components reported in \citet{tibaldo2018}, the 330\,MHz radio source and a smaller off-set radial disk, have $E>10\,$GeV spectral index values of $\Gamma_\gamma = 2.19 \pm 0.16$ and $\Gamma_\gamma = 0.9 \pm 0.3$, respectively. \RR{The differences found in this work compared to those of \citet{tibaldo2018} can be explained by the difference in data treatment. \citet{tibaldo2018}} do not make phase-cut selections to the Fermi--LAT data, \RR{but use events with energy above 10\,GeV. Furthermore, the source models in \citet{tibaldo2018} are constructed from the 2FHL and 3FHL catalogs \citep{2fhl,3fhl}.} 

\RRR{The larger Gaussian source has an indication for two spectral components (one soft \RR{with $\Gamma \sim 2.3$} and one hard \RR{with $\Gamma \sim 1.7$}) that connects to the TeV data, see the right panel of Figure~\ref{fig:gamma_sed}. The combined spectral shape of the Fermi--LAT GeV and HESS TeV data points to the presence of two electron populations or multiple emission components from the same electron population. The larger Gaussian source appears to characterize the majority of both low- and high-energy GeV PWN emission, with $E>40$\,GeV emission spatially correlated to the radio, X-ray, and TeV PWN counterparts \RRR{(Figure~\ref{fig:special_tsmap})}. FGES~J0832.0--4549 closely aligns to the $E>40$\,GeV emission from the Gaussian source (Figure~\ref{fig:special_tsmap}) which also matches the location and approximate size of the high-energy disk reported in \citet{tibaldo2018}. }

\RR{The smaller disk source is soft in nature \RR{with $\Gamma \sim 2.3$ and peaking below 10\,GeV}, which may indicate the presence of another electron population. Unlike the radial disk in \citet{tibaldo2018}, the compact disk source here does not appear to spatially correlate to the radio, X-ray, or the TeV PWN emission. The smaller disk source we find \RR{may correspond} to the point source reported in \citet{tibaldo2018}, see Figure~\ref{fig:special_tsmap}. It is also possible that the coincident point source, 4FGL~J0830.5--4451, is the counterpart to the point source in \citet{tibaldo2018}. In either case, the relation to the PWN is not confirmed. The smaller disk source also corresponds to FGES~J0830.3--4453.}

There is a new extended source reported in the second Fermi--LAT Galactic Extended Source Catalog \citep[2FGES,][]{2fges2024}, 2FGES~J0837.7--4534 that has a 2.85\,$^\circ$ radius which encompasses the entire Vela PWN region and overlaps with the nearby Vela Jr. SNR. The nature of this source is unclear and is considered a dubious detection in the 2FGES \citep{2fges2024}.

If both extended sources \RR{of the presented best-fit model (Model 3)} originate from the Vela--X PWN, this may indicate the presence of multiple particle populations contributing to the Fermi--LAT and TeV $\gamma-$ray emission, \RR{similar to the scenarios presented in \citet{dejager2008,hinton2011,grondin2013,abdalla_hess_2019}}. A \RR{plausible} scenario is one where the diffuse GeV and 330\,MHz radio emission is from an older population of electrons and the 843\,MHz, X-ray, and TeV emission is from a younger population of electrons. \RR{This is supported by the morphology of the high-energy component, visible in Figure~\ref{fig:special_tsmap}, concentrating near the 843\,MHz counterpart and the TeV emission peak, which is also the location of the X-ray cocoon.} How the off-set, compact radial disk source may be related to the PWN is not clear given the soft spectral nature and location. It is possible the PWN is interacting with SN ejecta and/or the interstellar medium in the region of the small disk source, initiated by the passage of the SNR reverse shock. In this scenario, one could expect thermal X-ray emission at the location of the small disk source. Indeed thermal X-rays in the northern region of the PWN have been detected \citep{lamassa2008,slane_investigating_2018}. The thermal X-ray emission in the northern region, recently analyzed in eROSITA observations, may be associated to the jet of the Vela pulsar \citep{erosita2023}. A thermal component arising from an anisotropic collision of the reverse shock with ejecta-rich material is also possible \citep{erosita2023}. A multi-wavelength investigation considering the X-ray and GeV emission may be able to rule out other possibilities, such as a coincident SNR shell interaction with local material.

\subsection{Vela pulsar}\label{sec:discuss_psr}

We have performed a phase-resolved analysis of the Vela pulsar Fermi--LAT emission using 13\,years of data, a substantial increase \RR{of Fermi--LAT}  data compared to \citet{abdo_psr_2010} and \citet{abdalla_hess_2019}\RR{, an increase of approximately 8 and 1 years, respectively}. Here, we make use of the new PLEC4 (\ref{appendix:PLEC4}) spectral model\RR{, over the PLEC2 (\ref{appendix:PLEC2}) spectral model,} 
\RR{which reduces} covariance issues between spectral parameters \citep{4fgldr3,3pc}. \R{Below we compare the PLEC4 peak energy, $E_p$, the spectral curvature, $d_p$ at $E_p$, and the asymptotic photon index $\Gamma_{0}$ (Equations~\ref{appendix:Ep}, \ref{appendix:dp}, and \ref{appendix:g100}) to those of the 3PC.}

\begin{figure}
\centering
\includegraphics[width=1.0\linewidth]{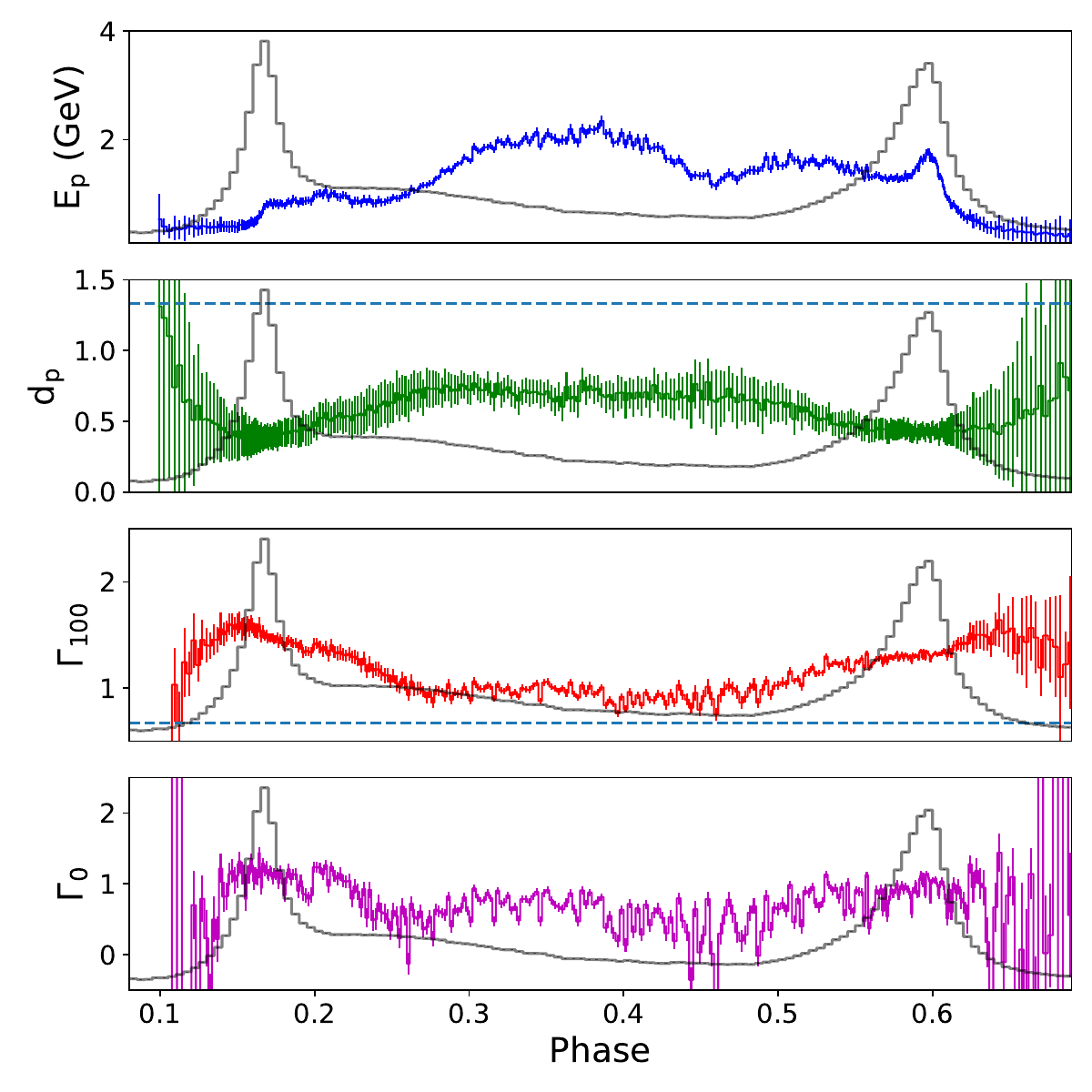}
\caption{Physical parameter evolution as a function of phase for the peak energy, $E_p$, the peak width, $d_p$, the spectral index at 100\,MeV, $\Gamma_{100}$, and the asymptotic spectral index, $\Gamma_0$. Overlaid on each panel is the pulse profile from phases 0.08 to 0.69. The middle two panels show the mono-energetic upper and lower limit, respectively, for synchrotron or curvature radiation as a blue dashed horizontal line.}
\label{fig:Physical}
\end{figure}

\begin{figure}
    \centering
    \includegraphics[width=1.0\linewidth]{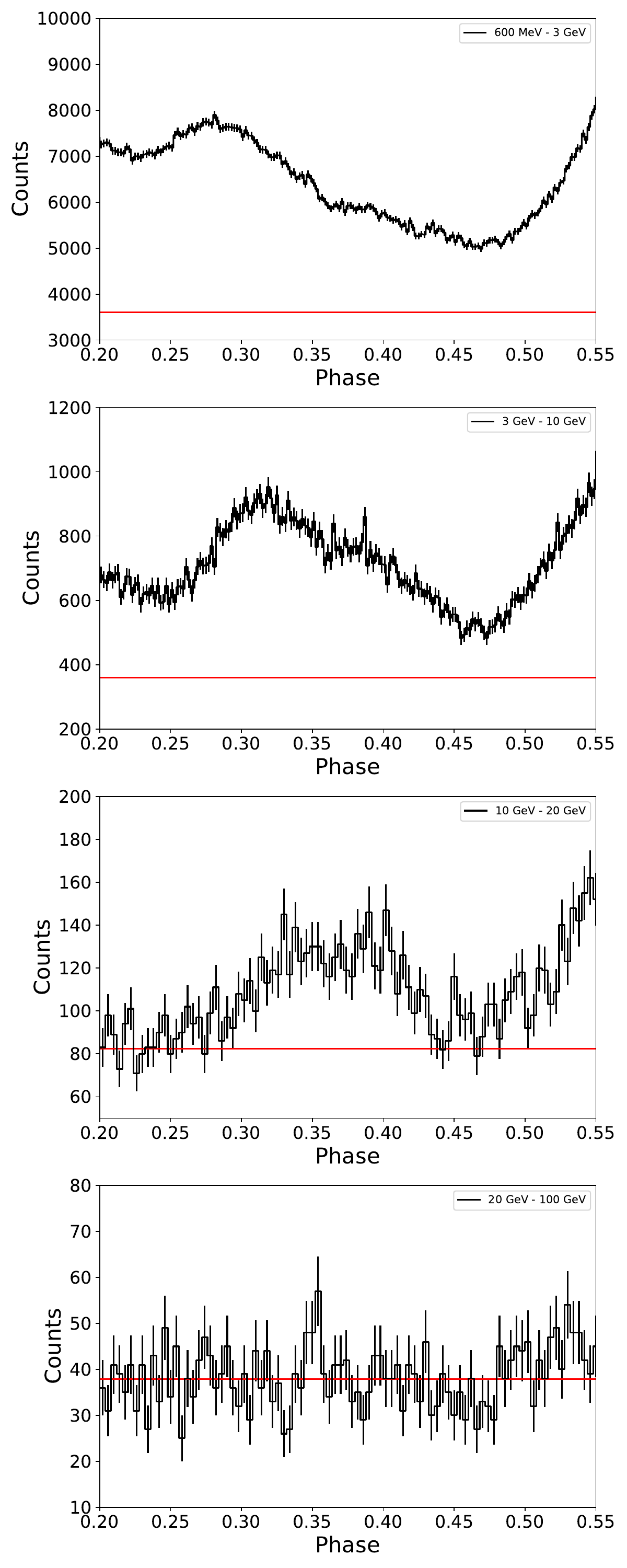}
    \caption{Energy-resolved inter-pulse structure. Energy ranges from top to bottom are: 600\,MeV--3\,GeV, 3--10\,GeV, 10--20\,GeV, 20--100\,GeV. The red horizontal line in each panel represents the average count flux during the ``off-pulse'' phases.}
    \label{fig:interpulse}
\end{figure}

In the PLEC4 model, $E_p$ is the SED's peak energy while $d_p$ and $\Gamma_{100}$ characterize the deviation from the mono-energetic spectrum (corresponding to the mono-energetic spectral curvature $d_p=4/3$ for either SR or CR) and the slope of the spectrum at 100\,MeV assuming a simple power-law (PL) model, respectively. Figure~\ref{fig:Physical} shows the evolution of $E_p$, $d_p$, and $\Gamma_{100}$ with respect to pulse phase. The peak energy is rather low  ($E_p\approx 800$\,MeV) before Peak 1, sharply increasing at Peak 1, and steadily increasing to its maximum ($E_p\approx 2$ GeV) at the phase $\phi \approx0.35$ before decreasing slightly ($E_p\approx 1.5$ GeV) during most of Peak 2. Typical phase-averaged $E_p$ values for young pulsars with a spin-down luminosity $\dot{E}=10^{36}$--$10^{37}$ erg s$^{-1}$  are in the range of $E_p=800-1500$\,MeV \citep[see Figure 19 in][]{3pc}. We note that the 3--20\,GeV inter-pulse ``bridge'' coincides with the phase range of the $E_p$ bridge ($\phi=0.25-0.4$, see Figure~\ref{fig:pulse_profiles}). A close up view of the inter-pulse structure is shown in Figure~\ref{fig:interpulse}. The peak width $d_p=0.6\sim0.7$  remains steady between the two peaks, decreasing to 
$d_p\approx0.4$ at both peaks.

The asymptotic power-law index, $\Gamma_0$, is notably systematically smaller than the PL index at 100\,MeV ($\Gamma_{100}$).  
It varies from $\sim$ 1.5
$\sim$ 1.2 at Peak 1 to $\approx 0.5$ during the inter-pulse (around $\phi\approx0.4-0.5$)
minimum (albeit showing large fluctuations) and increases to $\sim$ 1.0 at Peak 2. $d_p$ is significantly smaller than the maximum value of 4/3 for the mono-energetic SED, suggesting that the energies of electrons span a large range, especially during the Peak 1 and Peak 2 phases (see also below). The curvature radiation emission scenario likely results in a fairly narrow electron SED \citep[e.g.,][]{aharonian_discovery_2023}, especially in the strongly damped regime. Recent particle-in-cell (PIC) models simulating pulsar emission from first principles \citep{philippov_ab_2014,kalapotharakos_three-dimensional_2018,philippov_ab-initio_2018} allow for the possibility of producing the bulk (or significant portion) of the pulsed emission in the current sheet (outside the so-called Y-point, or the tip between the open and closed field lines) via \R{SR} \citep{cerutti_modeling_2016}, given that the synchrotron energy loss rate significantly exceeds the curvature loss rate for plausible electron Lorentz factors \citep{aharonian_discovery_2023}. Therefore, the resulting photon spectrum can be rather wide if the SED of electrons accelerated in the magnetically reconnecting regions is broad enough \citep[e.g.,][]{cerutti_dissipation_2020}. The lack of a noticeable break due to cooling at lower energies would imply that emission occurs in a fairly low magnetic field. The broad synchrotron spectrum can also form if the moving bulk of electrons, with a narrow SED, is radiating in regions with a strongly varying magnetic field superimposed along the same line of sight. It remains to be explored through modeling whether the synchrotron emission spectrum resembling a power-law with nearly the same slope can be formed in this case. Alternatively, \citet{harding_pulsar_2021} pointed out that in the CR (synchrocurvature) scenario, the broadness of the spectrum can be due to emission occurring over a range of radii near the current sheet where the curvature radius varies. While it is certainly possible to find SR/CR models that may explain the derived characteristics, it is not prudent to declare the emission mechanism until more robust broadband modeling is done.

We also compare our phase-resolved results to previous work \citep{abdo_psr_2010,3pc}. The best-fit $b=0.394$ value found here for the Vela pulsar is smaller than the 3PC or 4FGL--DR3 values, but is generally in good agreement with the LAT pulsar population having spin-down luminosities $\dot{E}=10^{36}$--$10^{37}$ erg s$^{-1}$ \citep[see Figure 16 in][]{3pc}.  We find a reasonably good agreement with our work and those of \citet{abdo_psr_2010} after fixing $b = 1$ (see bottom panels in  Figure~\ref{fig:comparison}). We note that the conversion equations (Appendix \ref{appendix:plec conversion}) are derived under the assumption of the same $b$ values for PLEC2 and PLEC4, hence a discrepancy can be expected if the $b$ values differ substantially.  \R{In general, $E_p\sim E_c$ when $b^2\approx d_p$ and $b>0$, but as shown in Figure~\ref{fig:comparison}
(right) and Figure~\ref{fig:Physical}, this is not the case for the Vela pulsar and is due to the difference between the \RR{parametrization} of $E_p$ and $E_c$ (i.e., $E_p$ does not fully encapsulate the shape of the SED without also considering $d_p$, $\Gamma_{100}$ and $b$.)}

Regarding the phase-resolved multi-wavelength SEDs, we find that, for Peak 1, the extrapolation  of the model fitted to LAT data matches the data at lower energies well down to the far-UV upper-limit (left panel; Figure~\ref{fig:Pk_sed}). This is not the case for Peak 2. We note that for Peak 2 (right panel; Figure~\ref{fig:Pk_sed}), in the far-UV, the position of the peak is already starting to shift in phase to the left compared to its position in the $\gamma-$ray pulse profile. Indeed, the phase ranges of Peak 2 in the far-UV and $\gamma-$ray are only slightly overlapping\footnote{This is reflected in the increased size of the far-UV \\ spectral flux error bar, which is significantly larger than \\ the uncertainty listed in \citet{romani_vela_2005}, because \\ the plotted upper bound corresponds to 68\% uncertainty \\ from $4.638\times10^{-9} \,{\rm MeV}\,{\rm cm}^{-2}\,{\rm s}^{-1}$. However, the lower \\ bound represents our estimate of the flux in the part of \\ far-UV Peak 2 overlapping with the phase interval \\ defined for Pk2 based on the LAT pulse profile.}. Not only is the Peak 1 far-UV flux measurement and upper-limit very close to the extrapolation of the LAT-fitted model but also that the slope of far-UV spectrum $\Gamma_{FUV, Pk1}=1.22\pm 0.13$ is compatible with $\Gamma_{0, {\rm Pk1}}=1.13\pm 0.04$. 

\begin{figure*}[!hp]
    \centering
    \includegraphics[scale=.22]{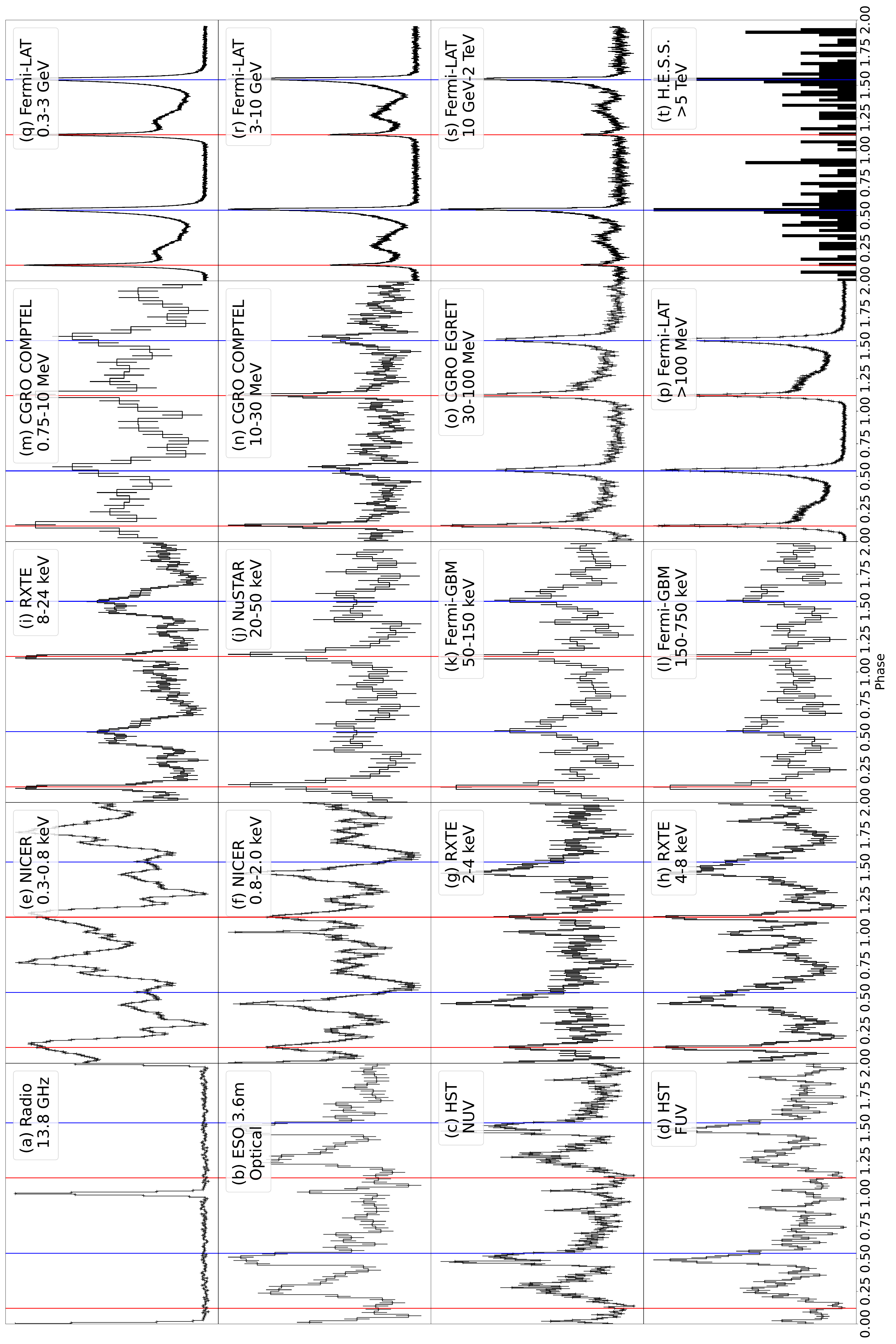}
    \caption{Pulse profiles across the multi-wavelength spectrum adapted from \citet{kuiper_soft_2015} (panels \emph{a}-\emph{d},\emph{g}-\emph{i},\emph{m}-\emph{p}), \citet{hess_collaboration_first_2018} (panel \emph{t}), and \citet{karg2023} (panel \emph{j}). We include the new Fermi--LAT (panels \emph{q}-\emph{s}; this work, see Section~\ref{sec:pulsar}), as well as pulse profiles from NICER (panels \emph{e}-\emph{f}) and Fermi--GBM-NaI (panels \emph{k}-\emph{l}) data (see Section~\ref{sec:MW pulse profiles}). Each subplot is labeled with the corresponding energy band. Red and blue vertical lines indicate $\gamma$-ray Peak 1 and Peak 2, respectively. Panel (\emph{e}) is considered to be the result of thermal emission \citep{kuiper_soft_2015}. Pulse profiles are extended to two cycles.}
    \label{fig:MW}
\end{figure*}

The extrapolation of the model fitted to the LAT spectrum of Peak 2 shows a significant deviation from X-ray data (which may indicate the appearance of another component at lower energies), while the Peak 1 spectrum shows much stronger agreement. In the far-UV, the phase of Peak 1 shifts more significantly from its $\gamma-$ray (and X-ray) counterpart when compared to Peak 2 (see Figure~\ref{fig:MW}), and the slope of the far-UV spectrum $\Gamma_{FUV, Pk1}=1.37 \pm 0.21$ is somewhat softer than $\Gamma_{0, {\rm Pk1}}=1.21\pm 0.02$. Therefore, it is not surprising that for the phase-integrated spectrum (which also includes the inter-pulse emission), the $\gamma-$ray model extrapolation may not agree with the spectrum measured at lower energies. Indeed, \citet{shibanov_vela_2003} and  \citet{mignani_first_2017} report a very different spectral index, $\alpha_\nu=0.1$ (near-IR, 1--5$\mu$m) and $\alpha_\nu = -0.93$ (sub-millimeter, 97.5\,GHz--343.5\,GHz) which only corresponds to LAT model extrapolation for Peak 1.
 
\subsection{Multi-wavelength Pulse Profiles} 
\label{sec:MW pulse profiles}
To provide a more complete MW picture, we provide an expanded updated comparison of MW pulsar profiles (cf.\ Figure 4 in \citealt{kuiper_soft_2015}). 

One important caveat of the combined pulse-profiles is that the previously published pulse profiles from \citet{kuiper_soft_2015} (13.8\,GHz, ESO 3.6m, HST FUV/NUV, RXTE PCA, CGRO COMPTEL, CGRO EGRET and Fermi--LAT $>100$ MeV) are created with phase 0 at the radio peak, based on the ephemeris introduced in \citet{kanbach_egret_1994}, while the updated Fermi--LAT (panels (\emph{q})-(\emph{s})), NuSTAR pulse profiles were generated using the Fermi--LAT ephemeris introduced in Section~\ref{sec:fermi}. The H.E.S.S. pulse profile is based on an ephemeris from the Parkes Radio Telescope \citep{abdalla_hess_2019}. Due to a difference in ephemerides, there was a noticeable shift ($\sim0.04$) between the archival Fermi--LAT (panel (\emph{p})) and updated Fermi--LAT (panel (\emph{q})) pulse profiles in Figure~\ref{fig:MW}. To account for this, we shift the pulse profiles that used the Fermi--LAT ephemeris (Figure~\ref{fig:MW}, panels: (\emph{j}), (\emph{q}-\emph{t})) by $\sim0.04$ to the left to match Peak 1 to the archival pulse profiles (Figure~\ref{fig:MW}, panels: (\emph{a}-\emph{i}), (\emph{k}-\emph{p})).

Overall, we find that there are peaks in the optical, nUV/fUV and soft X-rays that align with the radio peak. As we increase in energy, we note that peak 2 seems to shift closer to Peak 1 and the radio peak \citep{kanbach_egret_1994}. We attribute panels (e) to be the result of thermal emission from the surface of the NS. We see a two broad peaks throughout Figure~\ref{fig:MW} that are roughly aligned with our definition of Peak 1 and Peak 2. These peaks are consistent throughout the X-ray, $\gamma-$ray and GeV/TeV energies.    

It is conceivable that the same particles that produce the GeV pulse could produce the far-UV Peak 2 (panel \emph{d}) at a slightly different geometrical site in the magnetosphere, if these particles have moved to a region with a lower magnetic field (creating the far-UV pulse) from a region closer to the 
NS with a substantially higher magnetic field (producing the MeV--GeV emission) as the pulsar rotates. 

In the absence of multi-wavelength modeling (which is beyond the scope of this paper), we admit that this difference in phase between the GeV and UV peak could merely be a curious coincidence.  Additionally, as shown in Figure~\ref{fig:MW} (panels \emph{p}-\emph{t}), the disappearance of Peak 2 in the soft $\gamma-$ray band could be explained if the emission site's magnetic field places the characteristic synchrotron frequency in the hard X-ray band and is beamed away from the observer. Alternatively, the Peak 2 emission could be overtaken by a different spectral component dominating at hard X-rays (panels \emph{i}-\emph{n}) and lower energies. 

We note that Peak 1 exhibits a much more noticeable change in its position in far-UV compared to the GeV position ($\Delta\phi\approx0.15$ to the right, see panels (\emph{c}-\emph{d}) and (\emph{m}-\emph{s}), Figure~\ref{fig:MW}) and becomes noticeably smaller and broader than Peak 2. For this reason, for Peak 1 in Figure~\ref{fig:Pk_sed}, we choose to plot an upper limit on the pulsed flux scaled from the same phase range as the one used for the Peak 1 Fermi--LAT spectrum. The different behavior can be attributed to the fact that in the above mentioned first-principle PIC models \citep{philippov_ab_2014,kalapotharakos_three-dimensional_2018,philippov_ab-initio_2018}, positrons are moving outwards while electrons are moving inward (toward the NS) from the acceleration region beyond the Y-point.

The inter-pulse ``bridge'' could be attributed to an entirely different site \citep[e.g., a slot gap or separatrix region, the region between open and closed magnetic field lines, within the light cylinder,][]{harding_multi-tev_2018,barnard_probing_2022}, where particles are accelerated by the unscreened component (the component of the electric field that is not ``canceled out'' by a generated current) of the electric field parallel to the magnetic field. This ``bridge'' is also coincident with the position of Peak 1 in the optical, NUV, and FUV (panels \emph{b}-\emph{d}, Figure~\ref{fig:MW}), suggesting the same site may be responsible for this multi-wavelength emission. We note that the ``bridge'' behaves quite differently from Peak 1 and Peak 2 as it shifts to larger phases within the relatively narrow energy range (0.3--20\,GeV), while Peak 1 and Peak 2 remain at the same phase within much of the broader GeV energy range (see Figure~\ref{fig:pulse_profiles}). This suggests that as the pulsar rotates, we are looking deep (close to NS surface) into the magnetosphere where the magnetic field (for \R{SR}) and the field line curvature radius (for \R{CR}) are larger and smaller, respectively, resulting in higher energy emission.  

\section{Conclusions}\label{sec:conclude}
We have characterized the Vela pulsar and its PWN, Vela--X, in the Fermi--LAT band using 13\, years of data and compare our results to prior work. The Vela--X PWN has evidence of two extended source components above 1\,GeV, similar to previous findings \citep{fges2017,tibaldo2018}, but the association of the smaller extended source to the PWN is not certain. The SNR is known to interact with local molecular material, so a projected SNR origin is possible. The similar spectral indices for both extended sources, however, may support a common PWN origin. A multi-wavelength study considering molecular material in the region of the small extended source is required to determine a more likely scenario. \RR{Conversely, a broadband model exploring the underlying particle populations generating the Fermi--LAT GeV and HESS TeV data is warranted to develop a physical interpretation of the observed emission \citep[e.g.,][]{dejager2008,hinton2011,grondin2013,abdalla_hess_2019}.}

The 60\,MeV--100\,GeV Fermi--LAT analysis of the Vela pulsar provides a deeper look into the pulsed emission, the spectral characteristics, and the physical implications of possible particle acceleration sites and emission mechanisms. \R{The spectral results agree well with the phase-integrated results from the 3PC \citep{3pc} and expand upon both the 3PC and earlier work \citep[e.g.,][]{abdo_psr_2010} with phase-resolved likelihood analysis that shows changes in the behavior of the SED. We characterize the peak energy of the SED ($E_p$), the width at the peak energy ($d_p$), and the asymptotic spectral index ($\Gamma_{100}$), see Figure~\ref{fig:Physical}, which provides a more detailed description of the SED. Particularly of interest is that $E_p$ is significantly larger at the inter-pulse emission compared to the peak emission. The inverse is true for $\Gamma_{100}$, which is significantly larger than the inter-pulse emission.} Furthermore, the $\gamma-$ray SEDs of the two pulse peaks (Peak 1 and Peak 2) at $\phi_{\mathrm{PK1}} = 0.165$ and $\phi_{\mathrm{PK2}}= 0.595$ are extended and overlaid with archival phase-resolved broadband data from the UV to TeV energies. The extrapolation of the Fermi--LAT spectrum for Peak 1 is in good agreement with broadband data, but less so for Peak 2. Additionally, there is evidence of a change in emission mechanism during the inter-pulse ``bridge'' phases that may provide geometric and physical constraints on the inner pulsar magnetosphere. The results reported here can be used to test recent theoretical models to determine the structure of the magnetosphere. 

\begin{acknowledgments}

\RR{The authors thank the referee for useful comments and feedback. }We thank Matthew Kerr for an updated ephemeris from Fermi--LAT data and Alice Harding for useful
comments regarding our analyses.
The Fermi--LAT Collaboration acknowledges generous ongoing support
from a number of agencies and institutes that have supported both the
development and the operation of the LAT as well as scientific data analysis.
These include the National Aeronautics and Space Administration and the
Department of Energy in the United States, the Commissariat \`a l'Energie Atomique
and the Centre National de la Recherche Scientifique / Institut National de Physique
Nucl\'eaire et de Physique des Particules in France, the Agenzia Spaziale Italiana
and the Istituto Nazionale di Fisica Nucleare in Italy, the Ministry of Education,
Culture, Sports, Science and Technology (MEXT), High Energy Accelerator Research
Organization (KEK) and Japan Aerospace Exploration Agency (JAXA) in Japan, and
the K.~A.~Wallenberg Foundation, the Swedish Research Council and the
Swedish National Space Board in Sweden.
 
Additional support for science analysis during the operations phase is gratefully
acknowledged from the Istituto Nazionale di Astrofisica in Italy and the Centre
National d'\'Etudes Spatiales in France. This work performed in part under DOE
Contract DE-AC02-76SF00515. \end{acknowledgments}
\clearpage

\restartappendixnumbering

\appendix

\renewcommand\theequation{A.\arabic{equation}}
\renewcommand\thefigure{B.\arabic{figure}}    
\section{PLEC2, PLEC4 and 3PC conversions}

From the previous catalogs before 4FGL--DR3, Equation \ref{appendix:PLEC2} is most often used to characterize pulsars. In 4FGL--DR3 and the 3PC catalogs, Equation \ref{appendix:PLEC4} is used. For more information on the spectral models and the motivation see \citet{4fgldr3,3pc}.

\begin{equation}
\text{PLEC2: }
\frac{d N}{d E}=N\left(\frac{E}{E_0}\right)^{\Gamma_0} \exp \left(-\left(\frac{E}{E_c}\right)^b\right)
\label{appendix:PLEC2}
\end{equation}
\begin{equation}
\begin{aligned}
\label{appendix:PLEC4}
\text{PLEC4: }\frac{d N}{d E} & =N_0\left(\frac{E}{E_0}\right)^{-\Gamma+\frac{d}{b}} \exp \left[\frac{d}{b^2}\left(1-\left(\frac{E}{E_0}\right)^b\right)\right] \\
\end{aligned}
\end{equation}

The conversions between equation \ref{appendix:PLEC2} and equation \ref{appendix:PLEC4} are defined in equations~\ref{appendix:plec conversion}. 
\begin{equation}
\begin{aligned}
\label{appendix:plec conversion}
P L E C 2 & \Longleftrightarrow P L E C 4 \\
E_c & =E_0\left(\frac{b^2}{d}\right)^{\frac{1}{b}} \\
\Gamma_0 & =\Gamma-\frac{d}{b} \\
N & =N_0 \exp \left(\frac{d}{b^2}\right)
\end{aligned}
\end{equation}

The following equations~\ref{appendix:Ep}, \ref{appendix:dp}, and \ref{appendix:g100} are from the 3PC to evaluate the peak energy, $E_p$, spectral curvature at the energy peak, $d_p$, and the asymptotic spectral index at 100\,MeV, $\Gamma_{100}$.
\begin{equation}
\label{appendix:Ep}
E_{\mathrm{p}}={E_0}\left[1+\frac{b}{d}(2-\Gamma)\right]^{\frac{1}{b}}={E_0}\left[\frac{b}{d}\left(2-\Gamma_0\right)\right]^{\frac{1}{b}}
\end{equation}

\begin{equation}
\label{appendix:dp}
d_{\mathrm{p}}=d+b(2-\Gamma)=b\left(2-\Gamma_0\right)
\end{equation}

\begin{equation}
\label{appendix:g100}
\Gamma_{100}=\Gamma-\frac{d}{b}\left(1-E_{100}^b\right)=\Gamma_0+\frac{d}{b} E_{100}^b
\end{equation}

\section{Phase-Resolved Spectra}
\begin{figure*}[!h]

    \centering
    \begin{minipage}{0.32\textwidth}
        \includegraphics[width=\linewidth]{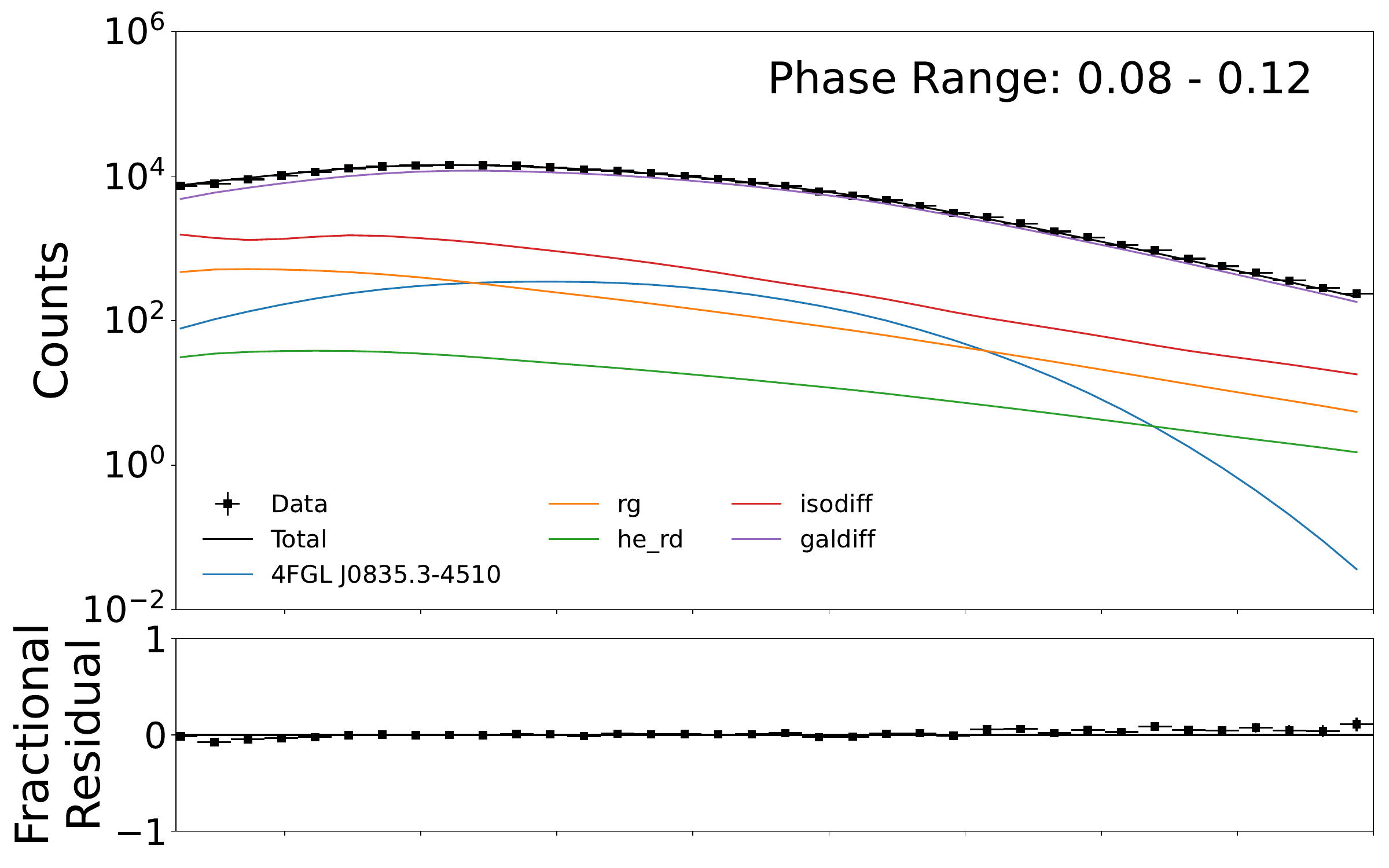}
    \end{minipage}
    \begin{minipage}{0.32\textwidth}
        \includegraphics[width=\linewidth]{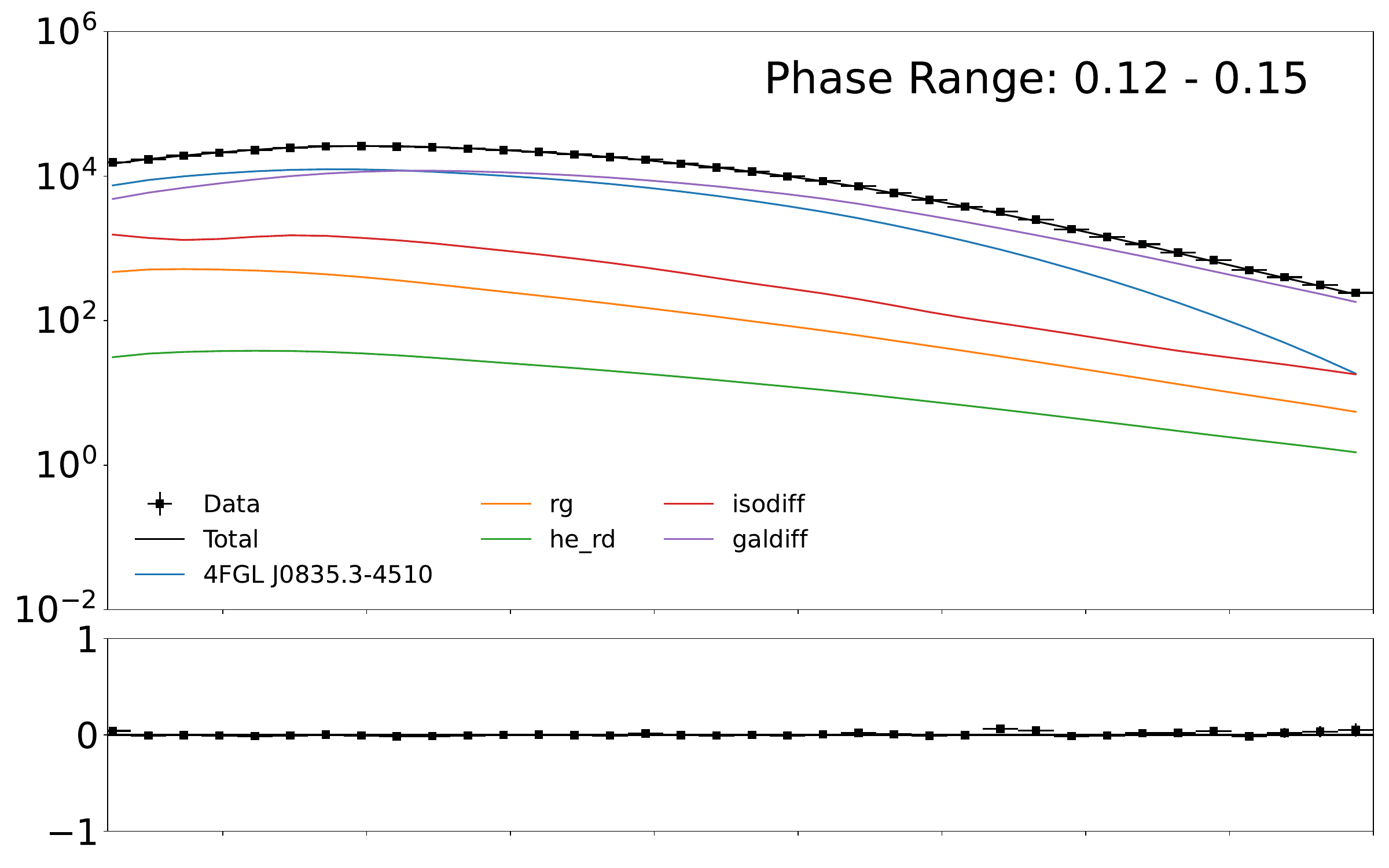}
    \end{minipage}%
    \begin{minipage}{0.32\textwidth}
        \includegraphics[width=\linewidth]{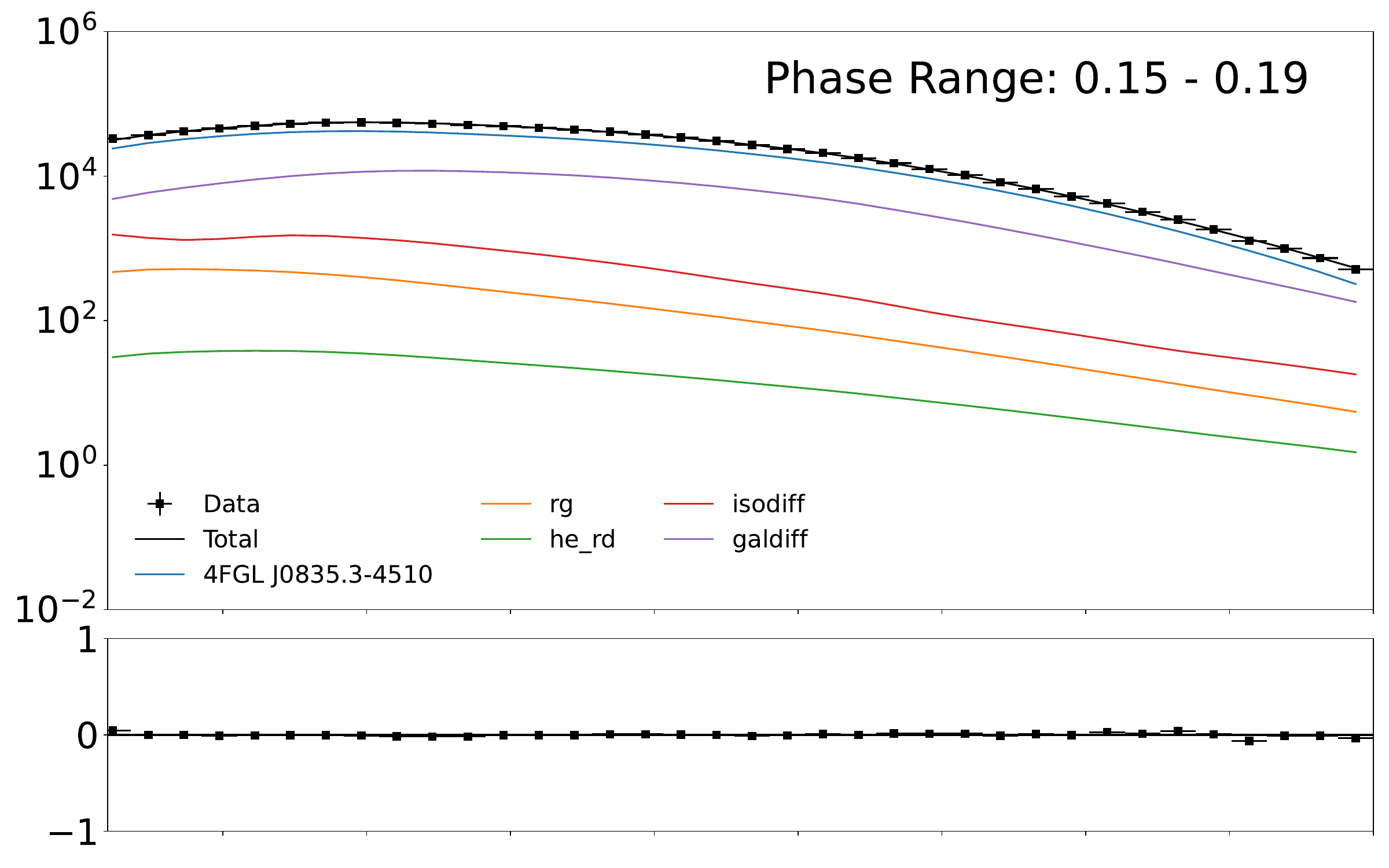}
    \end{minipage}

    \vspace{-2pt}

    \begin{minipage}{0.32\textwidth}
        \includegraphics[width=\linewidth]{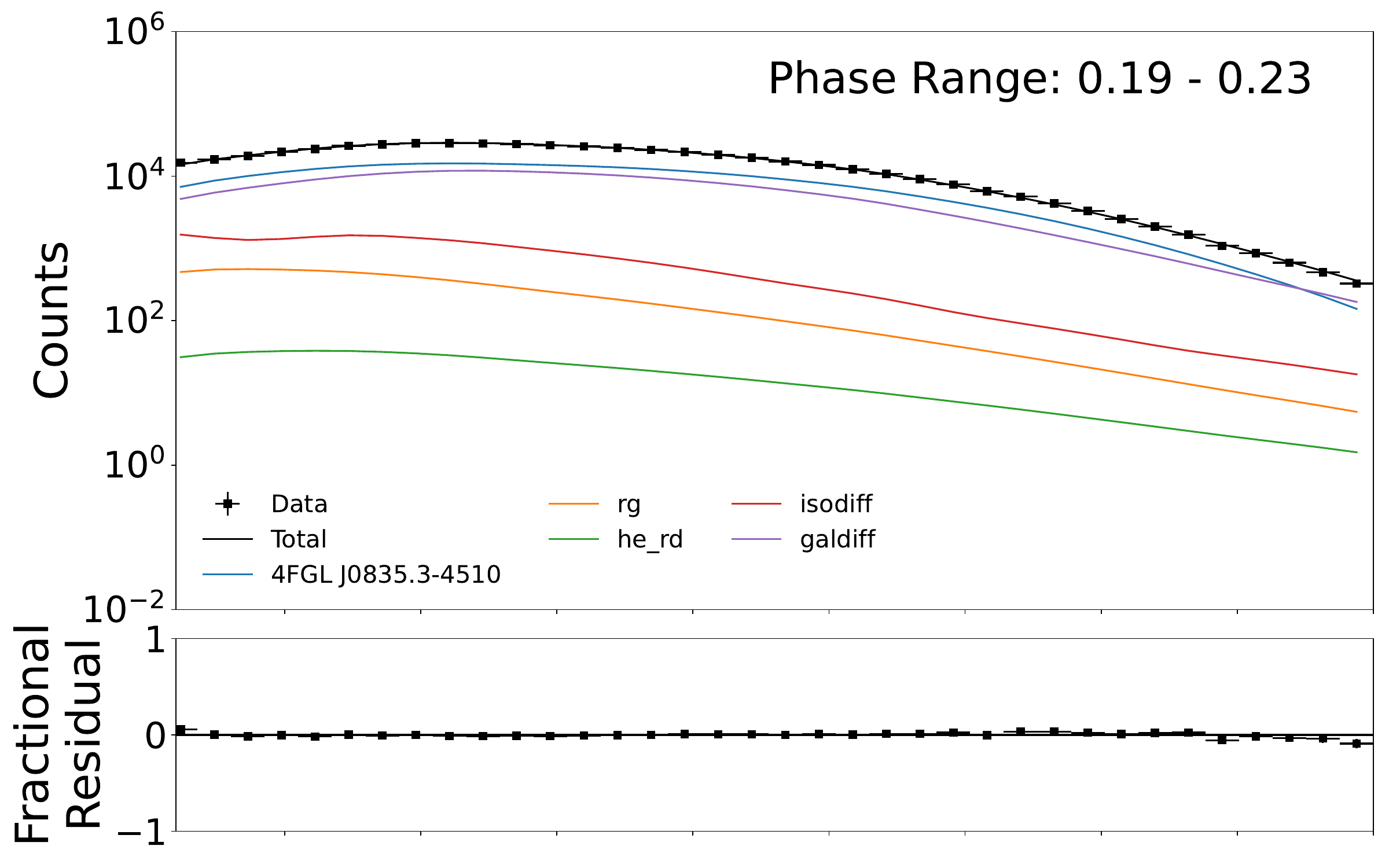}
    \end{minipage}
    \begin{minipage}{0.32\textwidth}
        \includegraphics[width=\linewidth]{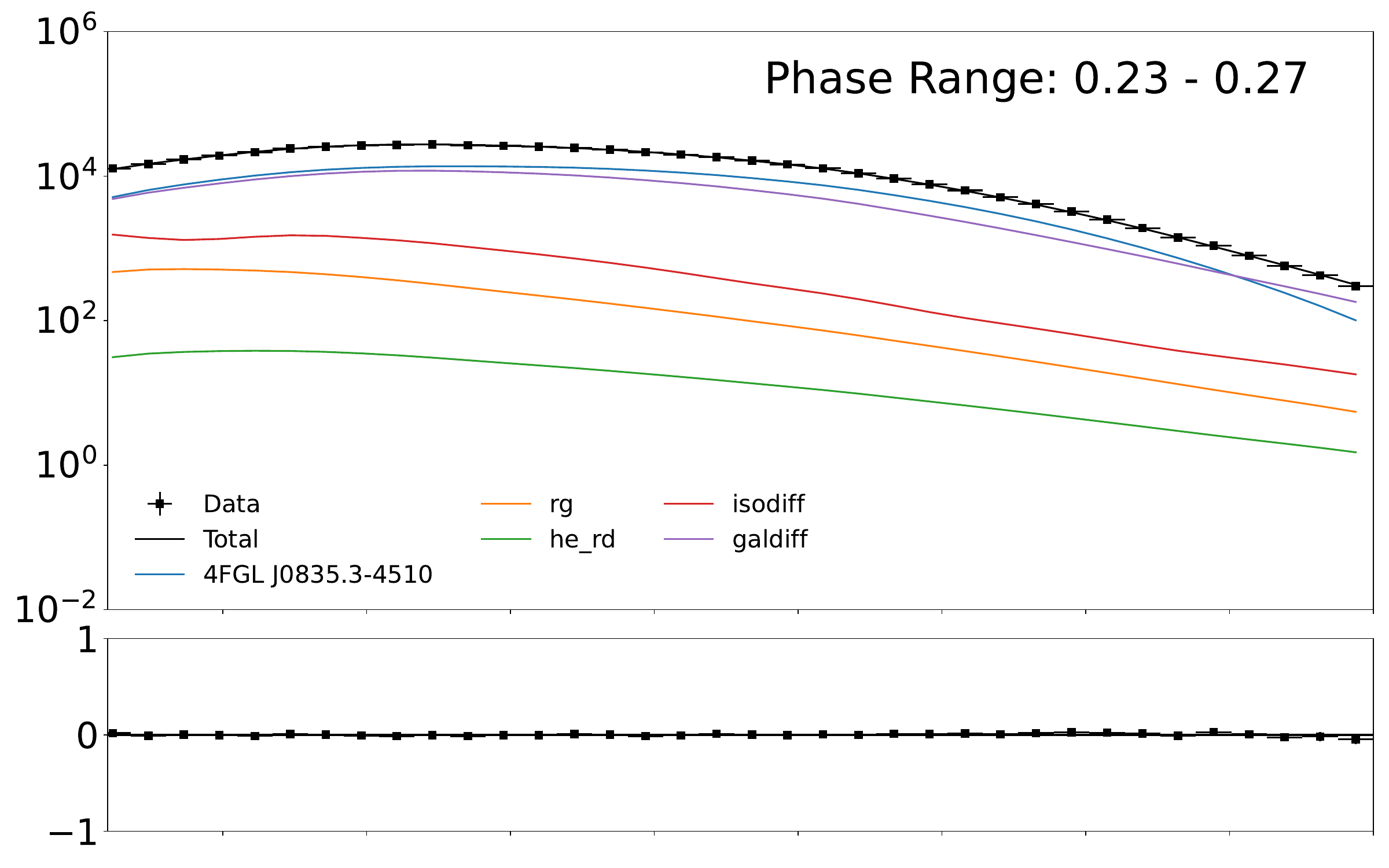}
    \end{minipage}%
    \begin{minipage}{0.32\textwidth}
        \includegraphics[width=\linewidth]{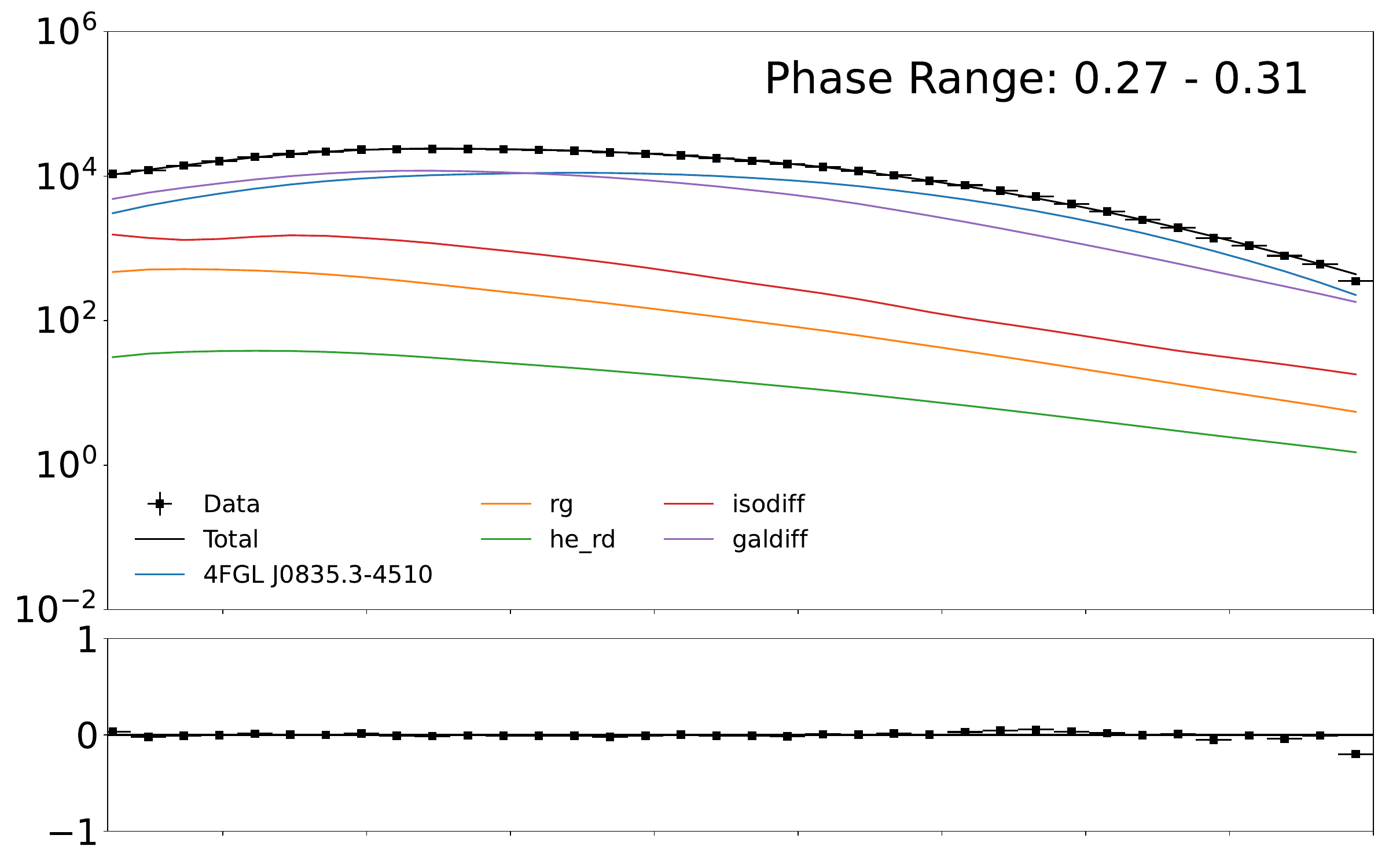}
    \end{minipage}

    \vspace{-2pt}

    \begin{minipage}{0.32\textwidth}
        \includegraphics[width=\linewidth]{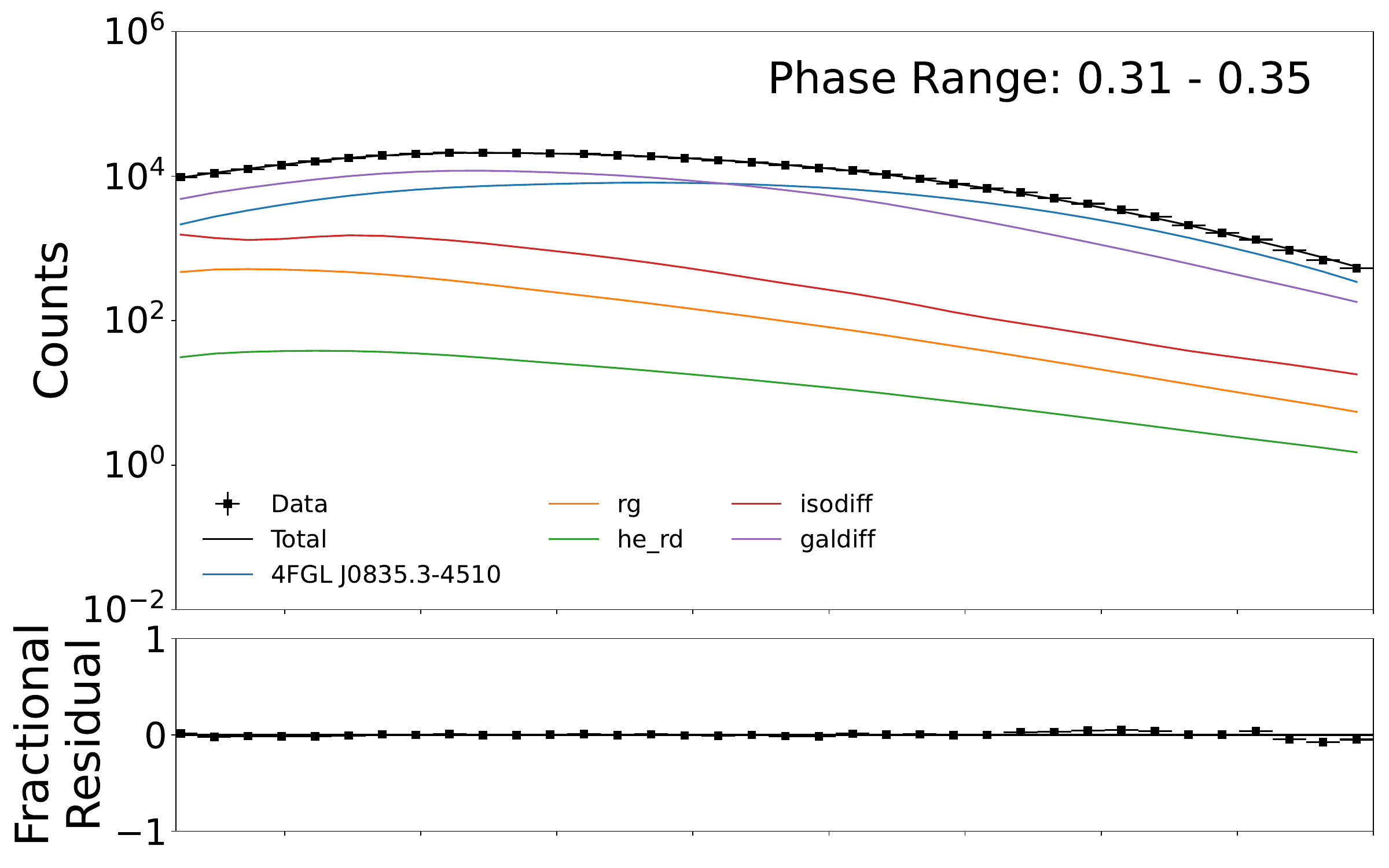}
    \end{minipage}%
    \begin{minipage}{0.32\textwidth}
        \includegraphics[width=\linewidth]{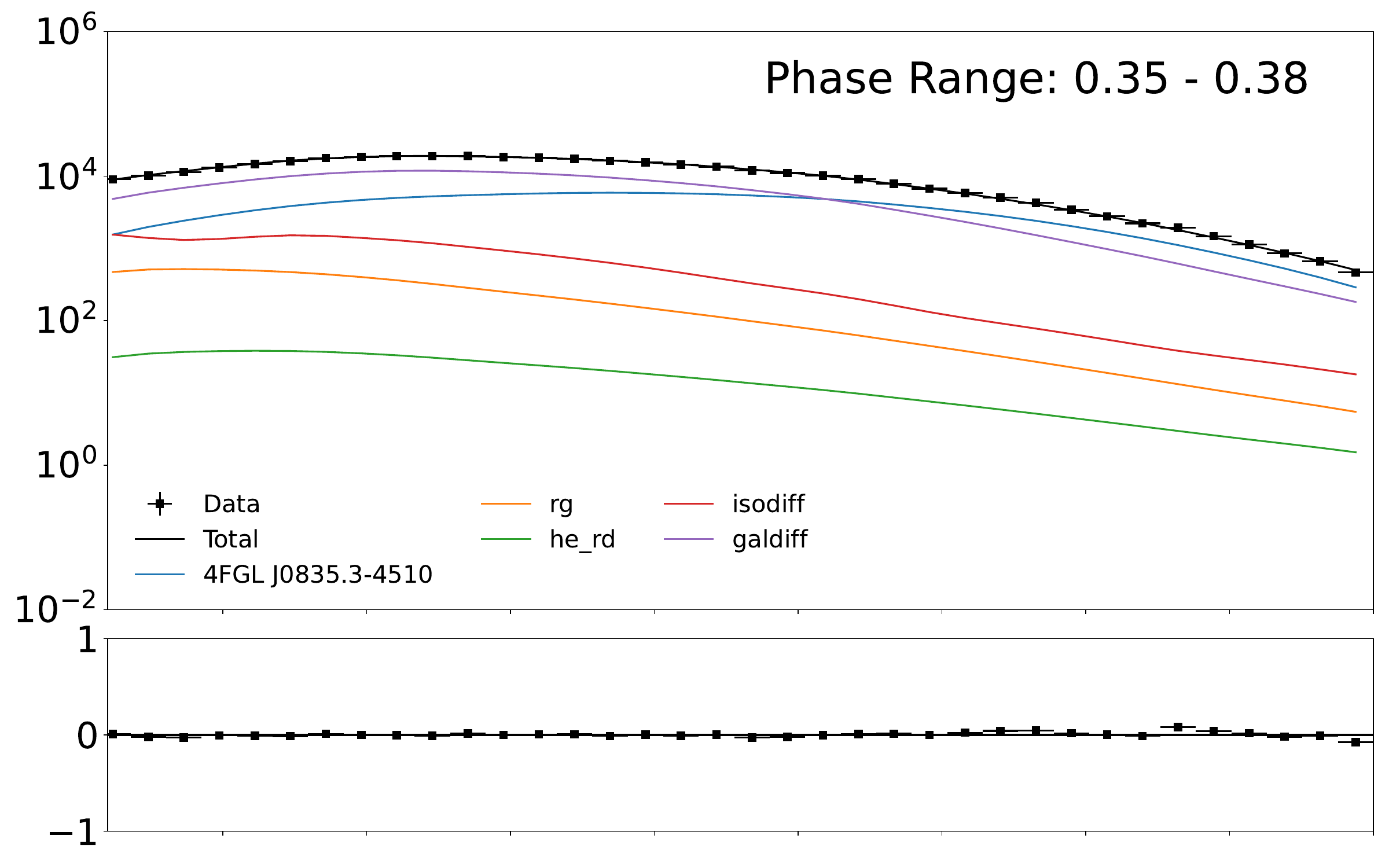}
    \end{minipage}%
    \begin{minipage}{0.32\textwidth}
        \includegraphics[width=\linewidth]{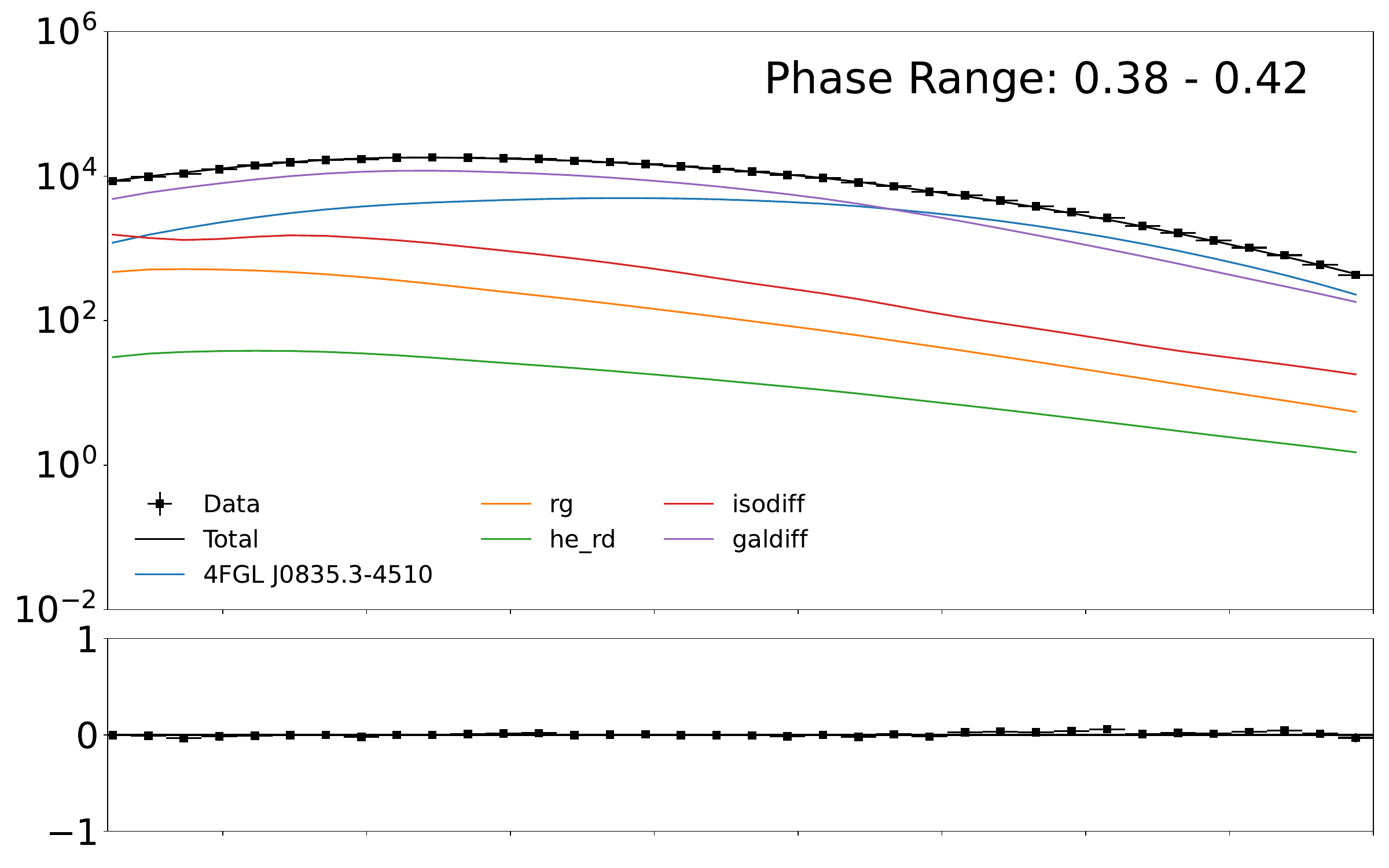}
    \end{minipage}

    \vspace{-2pt} 
    
    \begin{minipage}{0.32\textwidth}
        \includegraphics[width=\linewidth]{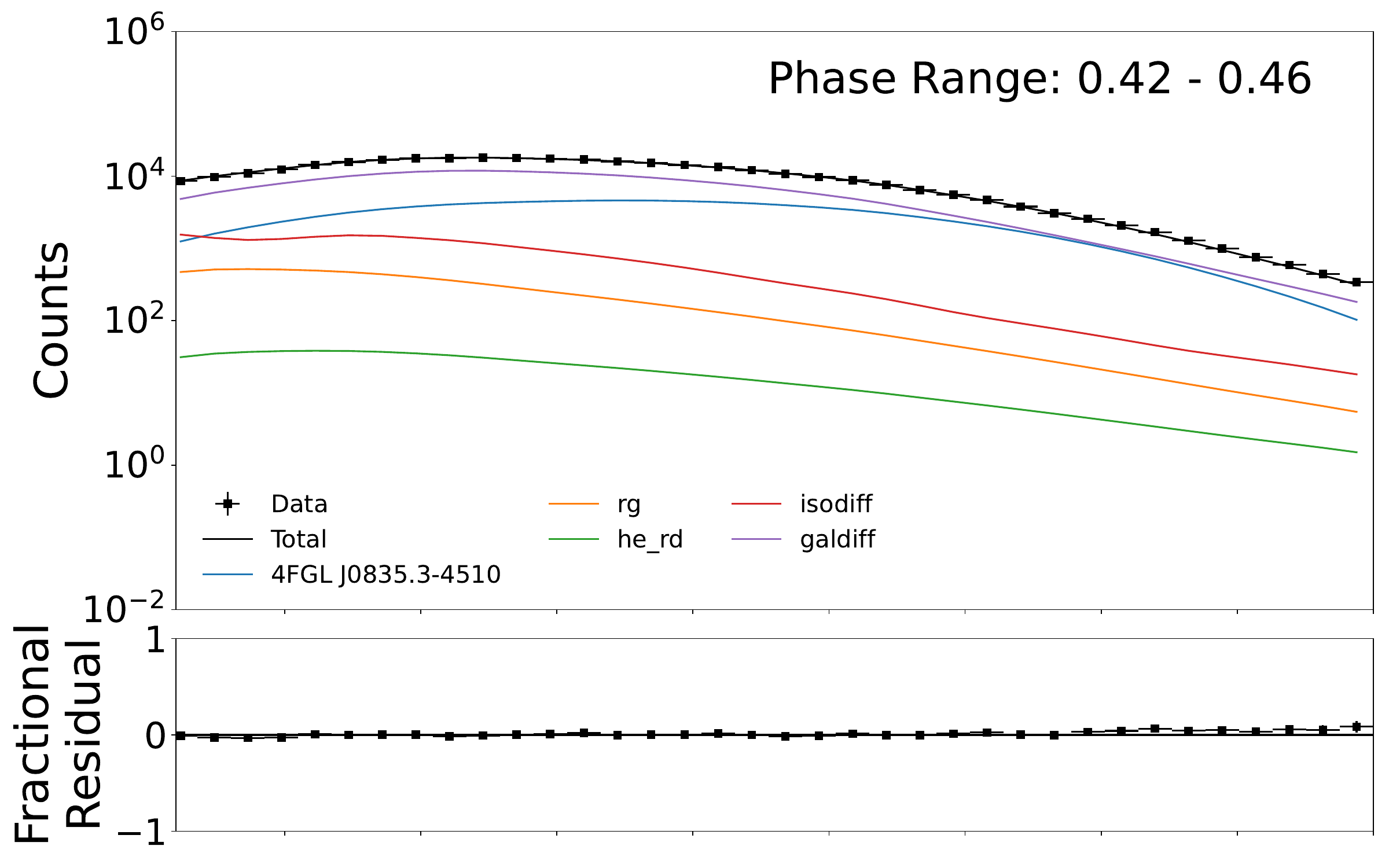}
    \end{minipage}%
    \begin{minipage}{0.32\textwidth}
        \includegraphics[width=\linewidth]{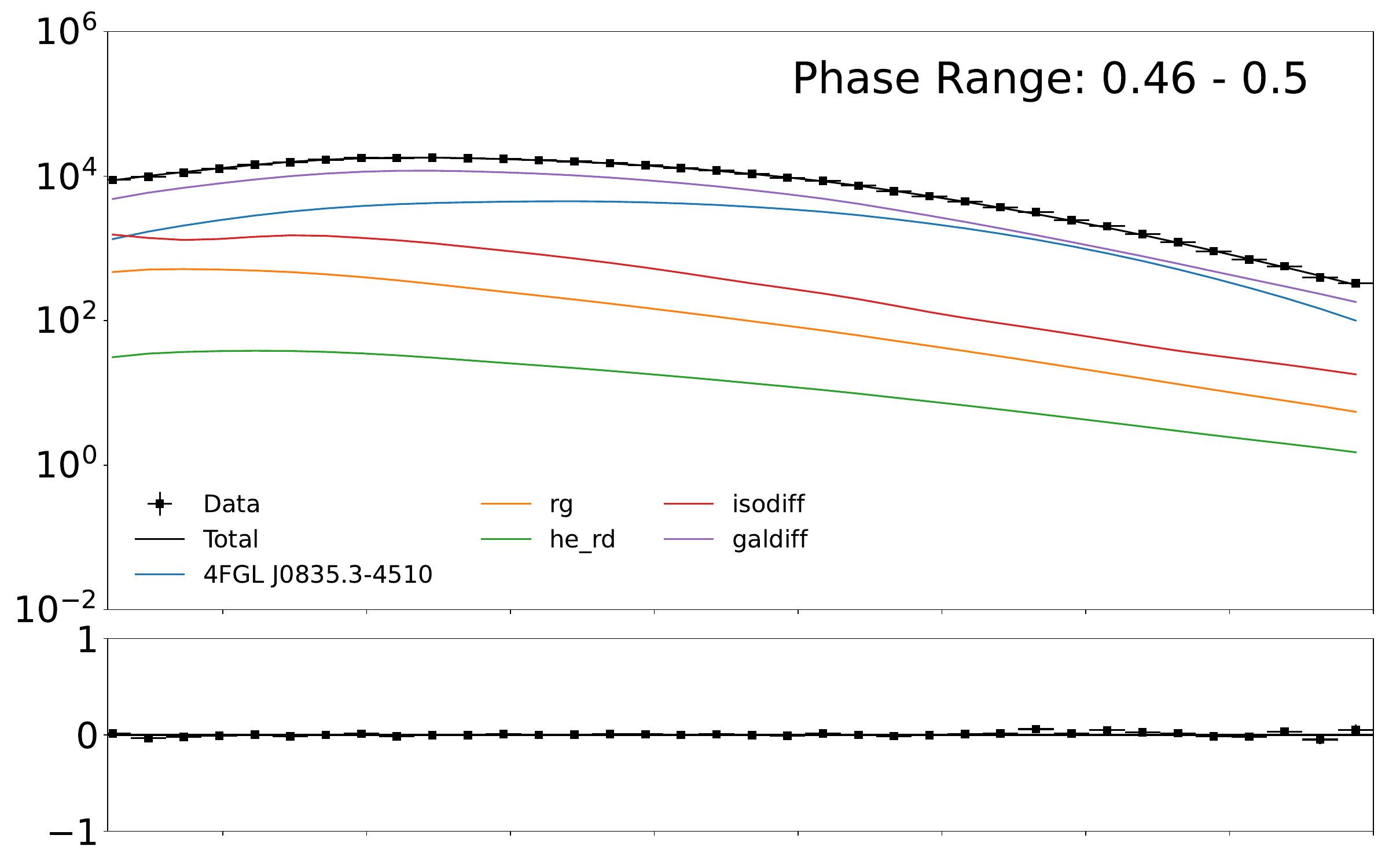}
    \end{minipage}%
    \begin{minipage}{0.32\textwidth}
        \includegraphics[width=\linewidth]{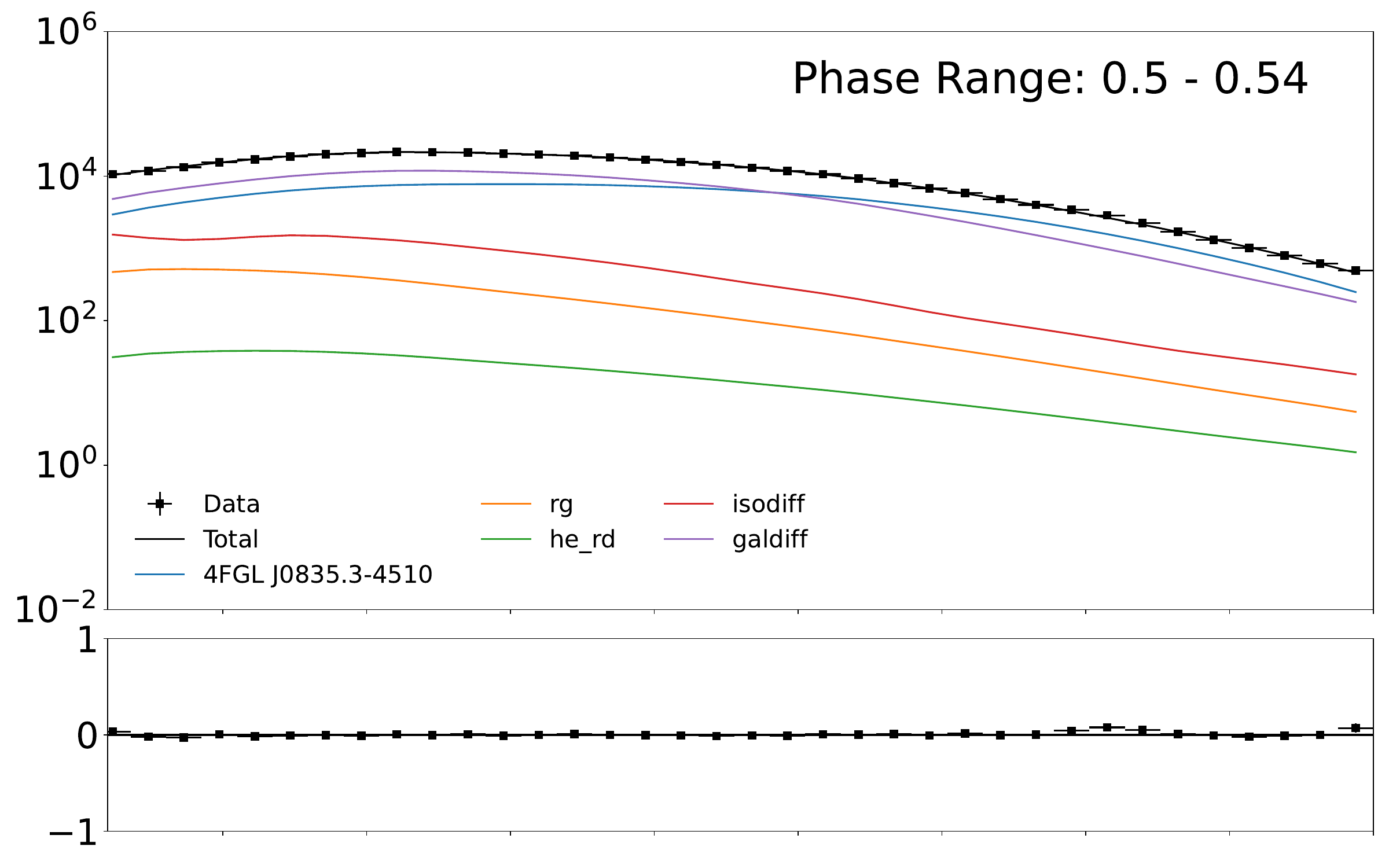}
    \end{minipage}

    \vspace{-2pt}

    \begin{minipage}{0.32\textwidth}
        \includegraphics[width=\linewidth]{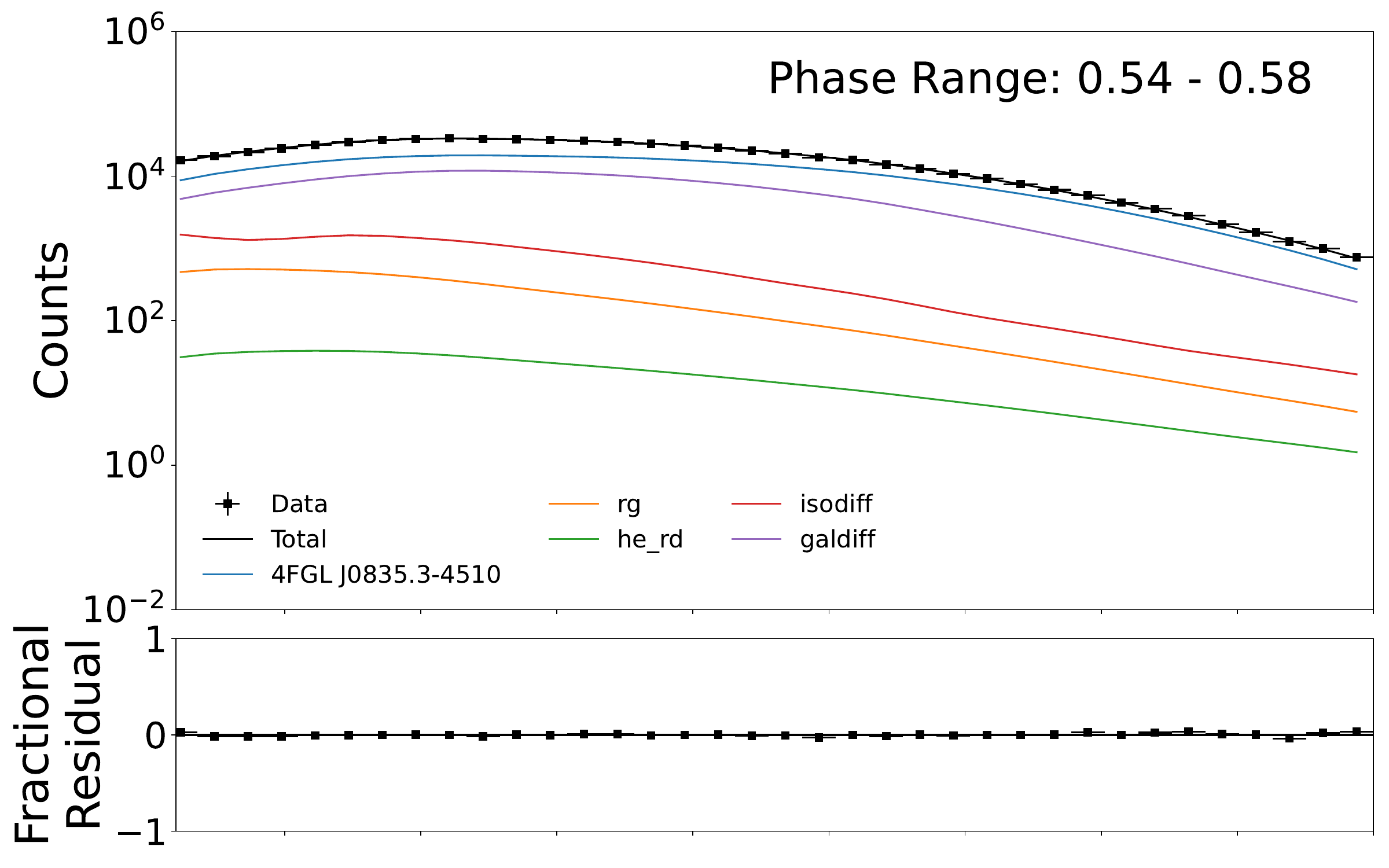}
    \end{minipage}%
    \begin{minipage}{0.32\textwidth}
        \includegraphics[width=\linewidth]{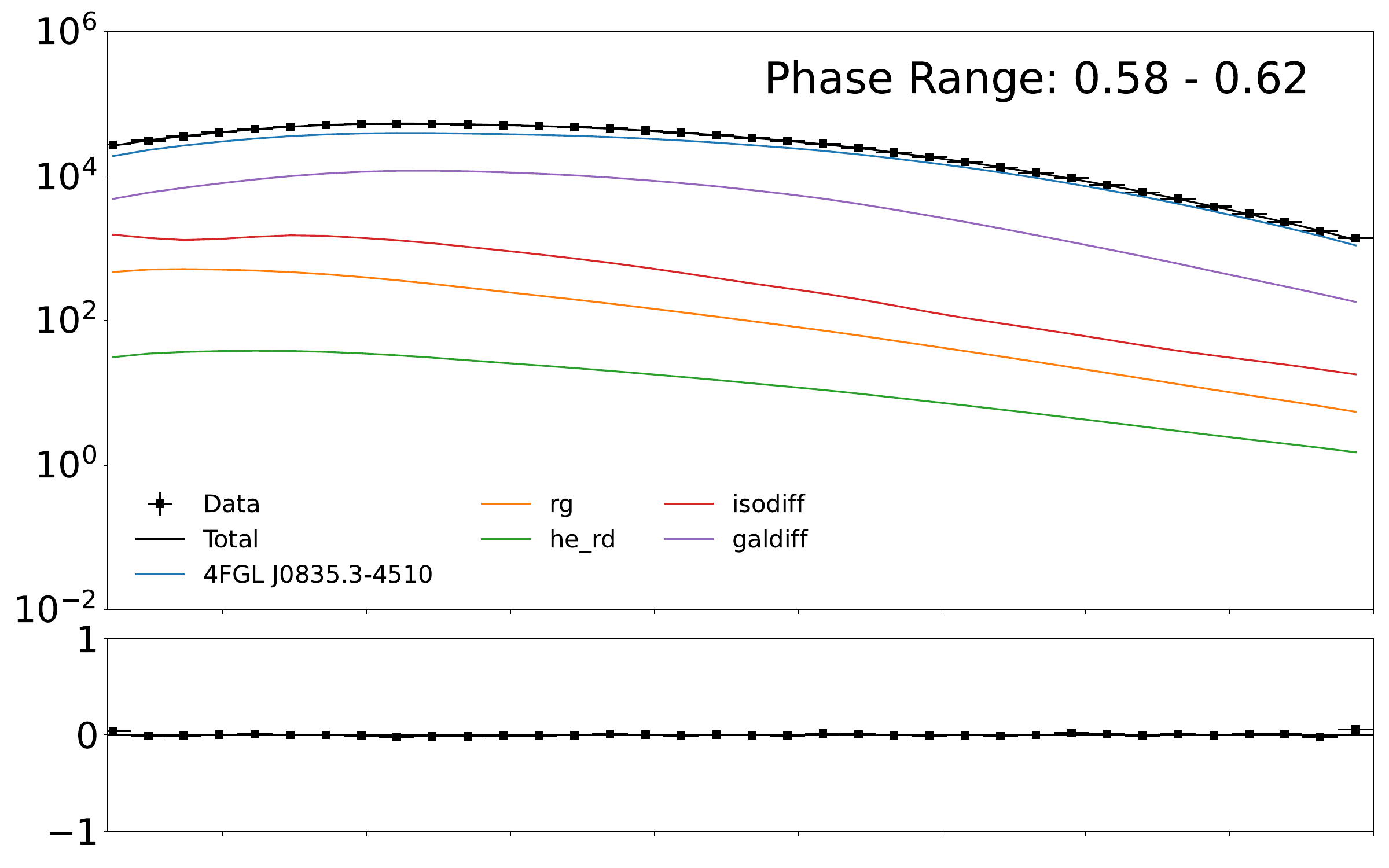}
    \end{minipage}%
    \begin{minipage}{0.32\textwidth}
        \includegraphics[width=\linewidth]{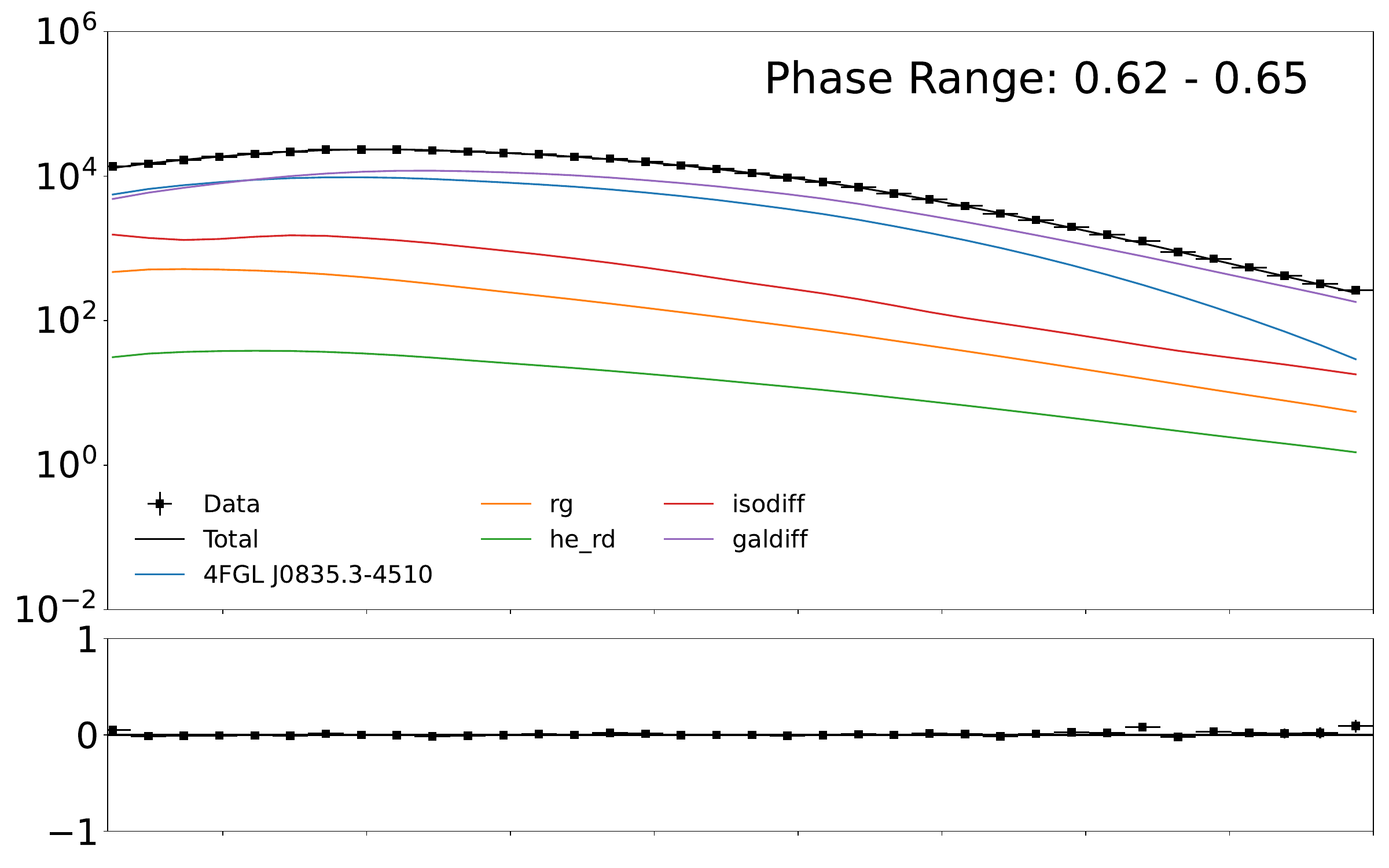}
    \end{minipage}

    \vspace{-2pt} 

    \begin{minipage}{0.32\textwidth}
        \includegraphics[width=\linewidth]{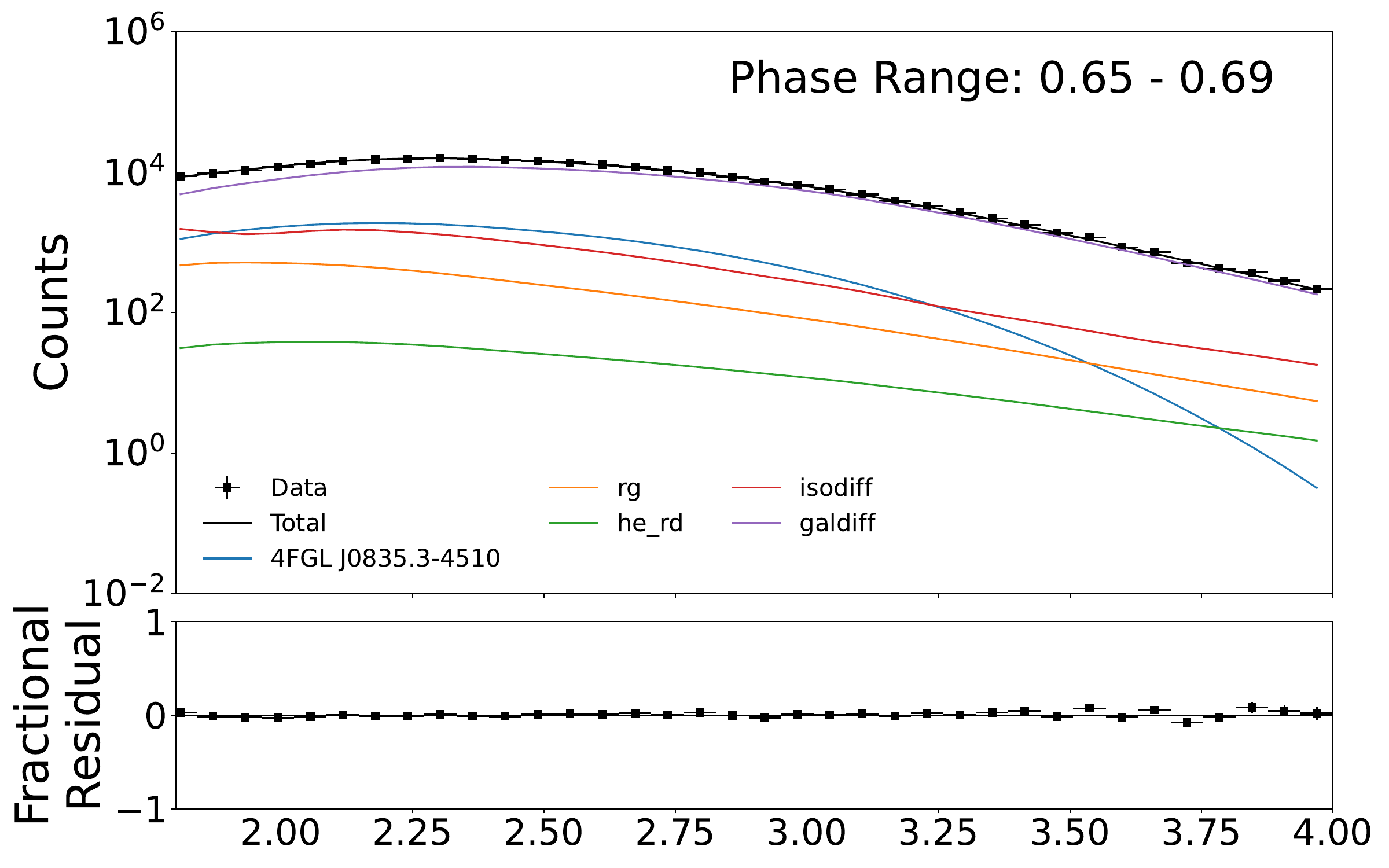}
    \end{minipage}%
    \begin{minipage}{0.32\textwidth}
        \includegraphics[width=\linewidth]{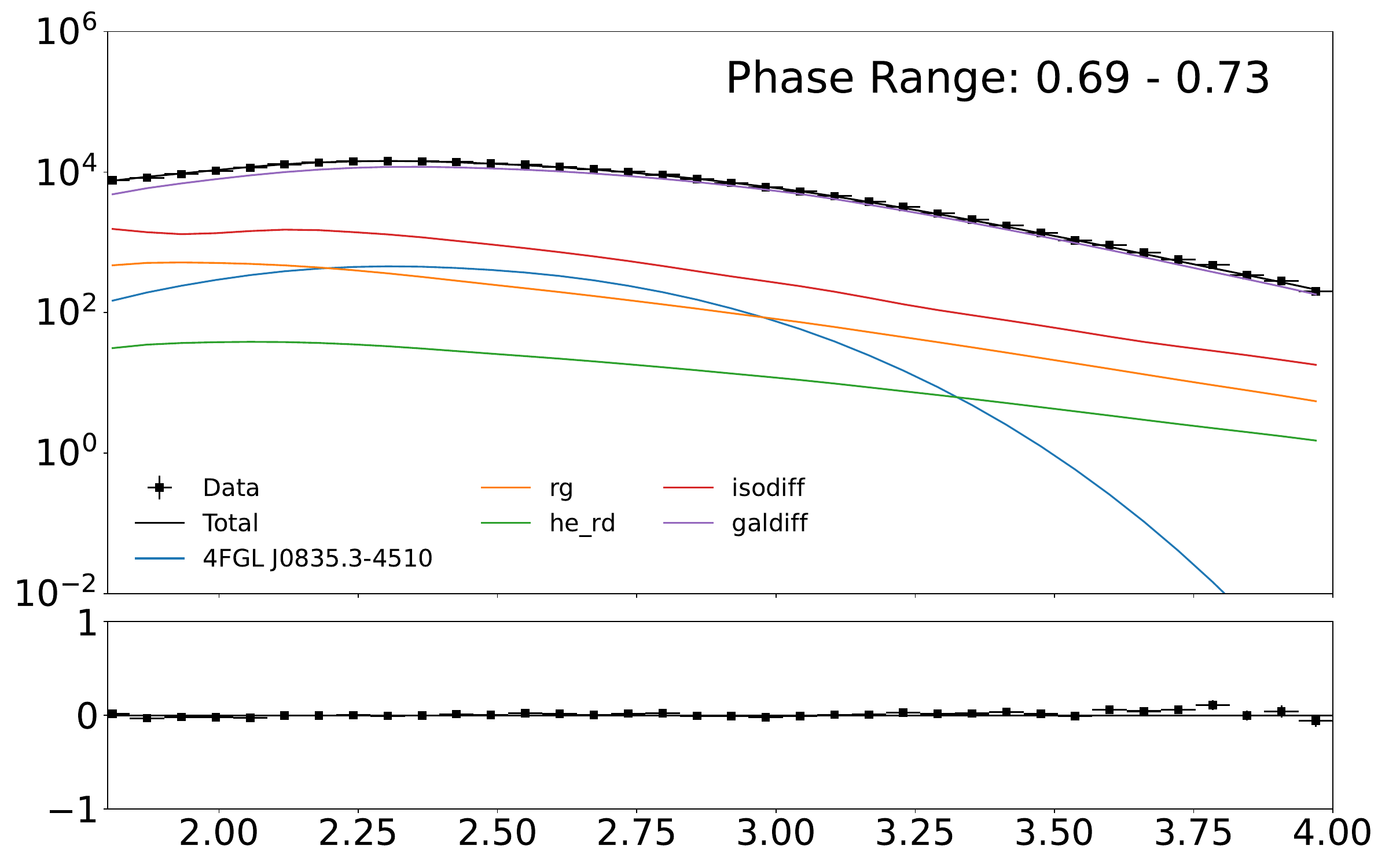}
    \end{minipage}%
    \begin{minipage}{0.32\textwidth}
        \includegraphics[width=\linewidth]{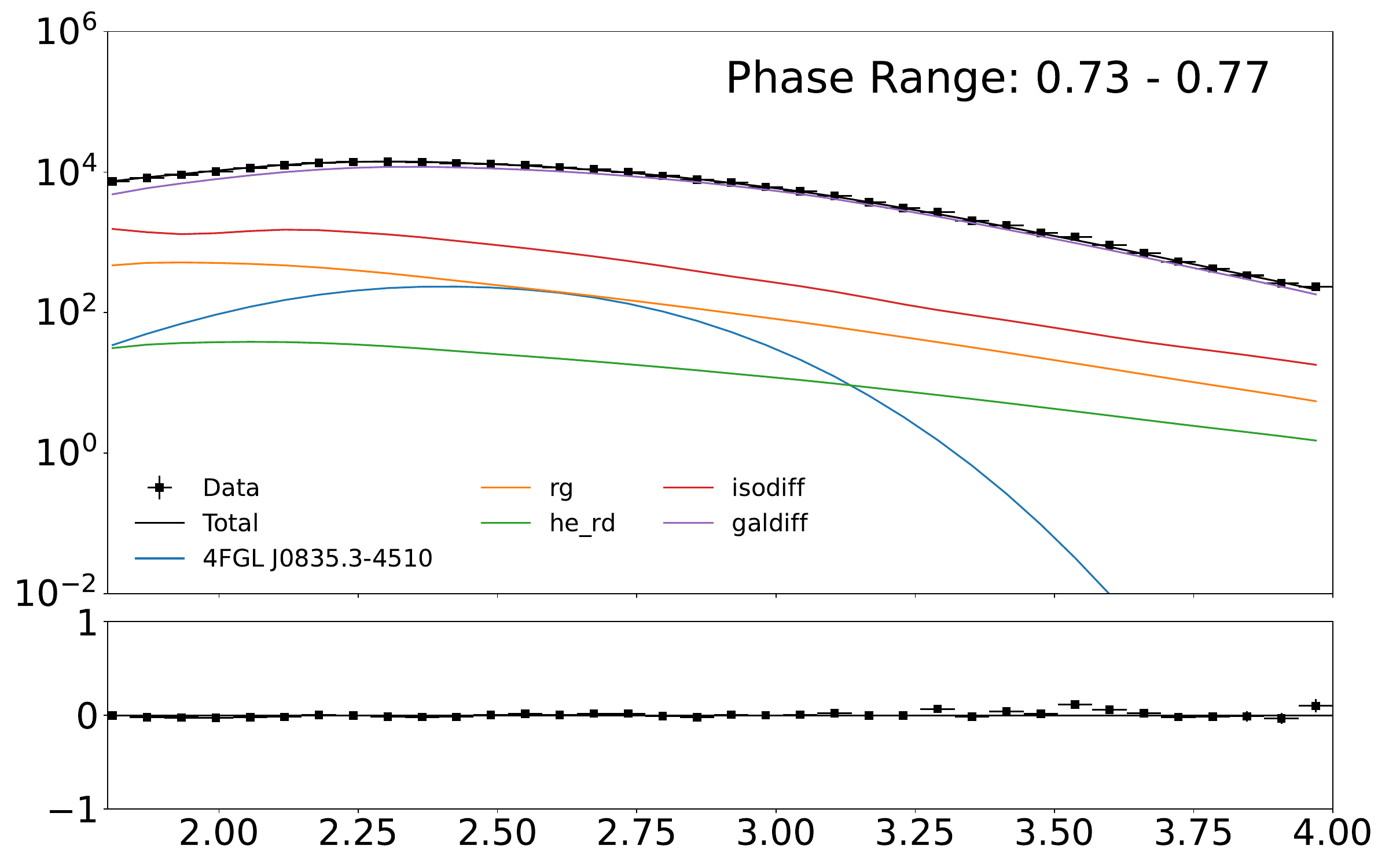}
    \end{minipage}

        \vspace{-60pt} %
    
    \begin{minipage}{0.32\textwidth}
        \includegraphics[width=\linewidth]{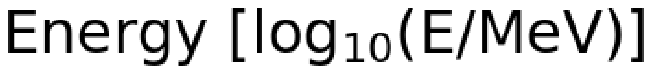}
    \end{minipage}%
    \begin{minipage}{0.32\textwidth}
        \includegraphics[width=\linewidth]{B_label.pdf}
    \end{minipage}%
    \begin{minipage}{0.32\textwidth}
        \includegraphics[width=\linewidth]{B_label.pdf}
    \end{minipage}
    
    \vspace{-40pt} %
    
    \caption{Count spectra of the Vela pulsar for the phase range [0.08, 0.77] for $b=0.394$. Spectral data are shown in black. The five most significant sources are shown (including the isotropic and Galactic diffuse backgrounds as well as our radial Gaussian and radial disk PWN components from Section~\ref{sec:pwn}). The bottom panel of each count spectrum show the fractional residuals. The phase bin range is indicated in each plot.  
    }
\label{app:count_fixedb}
\end{figure*}

\begin{figure*}[!h]
    \centering
    \begin{minipage}{0.32\textwidth}
        \includegraphics[width=\linewidth]{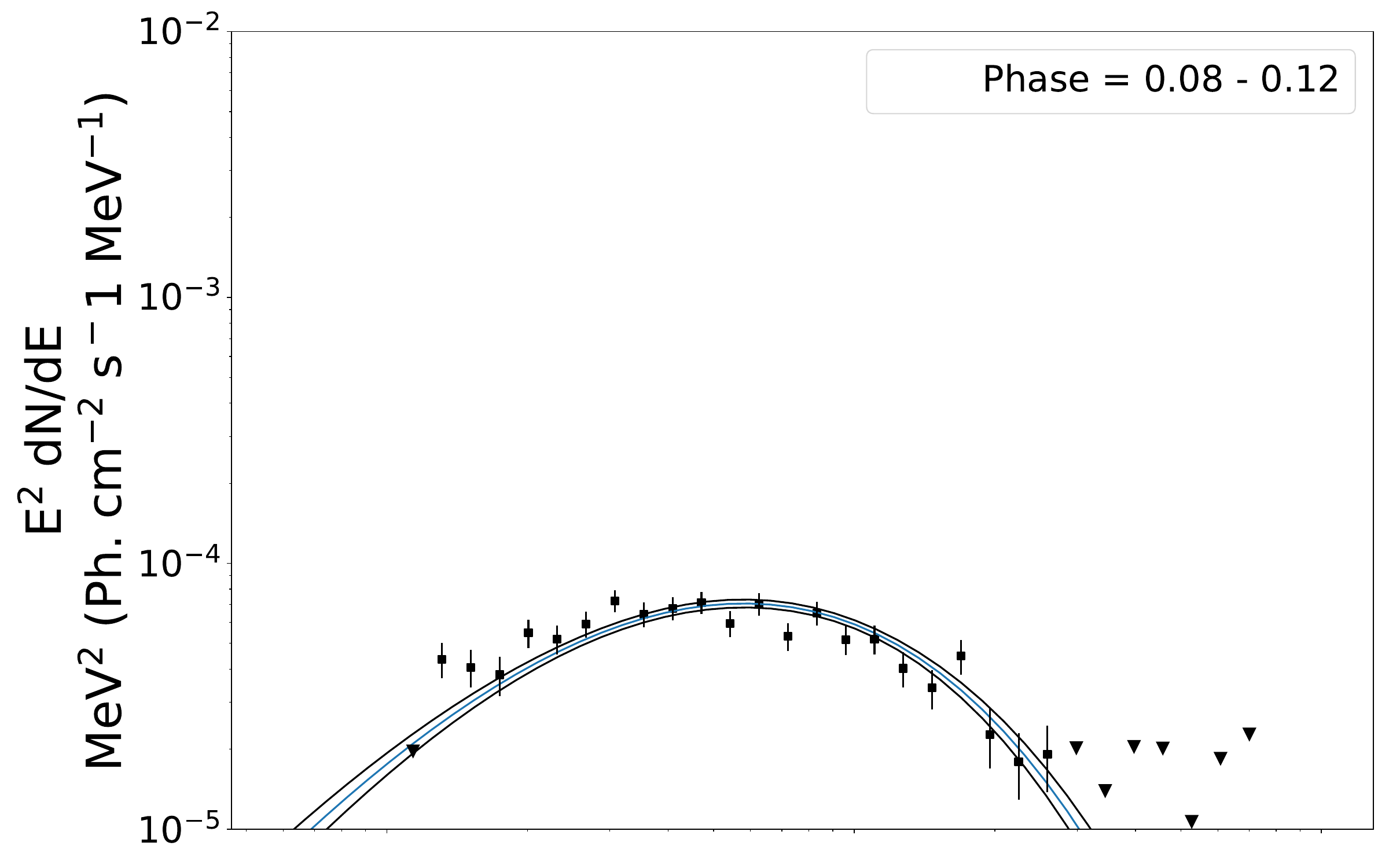}
    \end{minipage}%
    \begin{minipage}{0.32\textwidth}
        \includegraphics[width=\linewidth]{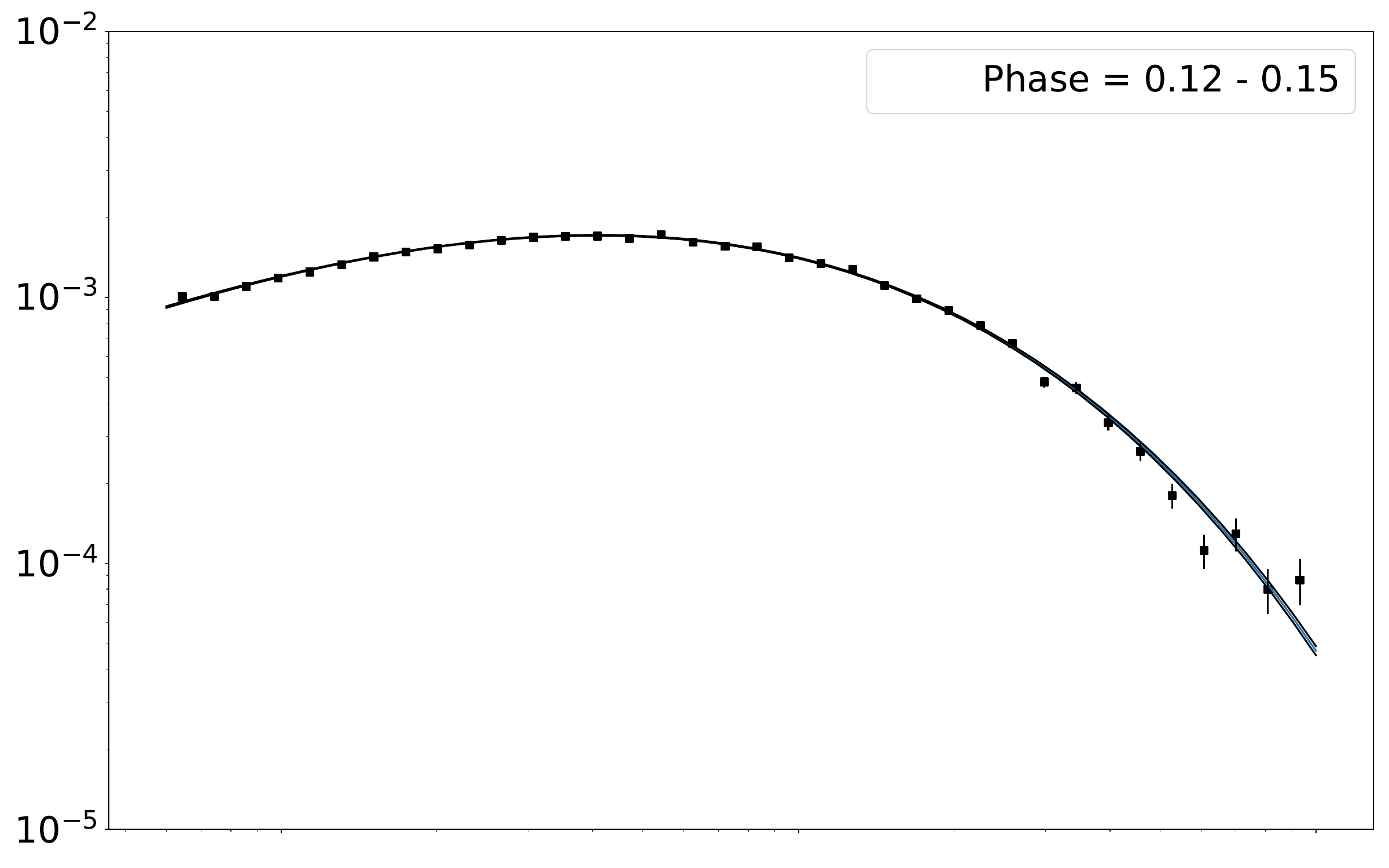}
    \end{minipage}%
    \begin{minipage}{0.32\textwidth}
        \includegraphics[width=\linewidth]{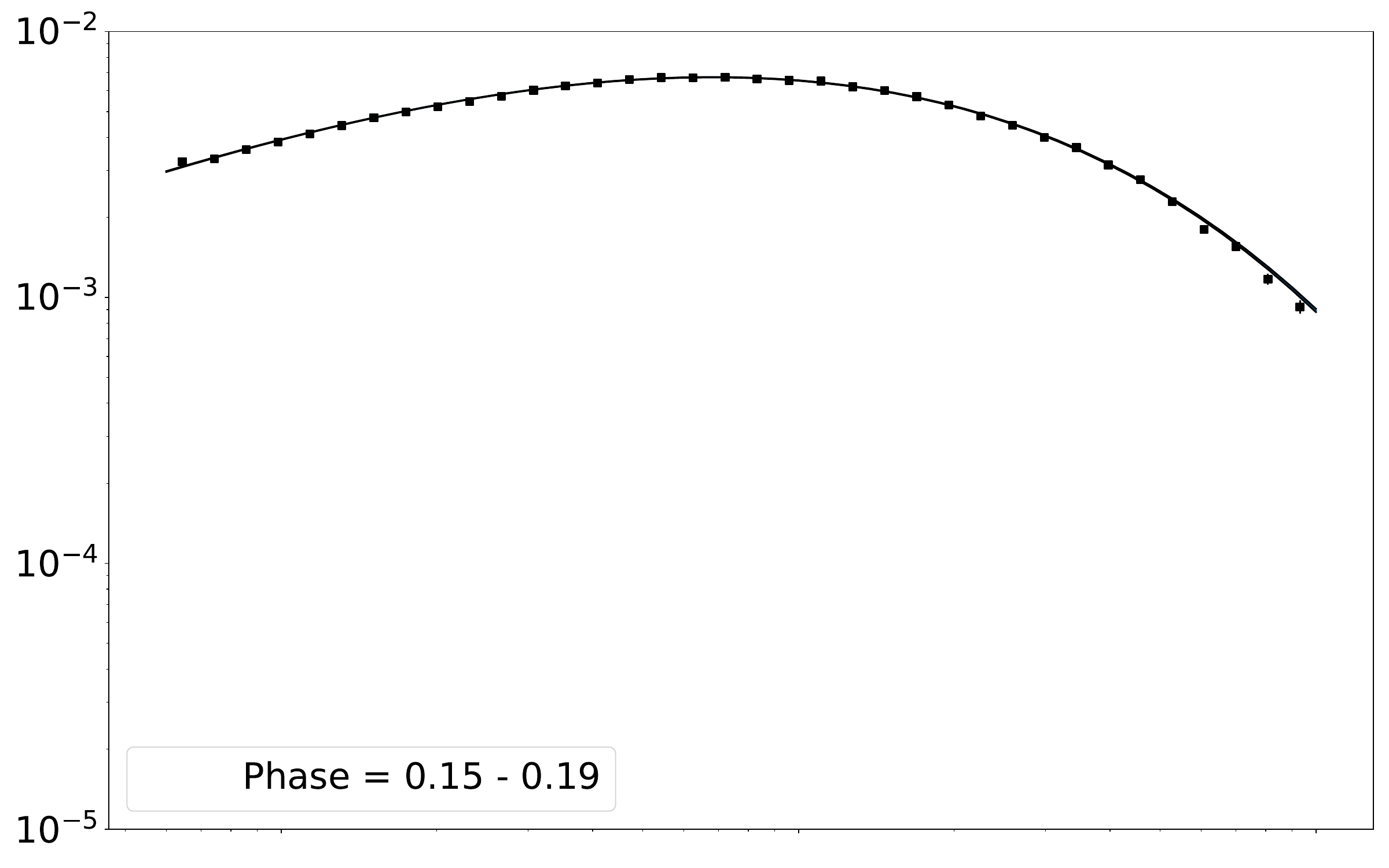}
    \end{minipage}

    \vspace{-2pt} 

    \begin{minipage}{0.32\textwidth}
        \includegraphics[width=\linewidth]{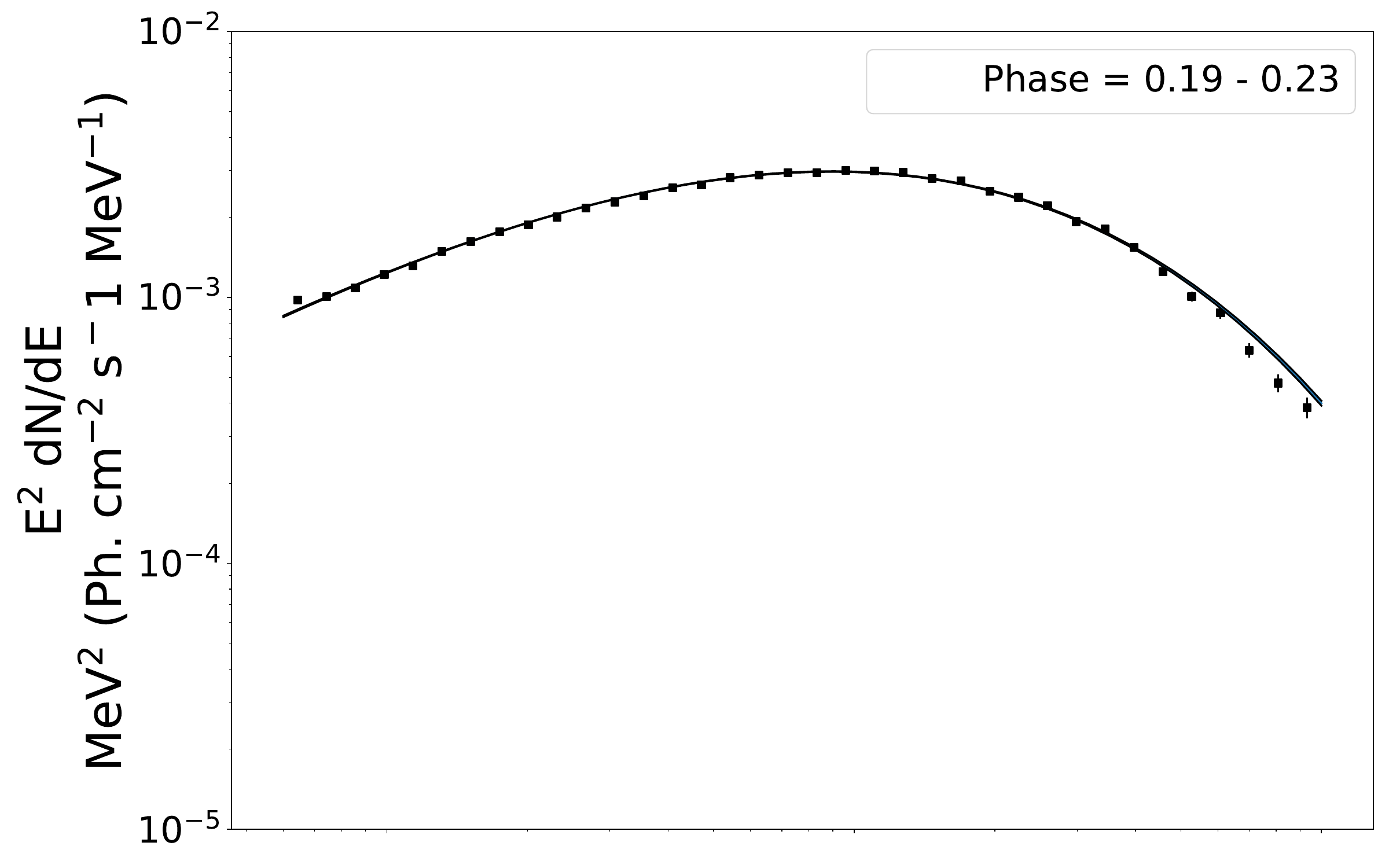}
    \end{minipage}%
    \begin{minipage}{0.32\textwidth}
        \includegraphics[width=\linewidth]{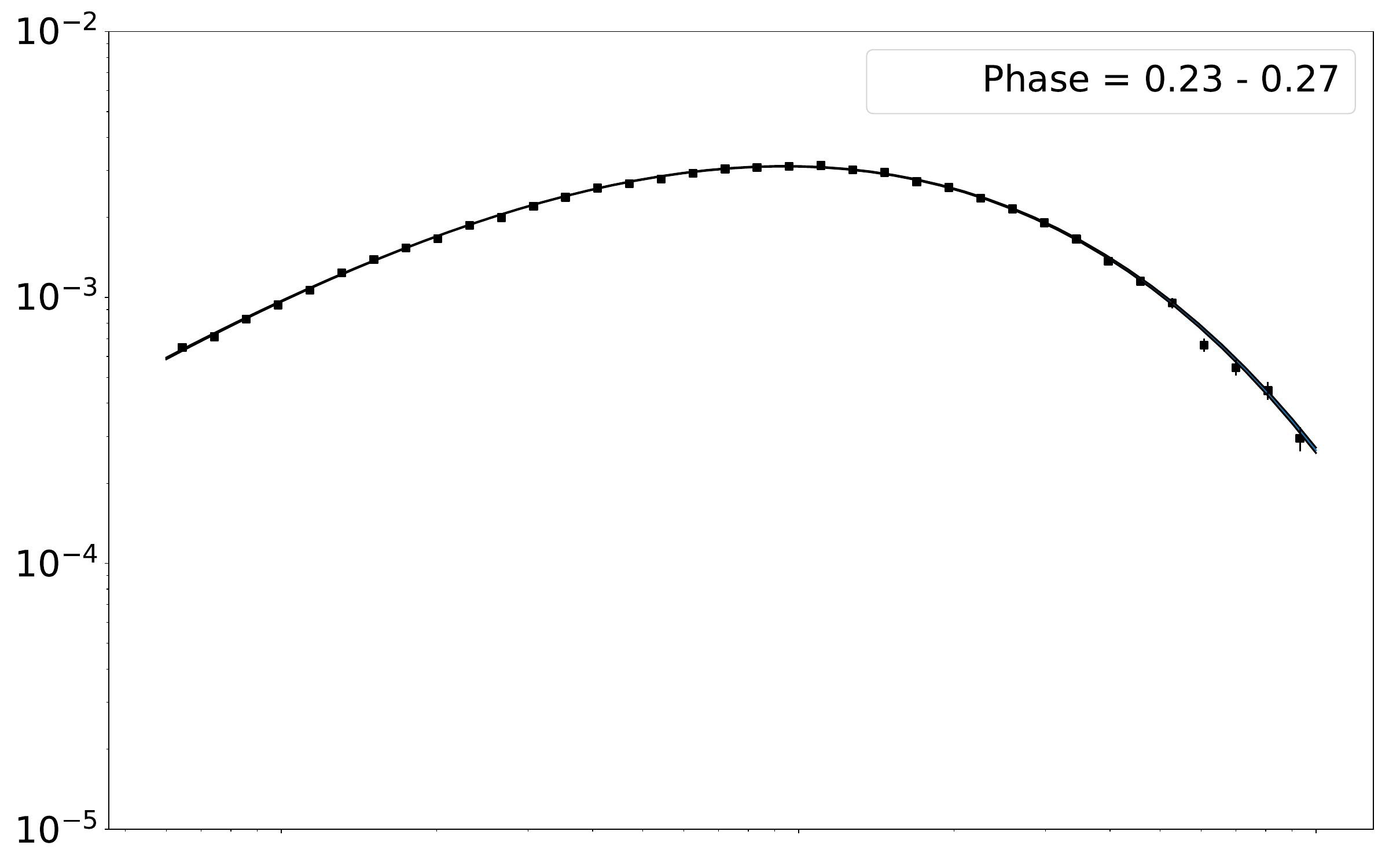}
    \end{minipage}%
    \begin{minipage}{0.32\textwidth}
        \includegraphics[width=\linewidth]{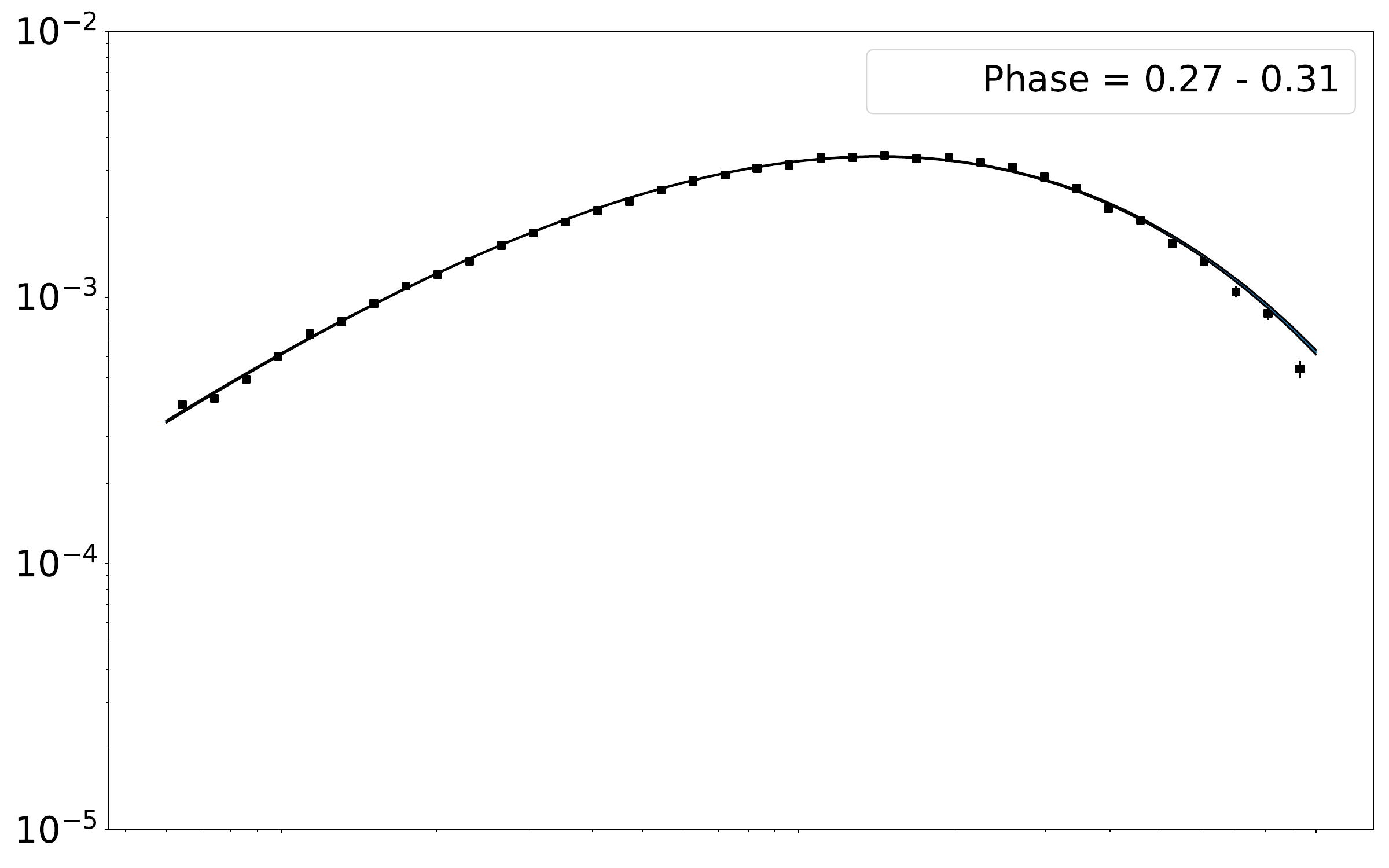}
    \end{minipage}

    \vspace{-2pt}

    \begin{minipage}{0.32\textwidth}
        \includegraphics[width=\linewidth]{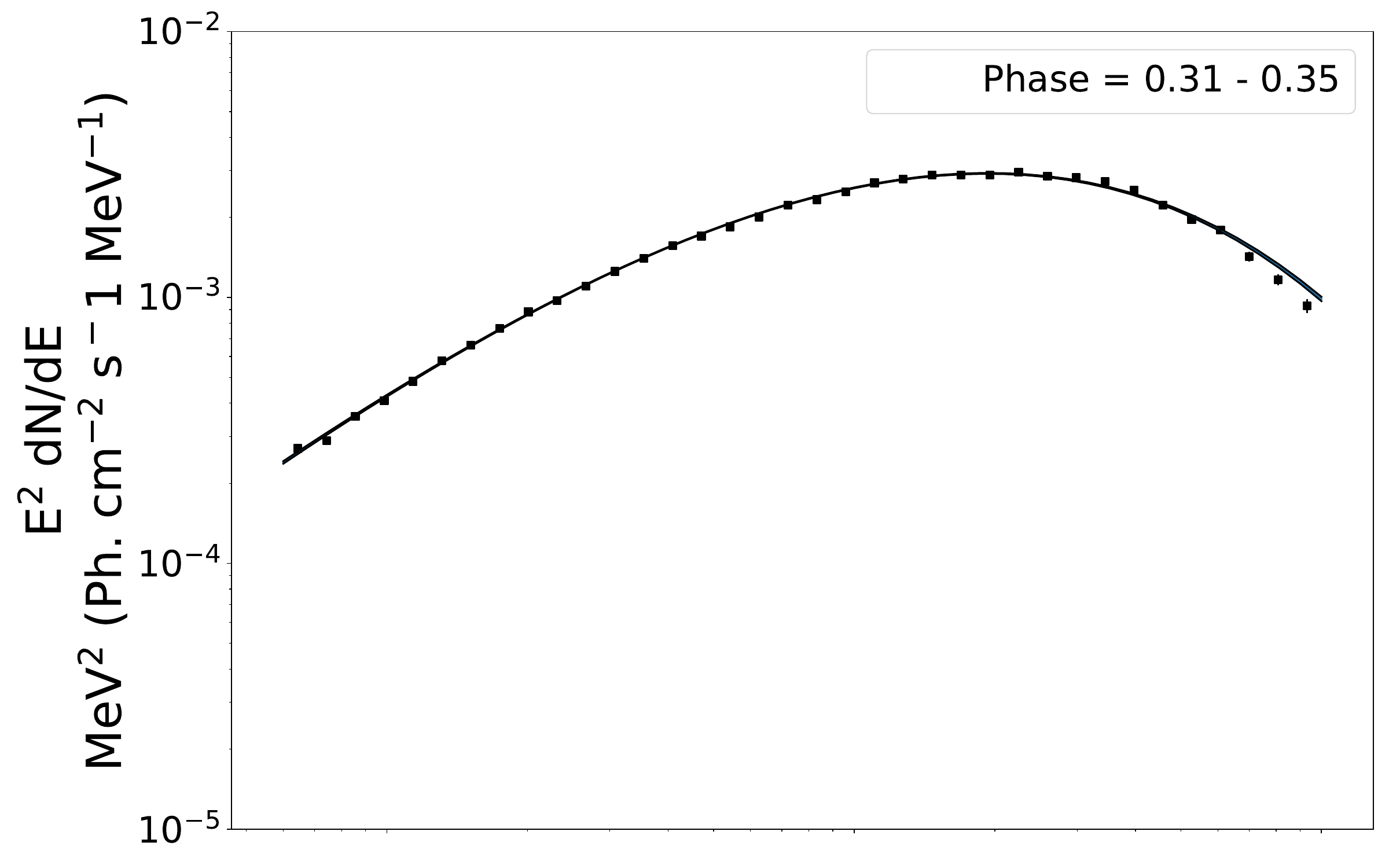}
    \end{minipage}%
    \begin{minipage}{0.32\textwidth}
        \includegraphics[width=\linewidth]{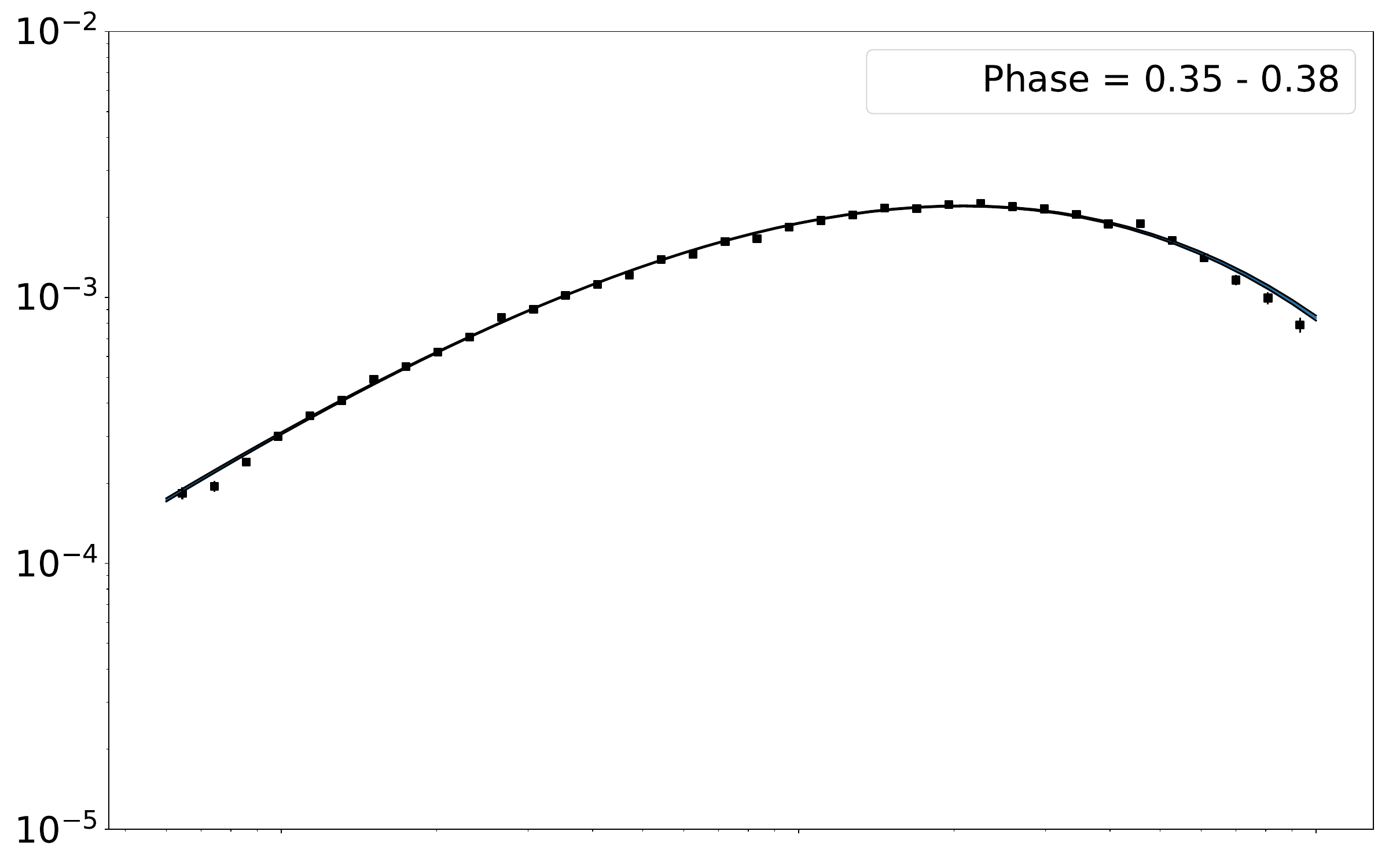}
    \end{minipage}%
    \begin{minipage}{0.32\textwidth}
        \includegraphics[width=\linewidth]{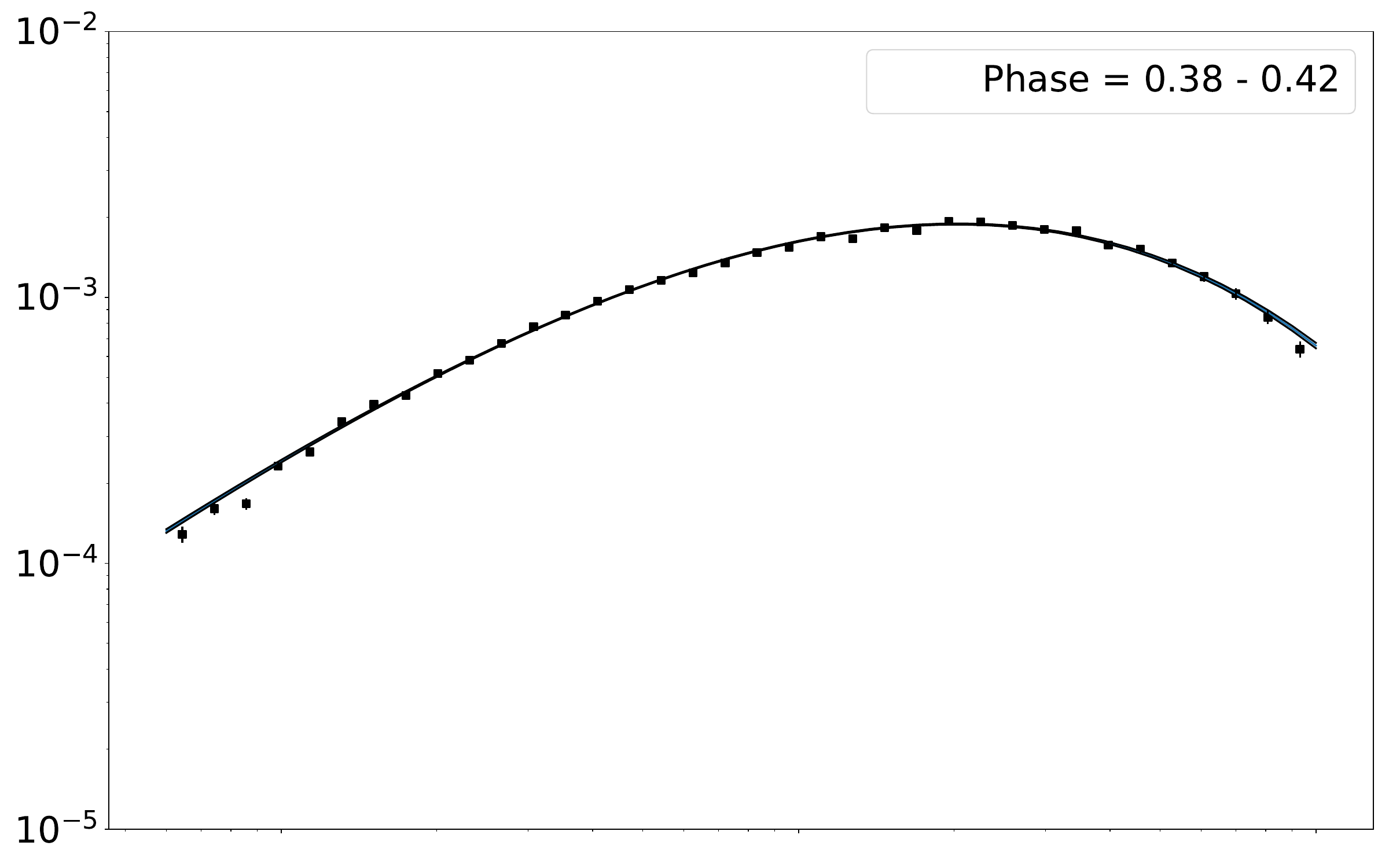}
    \end{minipage}

    \vspace{-2pt} 

    \begin{minipage}{0.32\textwidth}
        \includegraphics[width=\linewidth]{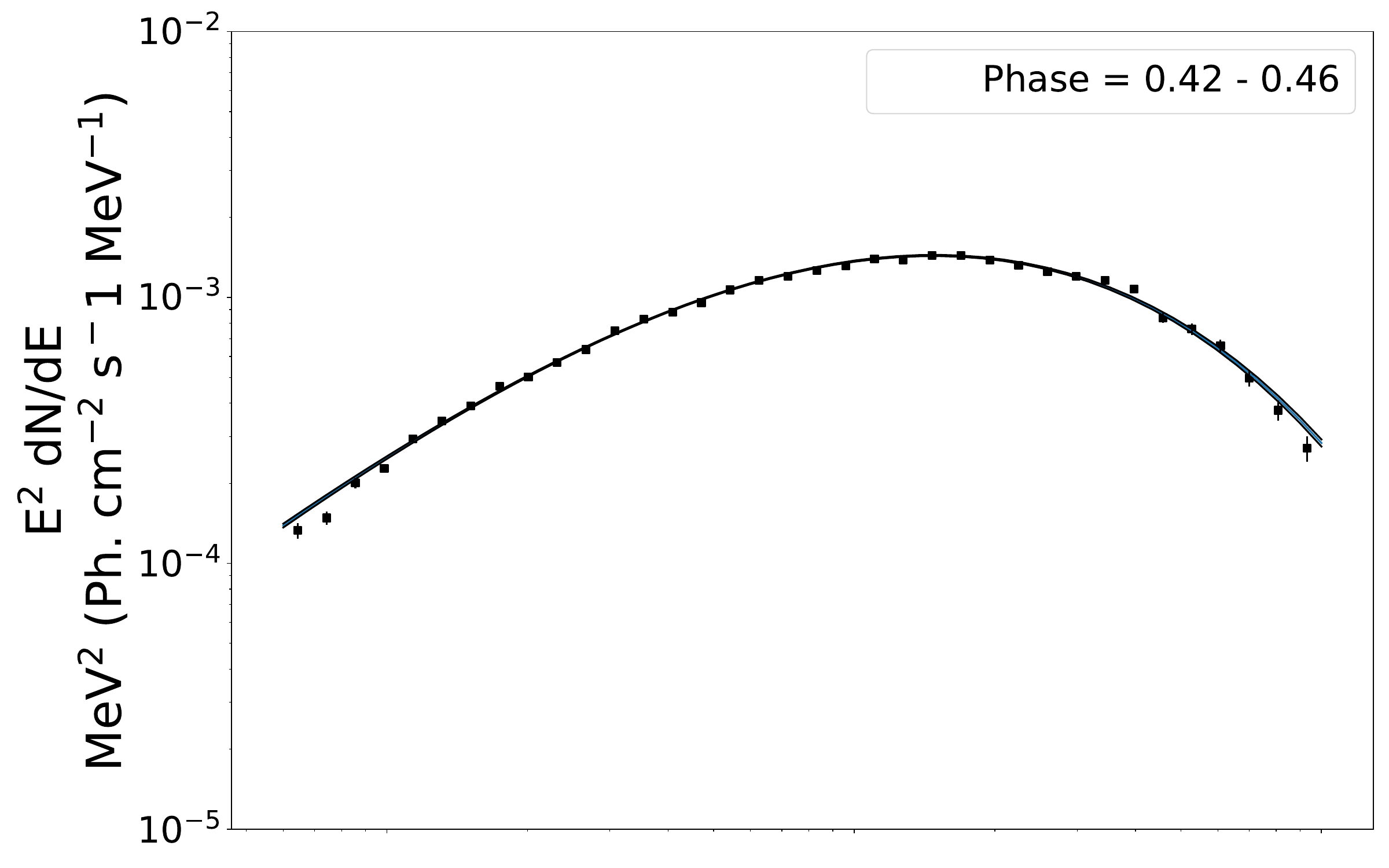}
    \end{minipage}%
    \begin{minipage}{0.32\textwidth}
        \includegraphics[width=\linewidth]{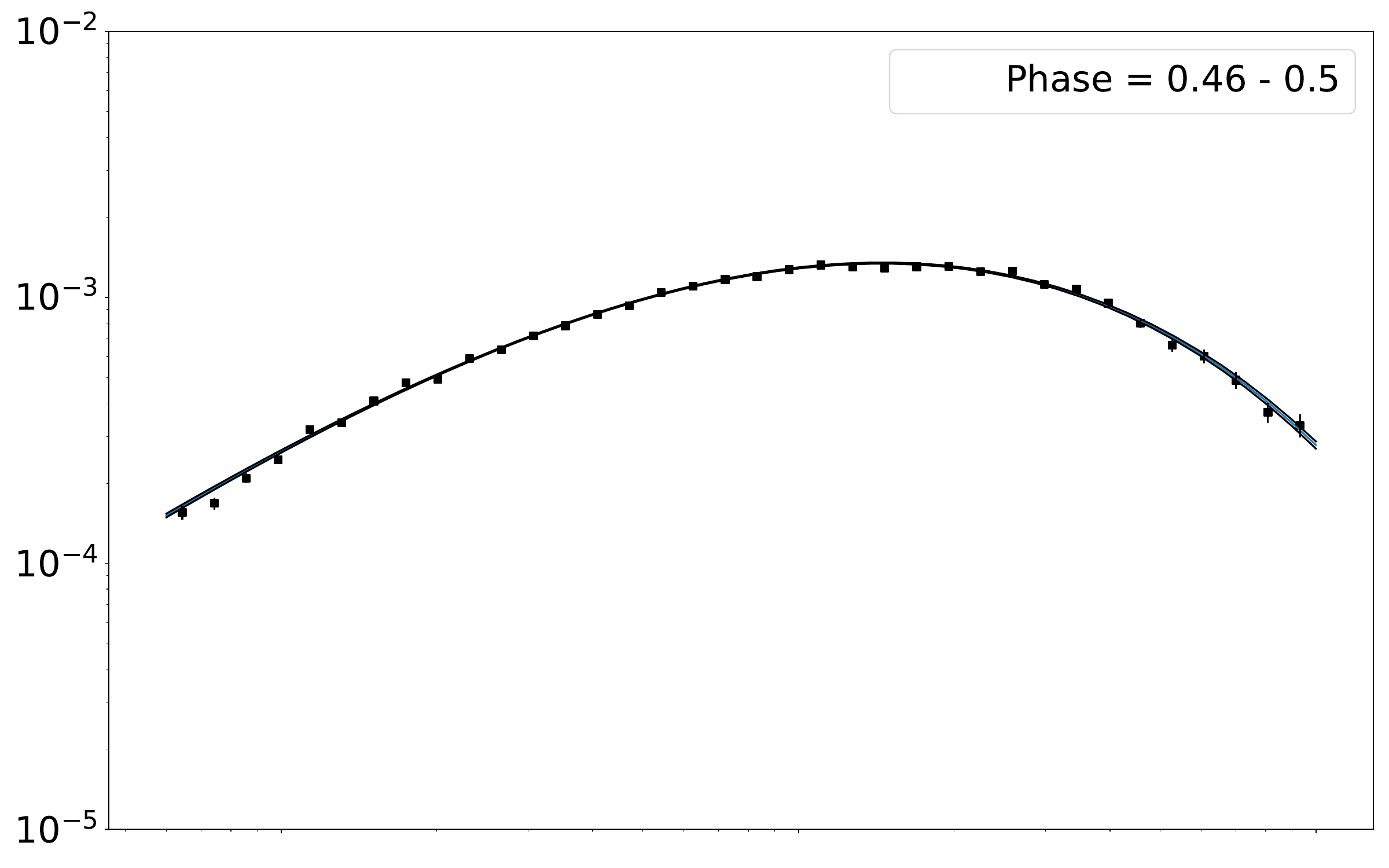}
    \end{minipage}%
    \begin{minipage}{0.32\textwidth}
        \includegraphics[width=\linewidth]{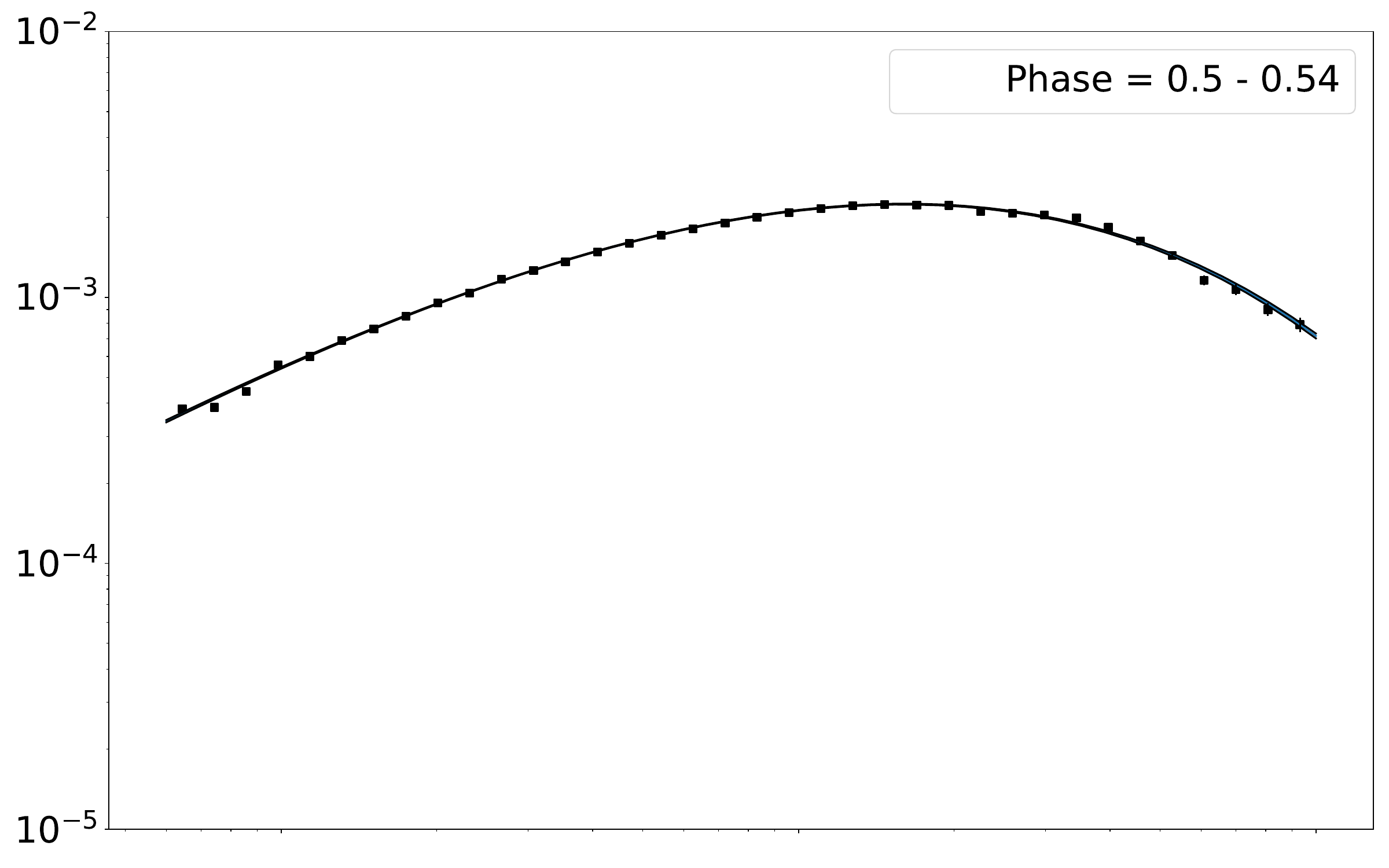}
    \end{minipage}

    \vspace{-2pt} 

    \begin{minipage}{0.32\textwidth}
        \includegraphics[width=\linewidth]{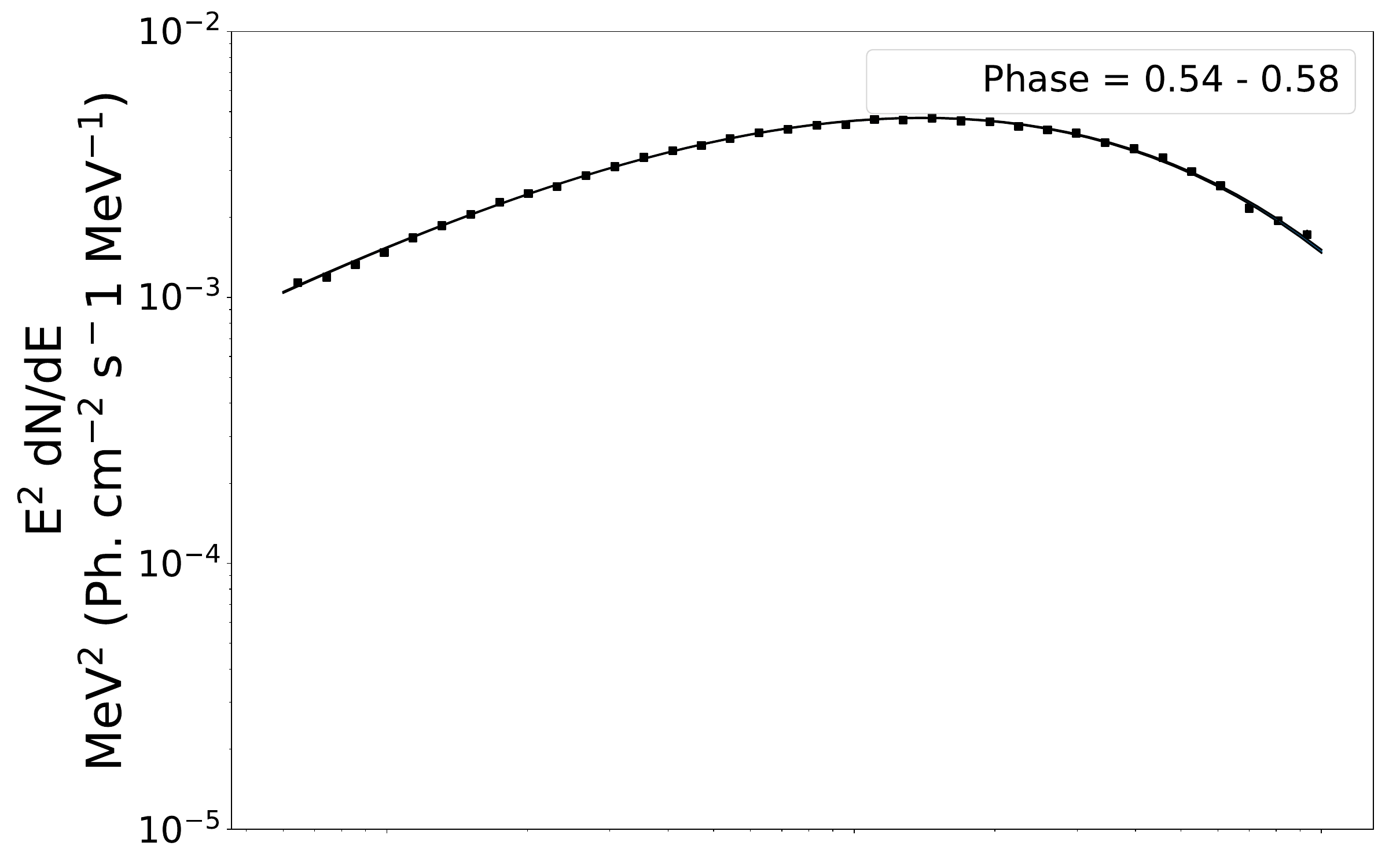}
    \end{minipage}%
    \begin{minipage}{0.32\textwidth}
        \includegraphics[width=\linewidth]{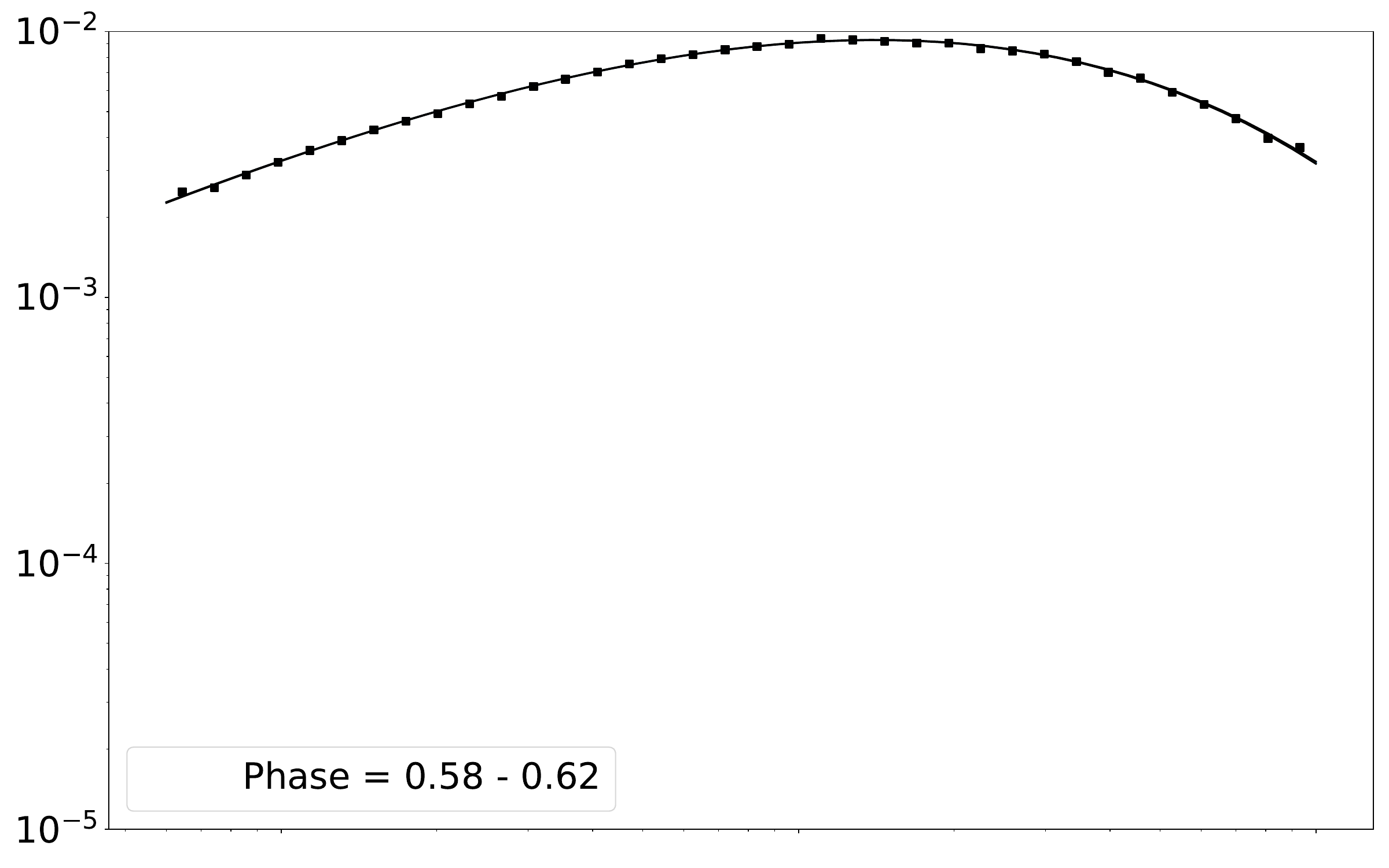}
    \end{minipage}%
    \begin{minipage}{0.32\textwidth}
        \includegraphics[width=\linewidth]{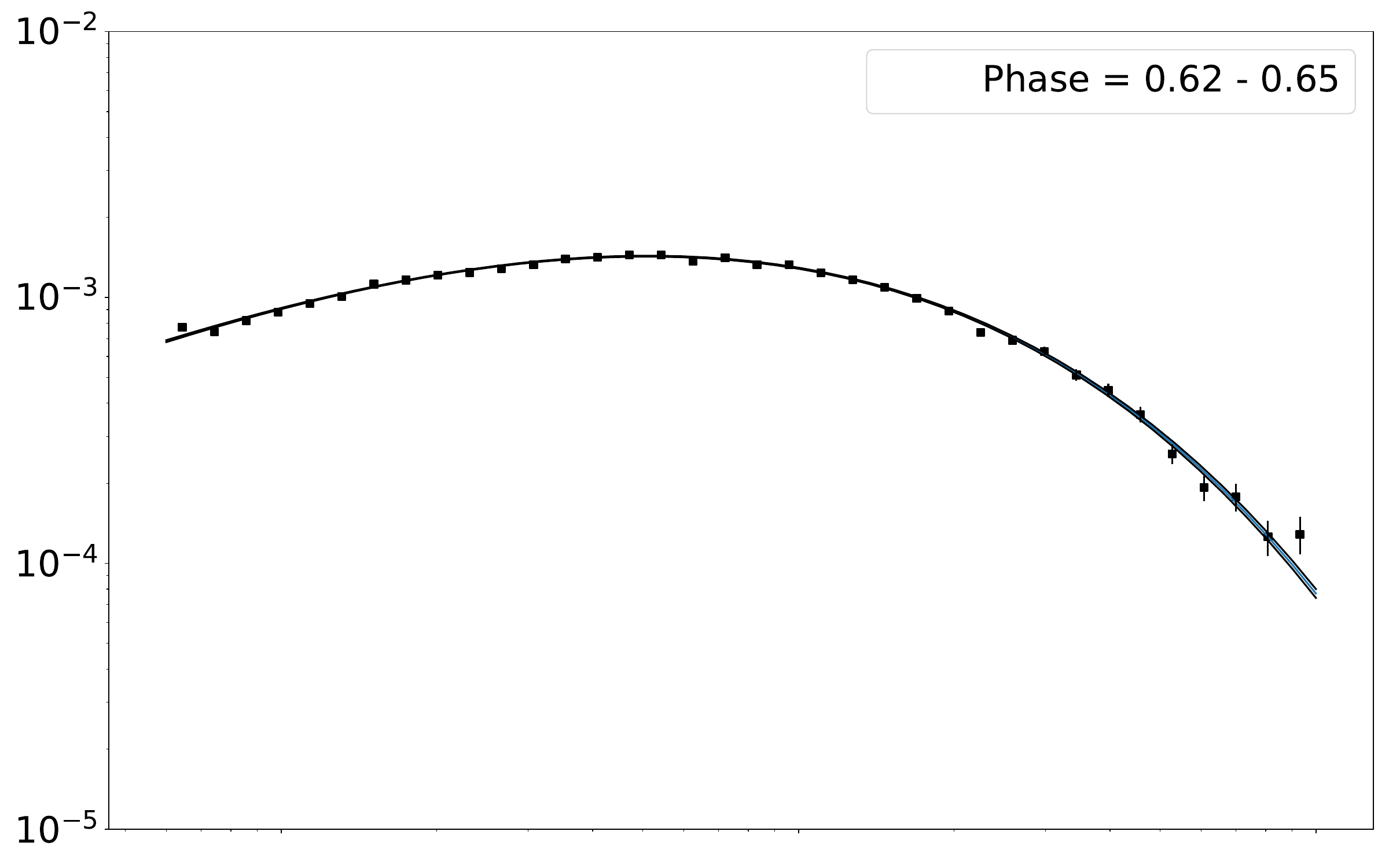}
    \end{minipage}
    
    \vspace{-2pt} 
    
    \begin{minipage}{0.32\textwidth}
        \includegraphics[width=\linewidth]{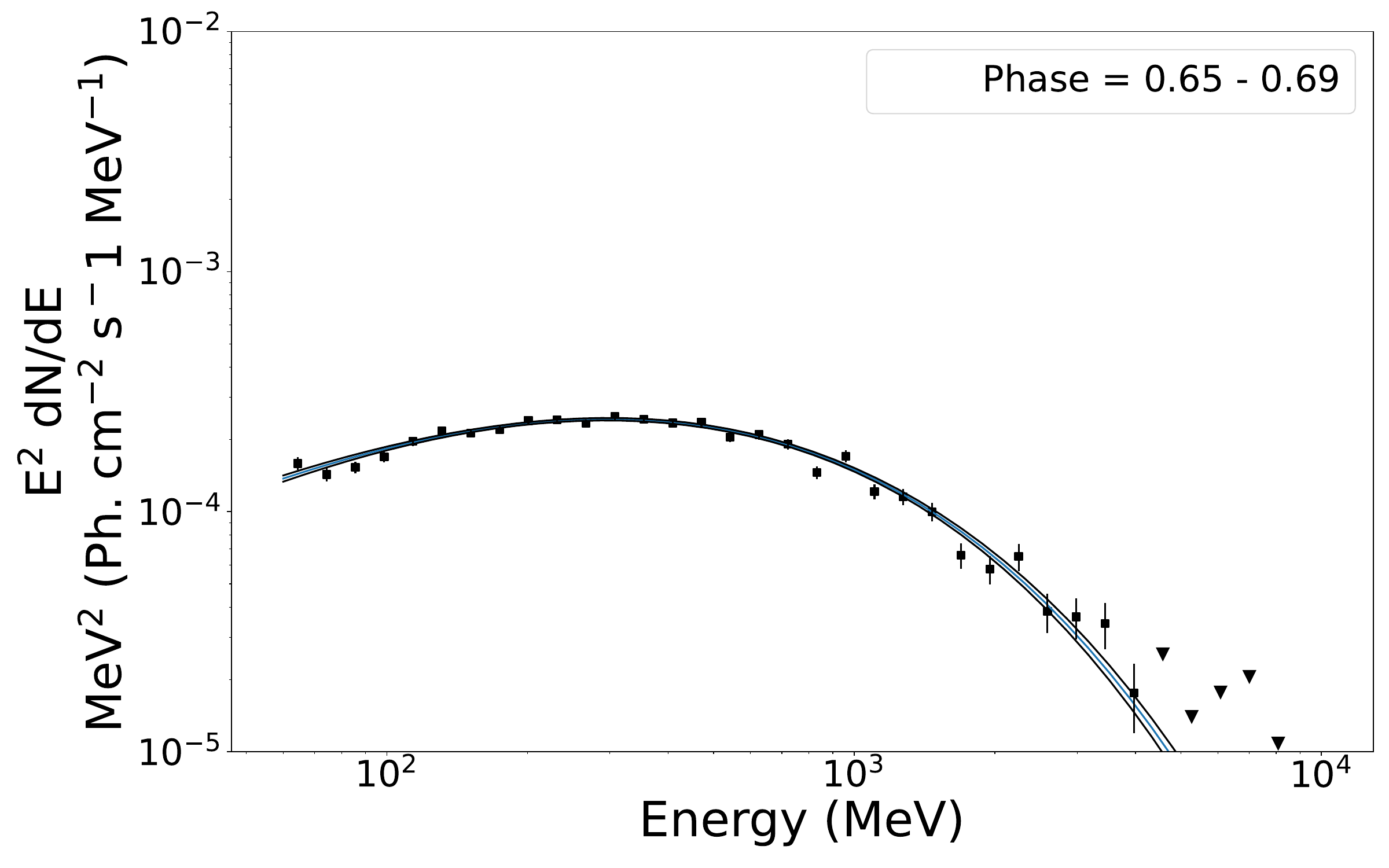}
    \end{minipage}%
    \begin{minipage}{0.32\textwidth}
        \includegraphics[width=\linewidth]{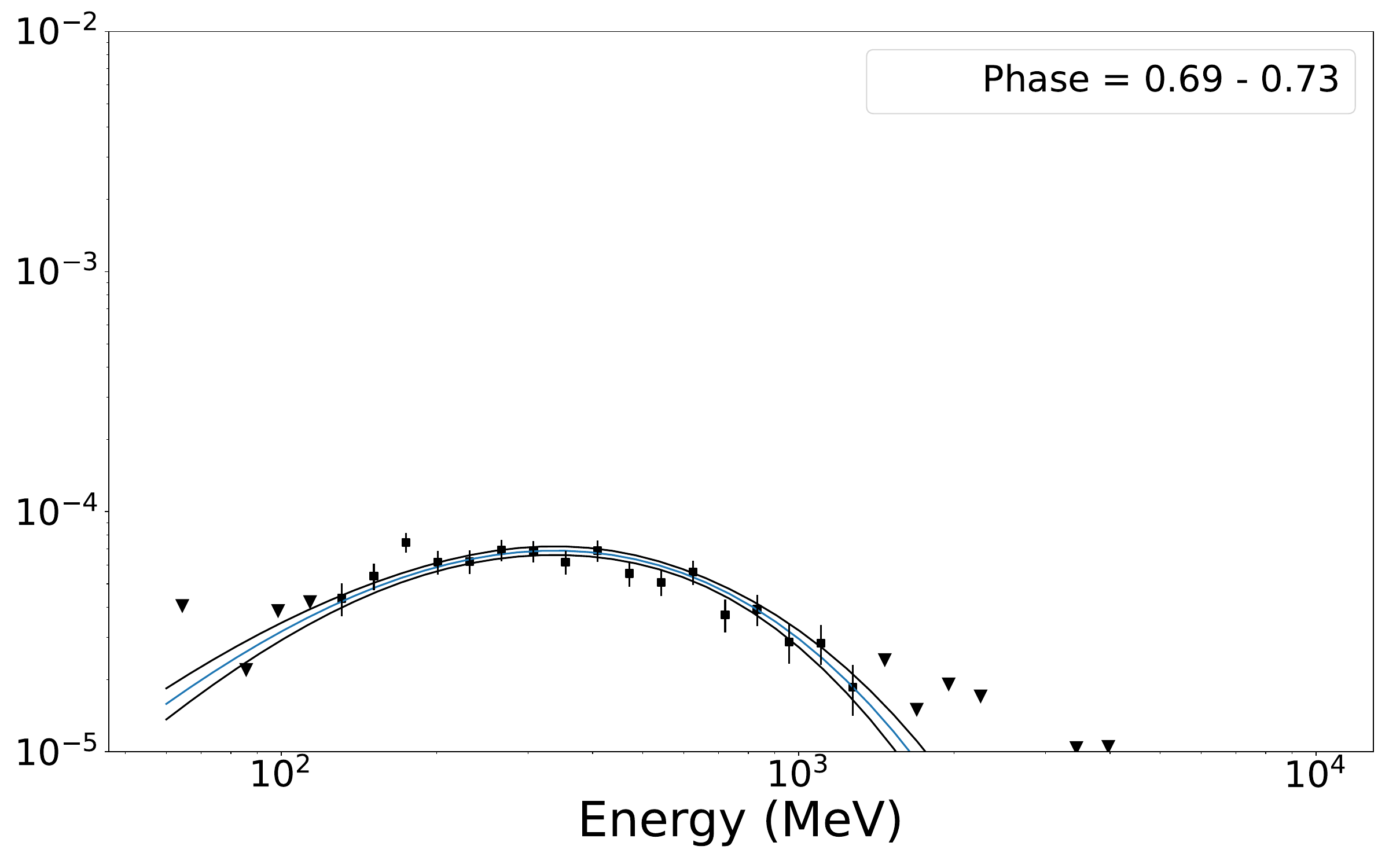}
    \end{minipage}%
    \begin{minipage}{0.32\textwidth}
        \includegraphics[width=\linewidth]{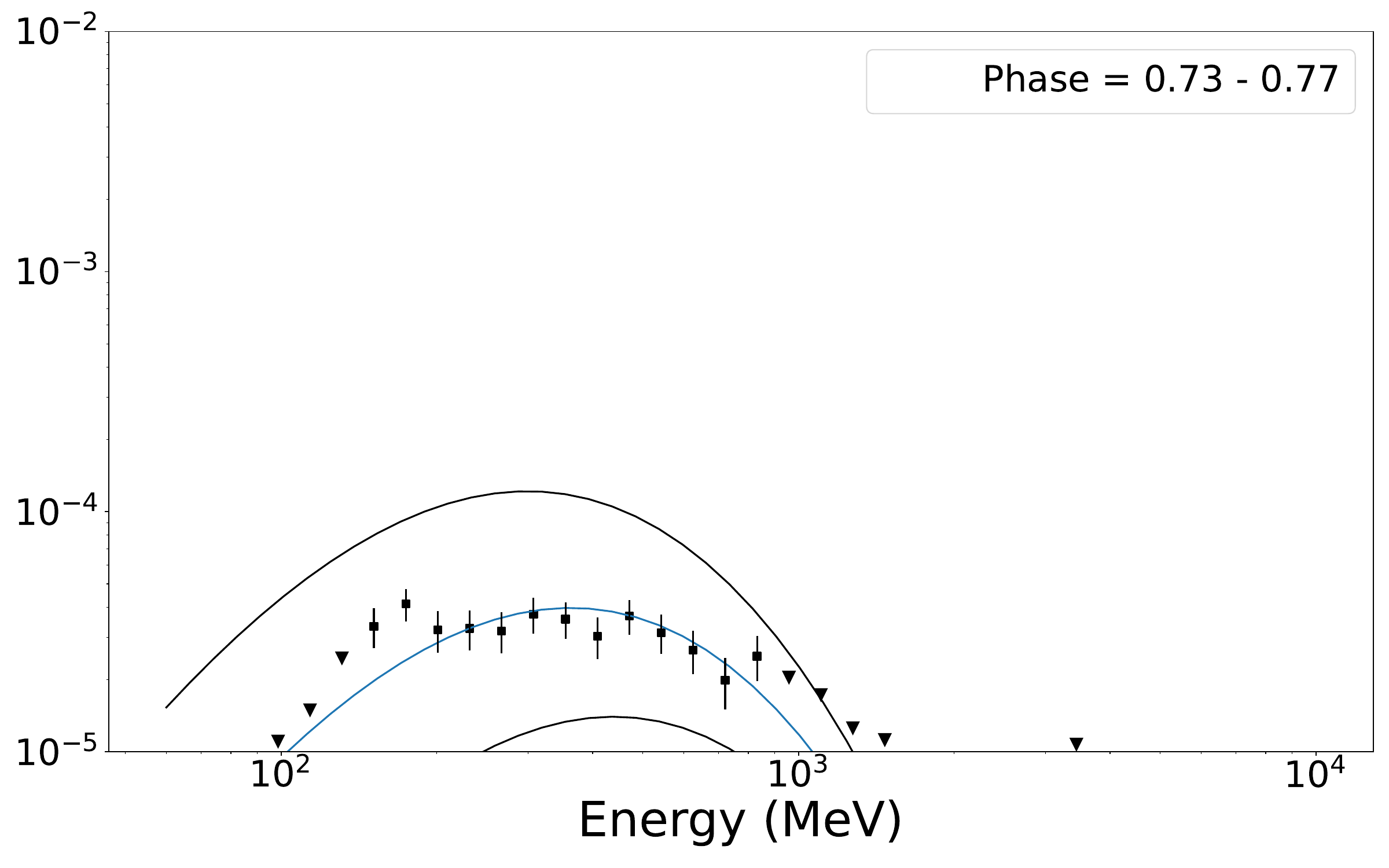}
    \end{minipage}
    
    \caption{The $E^2 dN/dE$ spectrum for the Vela pulsar in the phase range [0.08, 0.77] for $b=0.394$. The phase bin is indicated in each plot. Spectral data are indicated by black crosses, while upper-limits are indicated by black triangles. The best-fit model is represented by a solid blue line and the $95\%$ confidence interval is represented by black lines.}
\label{app:e2_fixedb}
\end{figure*}

\begin{figure*}[!h]

    \centering
    
    \begin{minipage}{0.32\textwidth}
        \includegraphics[width=\linewidth]{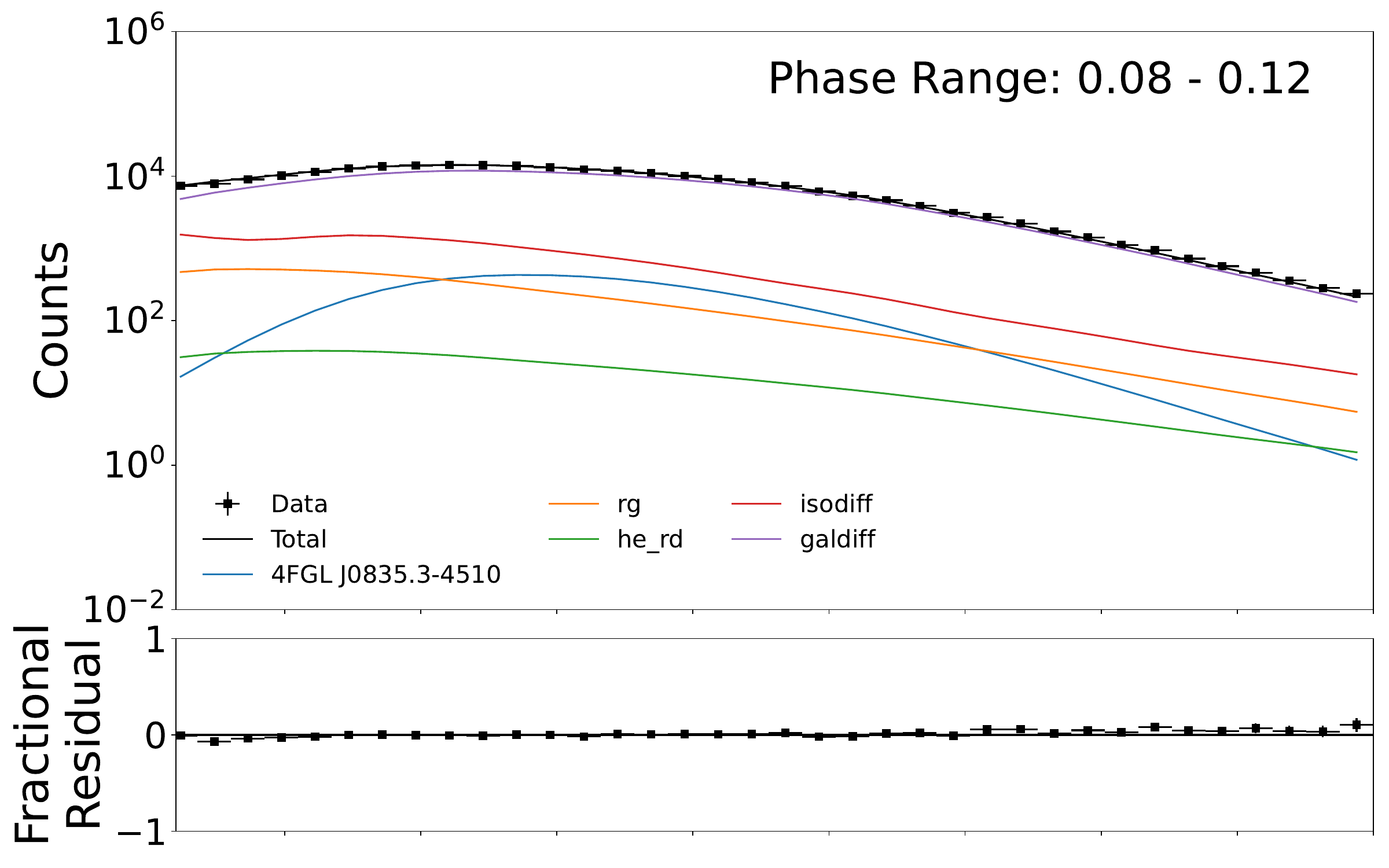}
    \end{minipage}%
    \begin{minipage}{0.32\textwidth}
        \includegraphics[width=\linewidth]{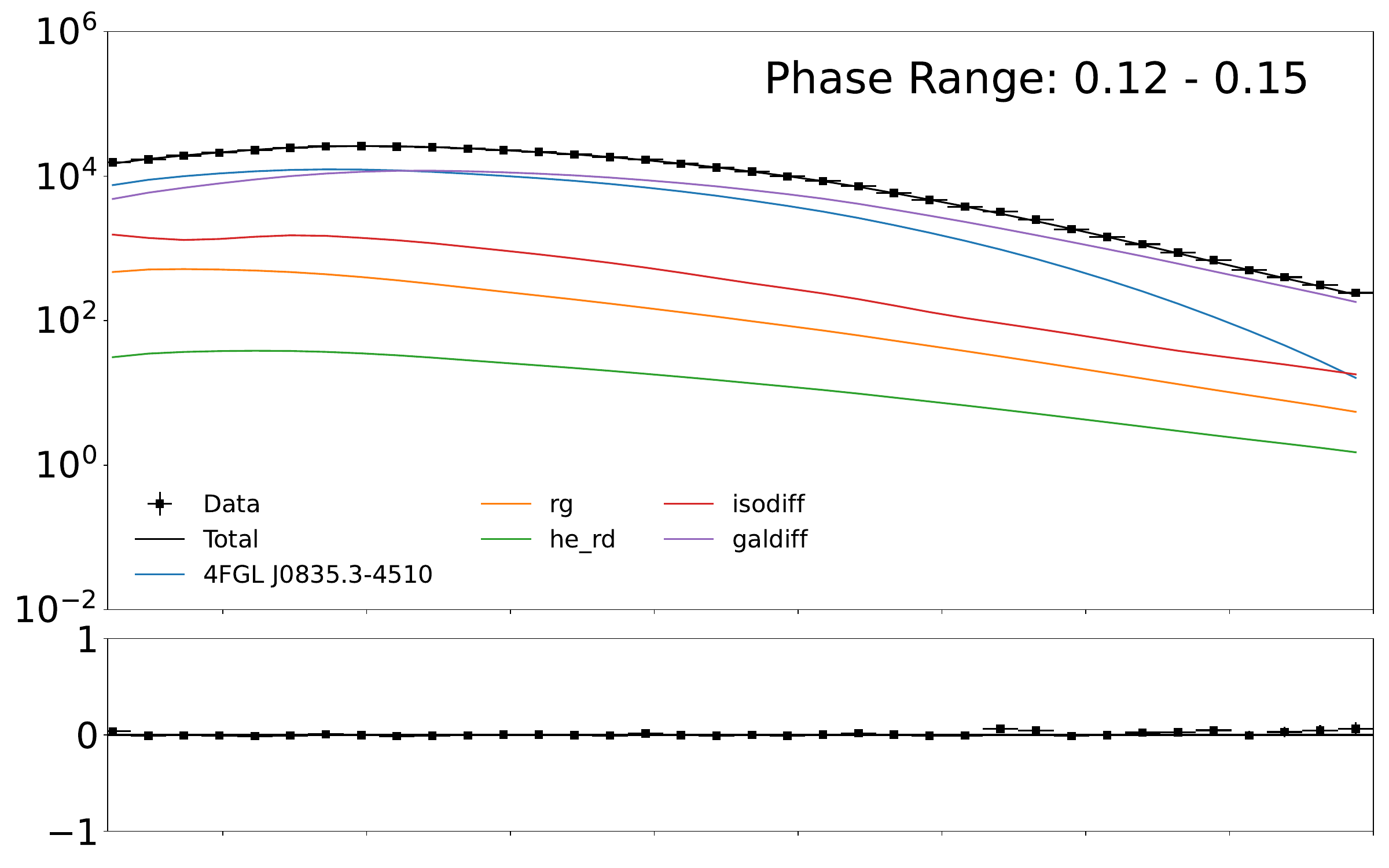}
    \end{minipage}%
    \begin{minipage}{0.32\textwidth}
        \includegraphics[width=\linewidth]{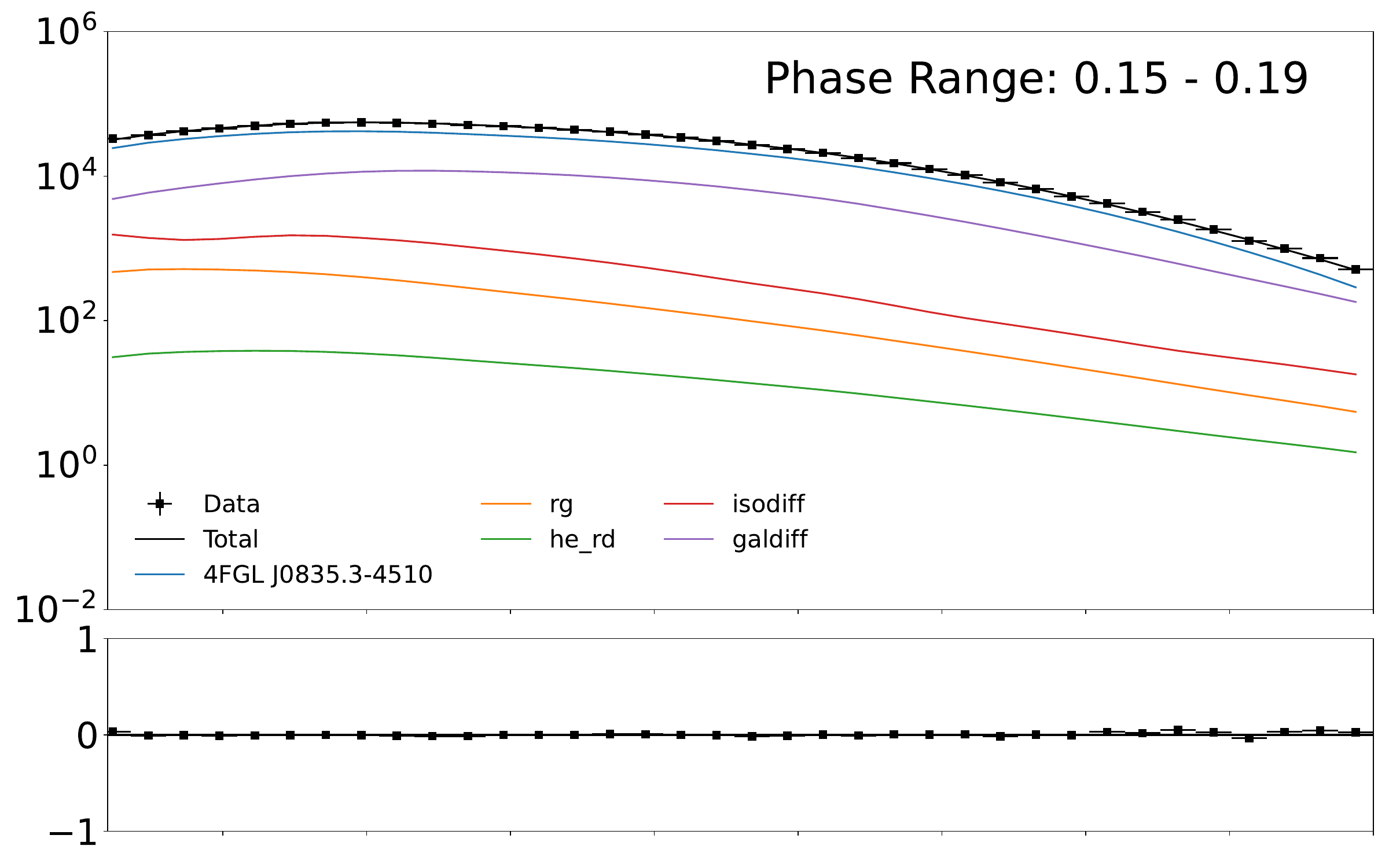}
    \end{minipage}

    \vspace{-2pt}

    \begin{minipage}{0.32\textwidth}
        \includegraphics[width=\linewidth]{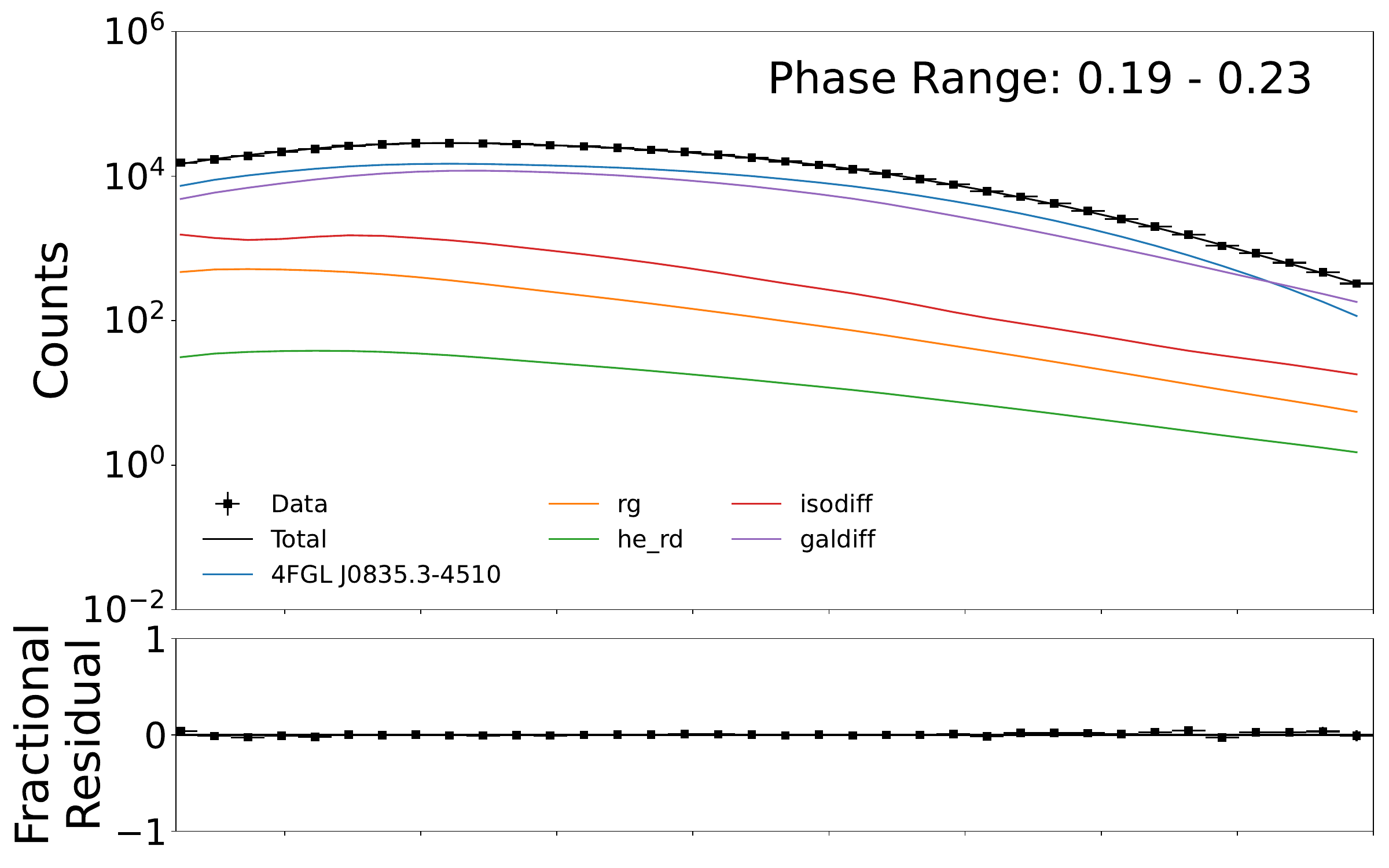}
    \end{minipage}%
    \begin{minipage}{0.32\textwidth}
        \includegraphics[width=\linewidth]{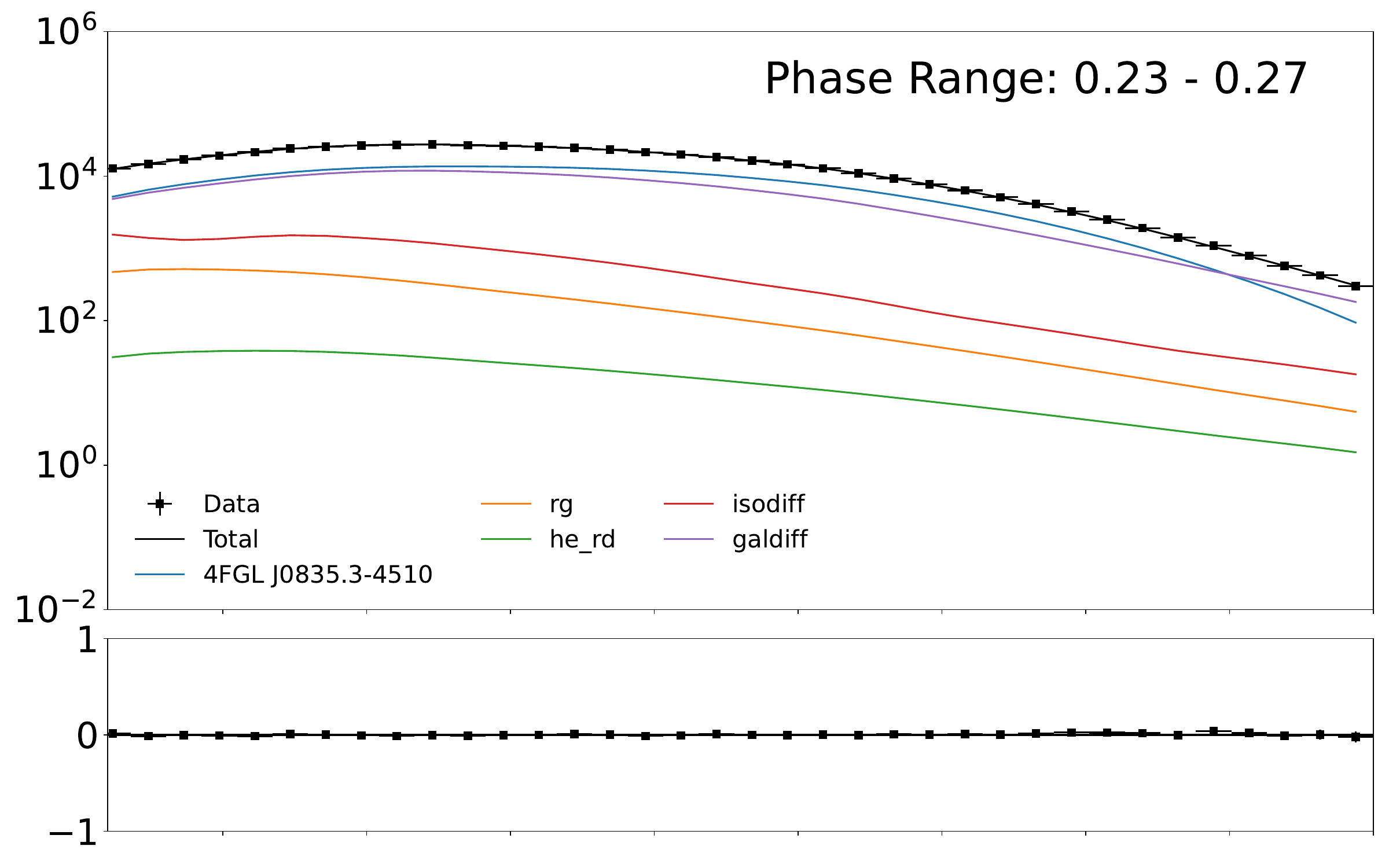}
    \end{minipage}%
    \begin{minipage}{0.32\textwidth}
        \includegraphics[width=\linewidth]{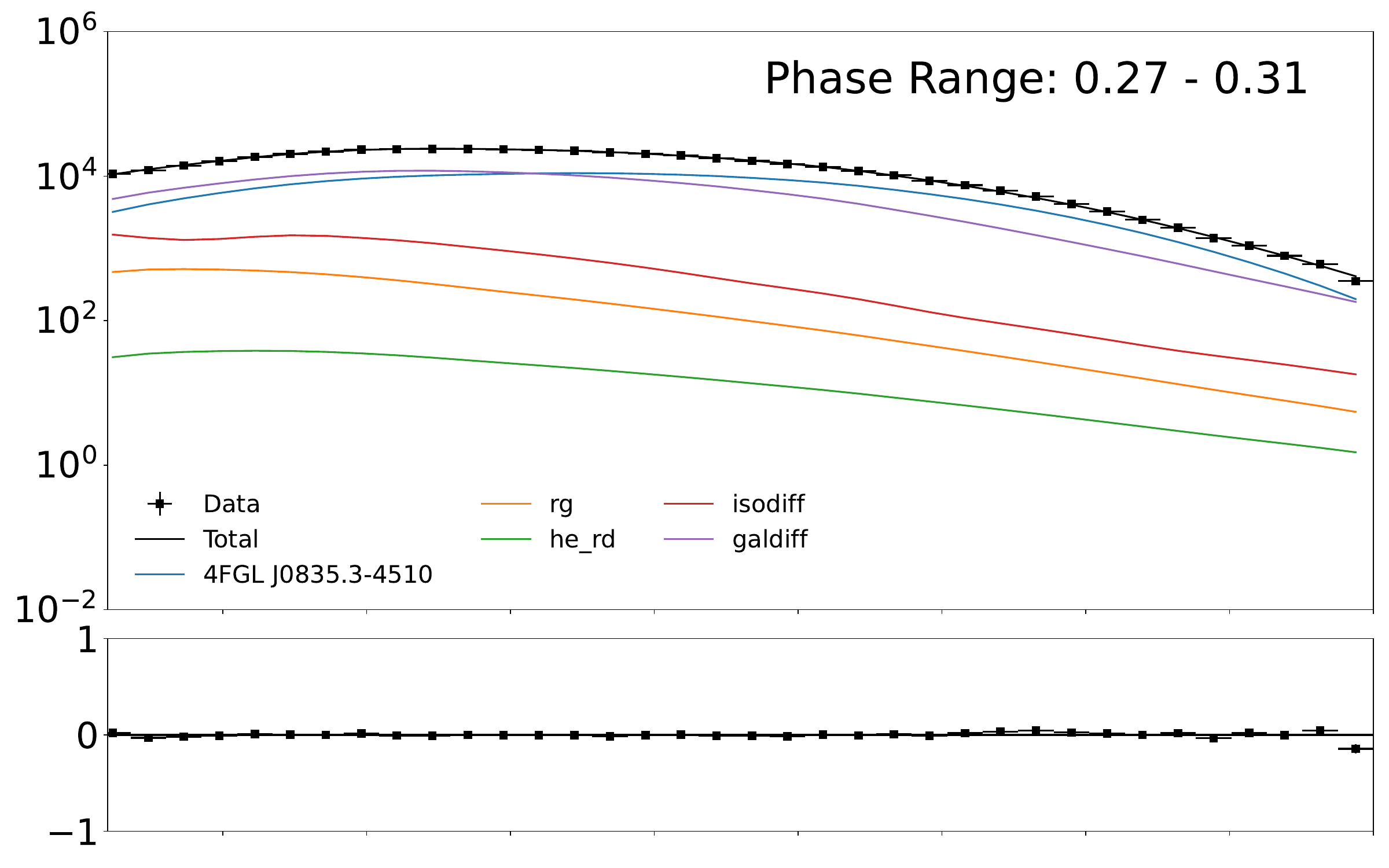}
    \end{minipage}

    \vspace{-2pt}

    \begin{minipage}{0.32\textwidth}
        \includegraphics[width=\linewidth]{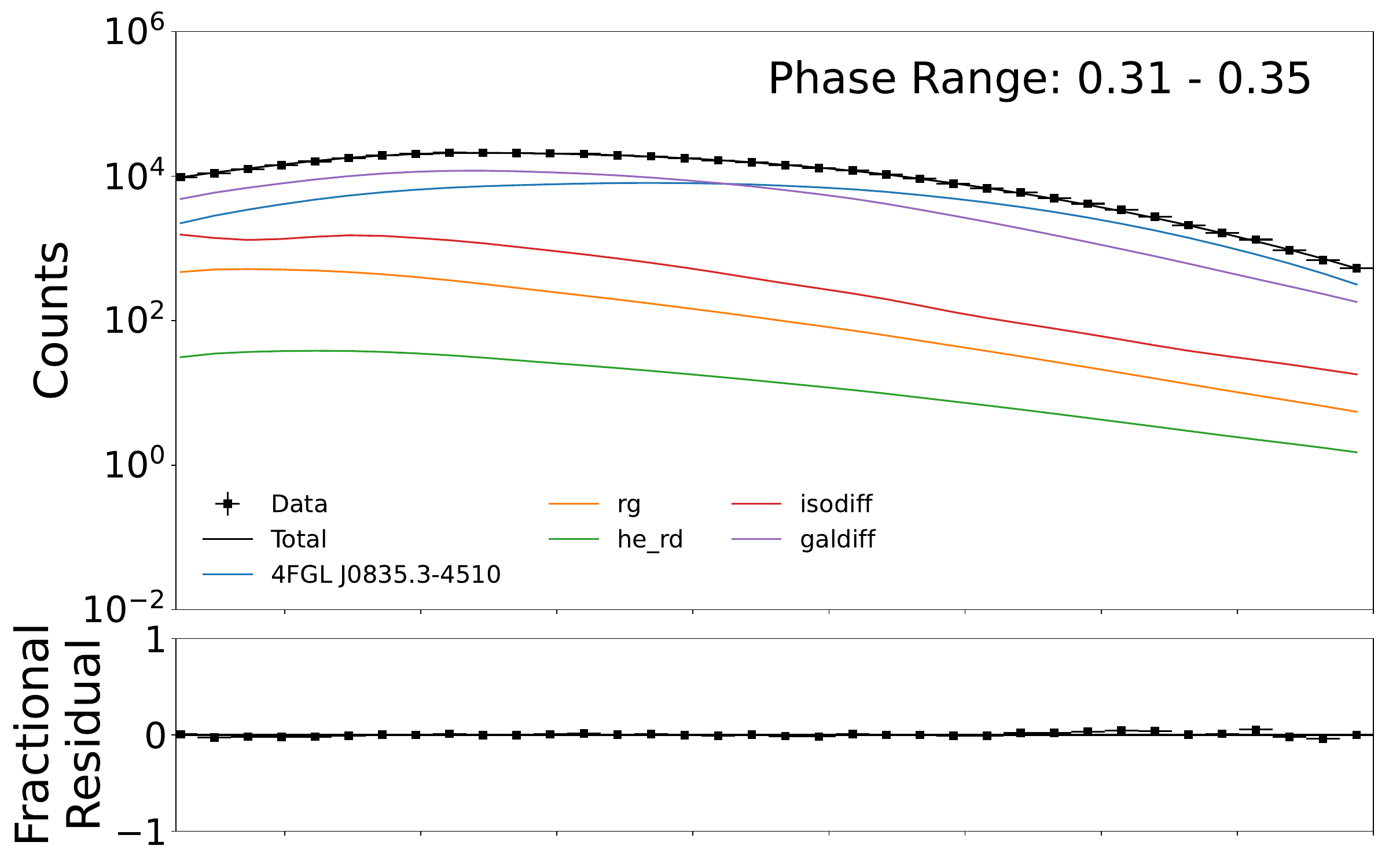}
    \end{minipage}%
    \begin{minipage}{0.32\textwidth}
        \includegraphics[width=\linewidth]{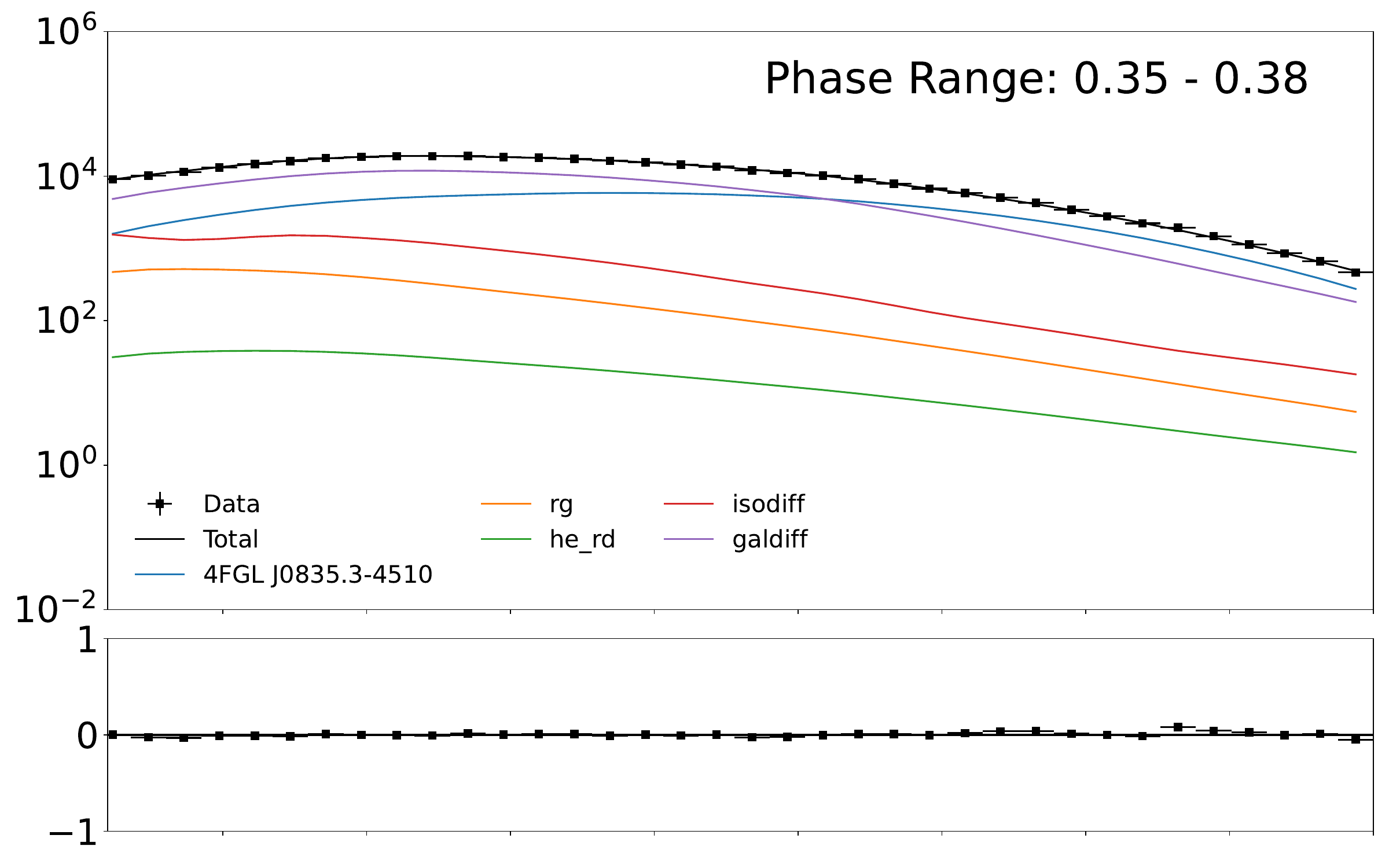}
    \end{minipage}%
    \begin{minipage}{0.32\textwidth}
        \includegraphics[width=\linewidth]{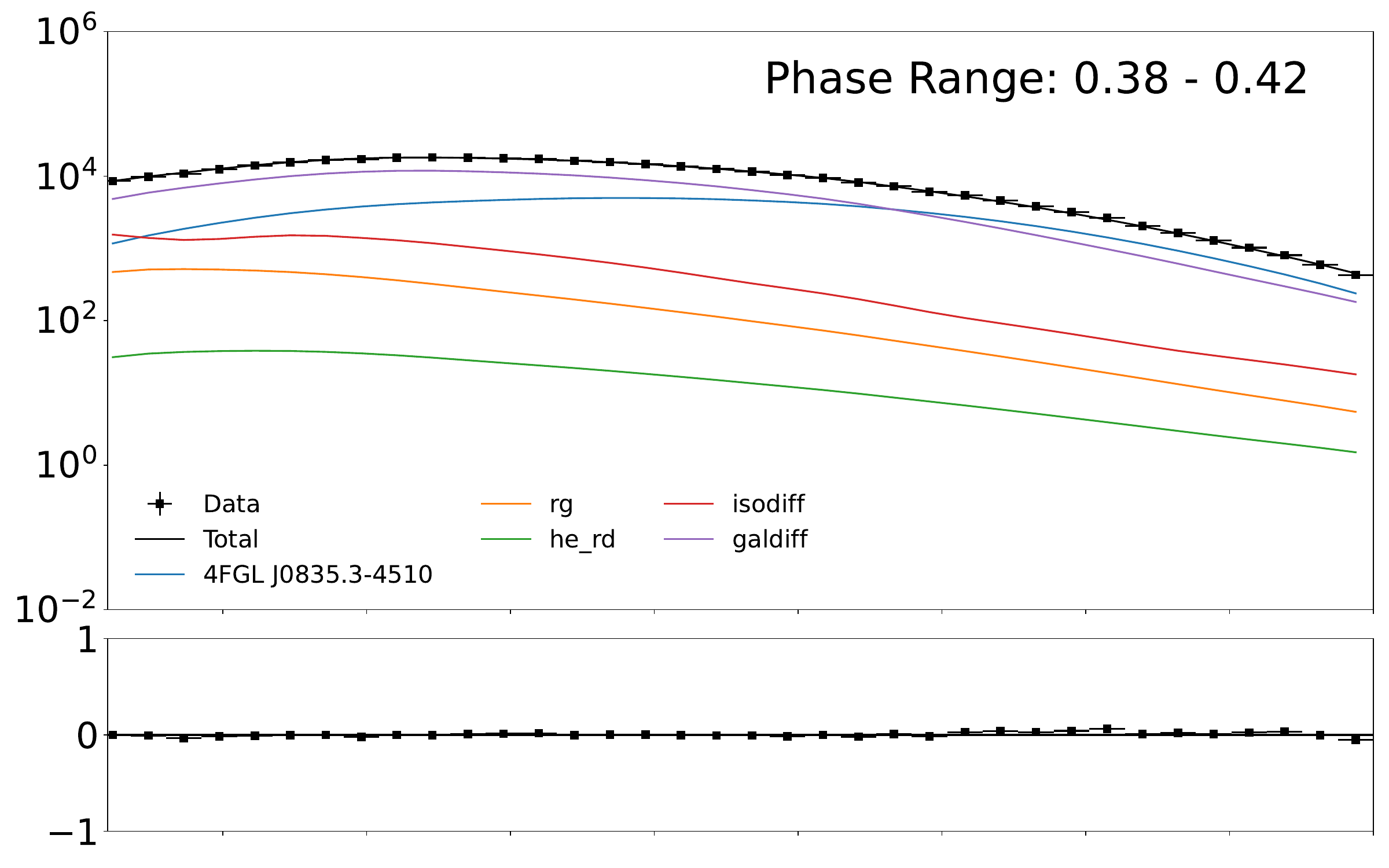}
    \end{minipage}

    \vspace{-2pt}
    
    \begin{minipage}{0.32\textwidth}
        \includegraphics[width=\linewidth]{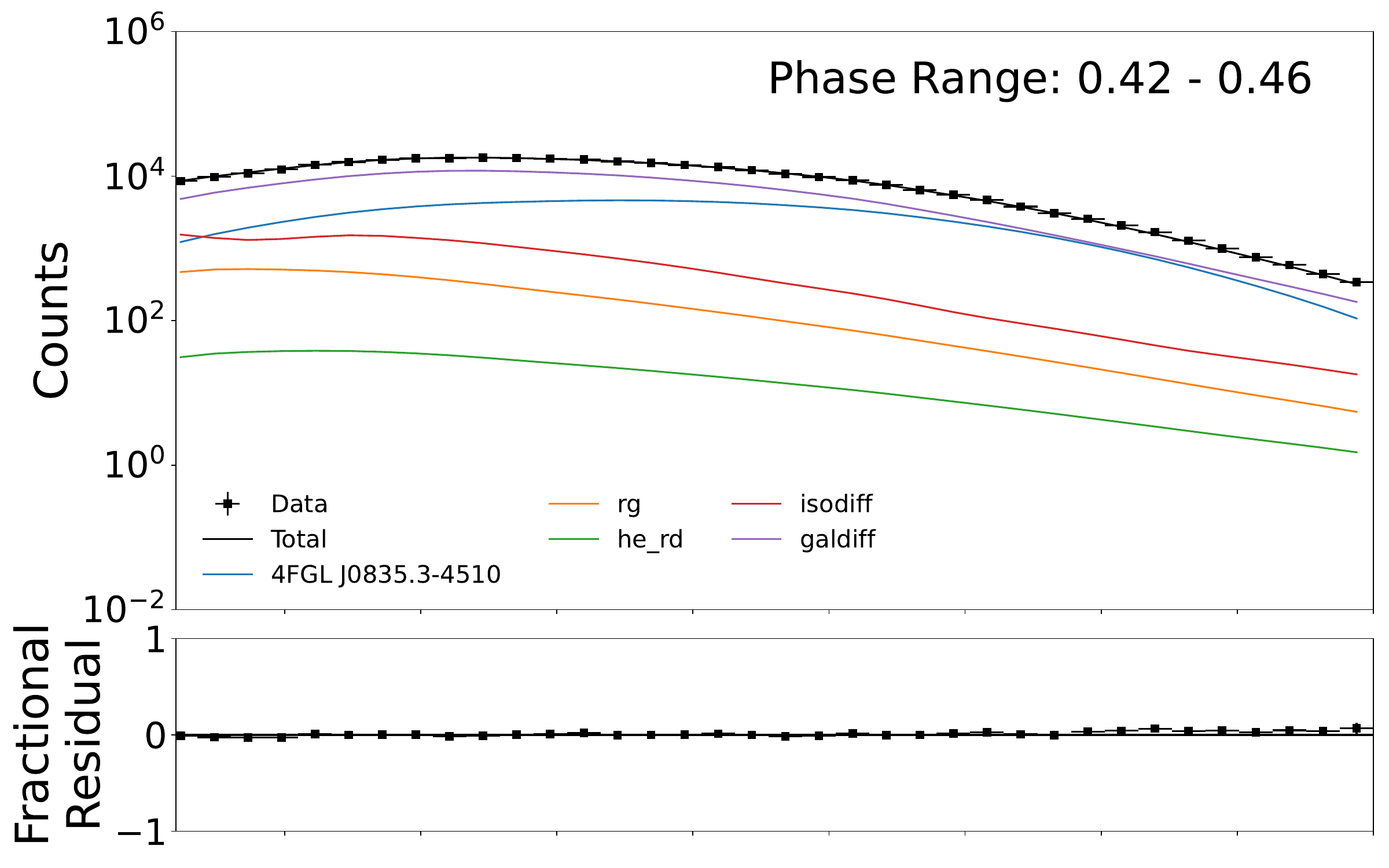}
    \end{minipage}%
    \begin{minipage}{0.32\textwidth}
        \includegraphics[width=\linewidth]{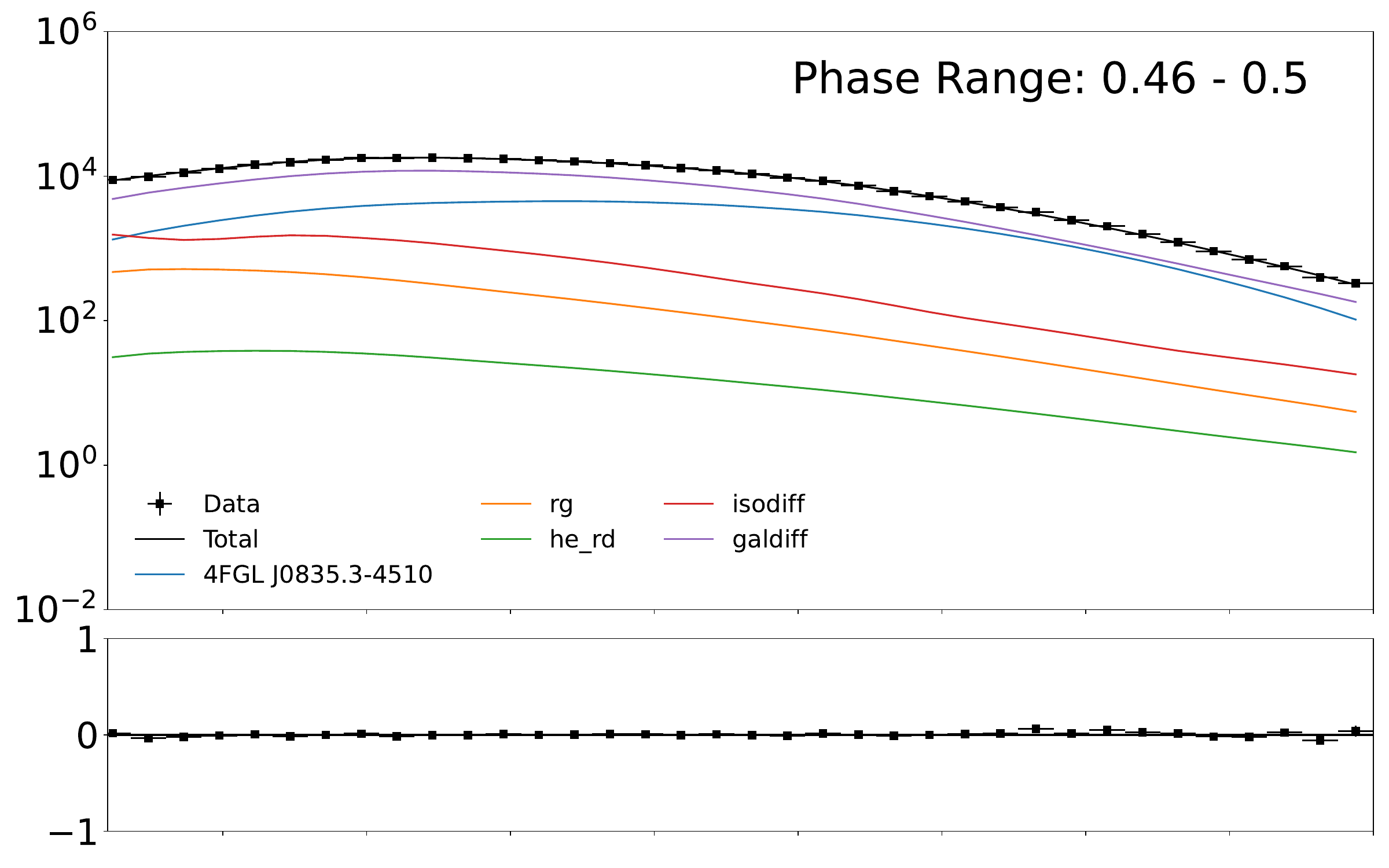}
    \end{minipage}%
    \begin{minipage}{0.32\textwidth}
        \includegraphics[width=\linewidth]{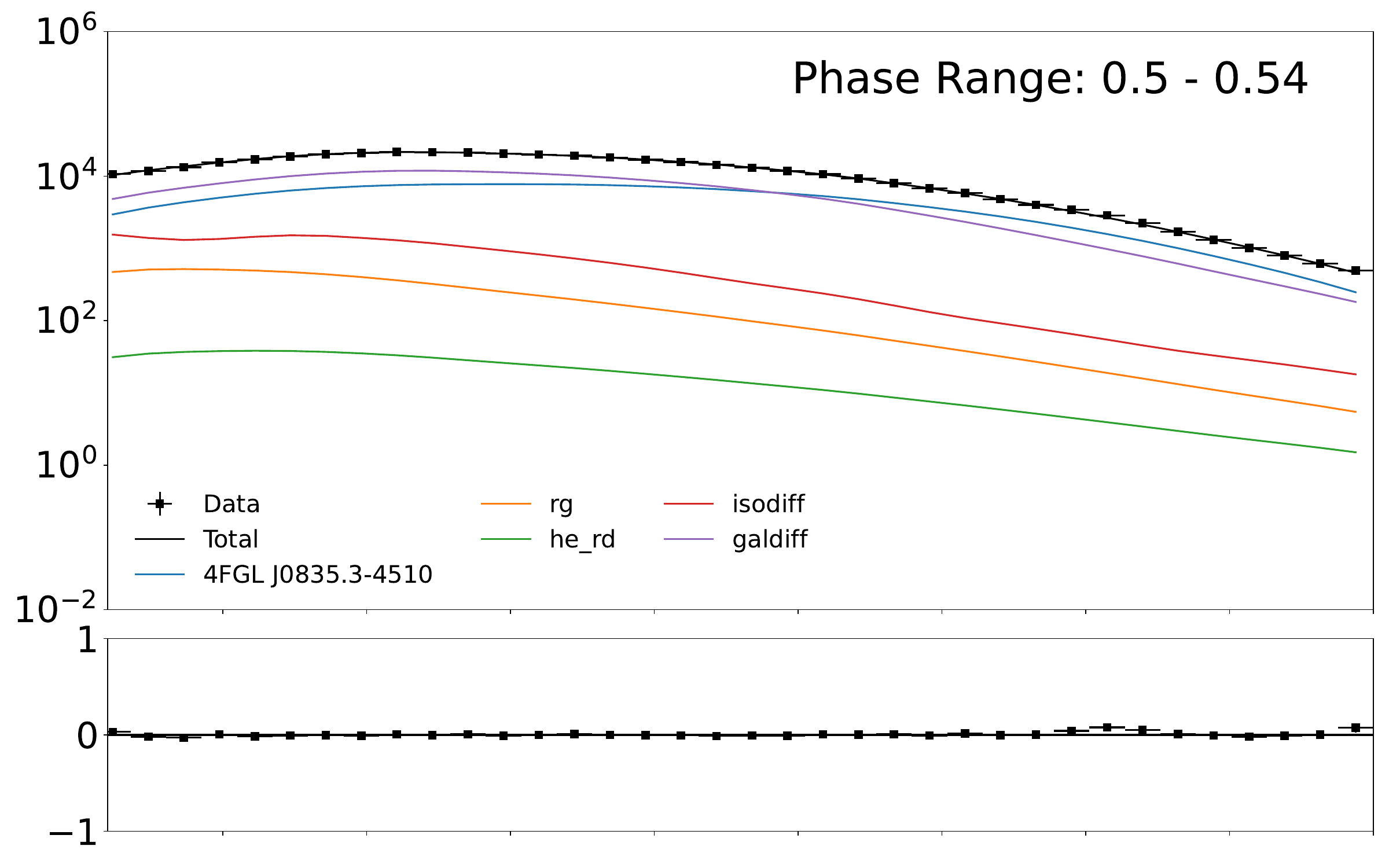}
    \end{minipage}

    \vspace{-2pt}

    \begin{minipage}{0.32\textwidth}
        \includegraphics[width=\linewidth]{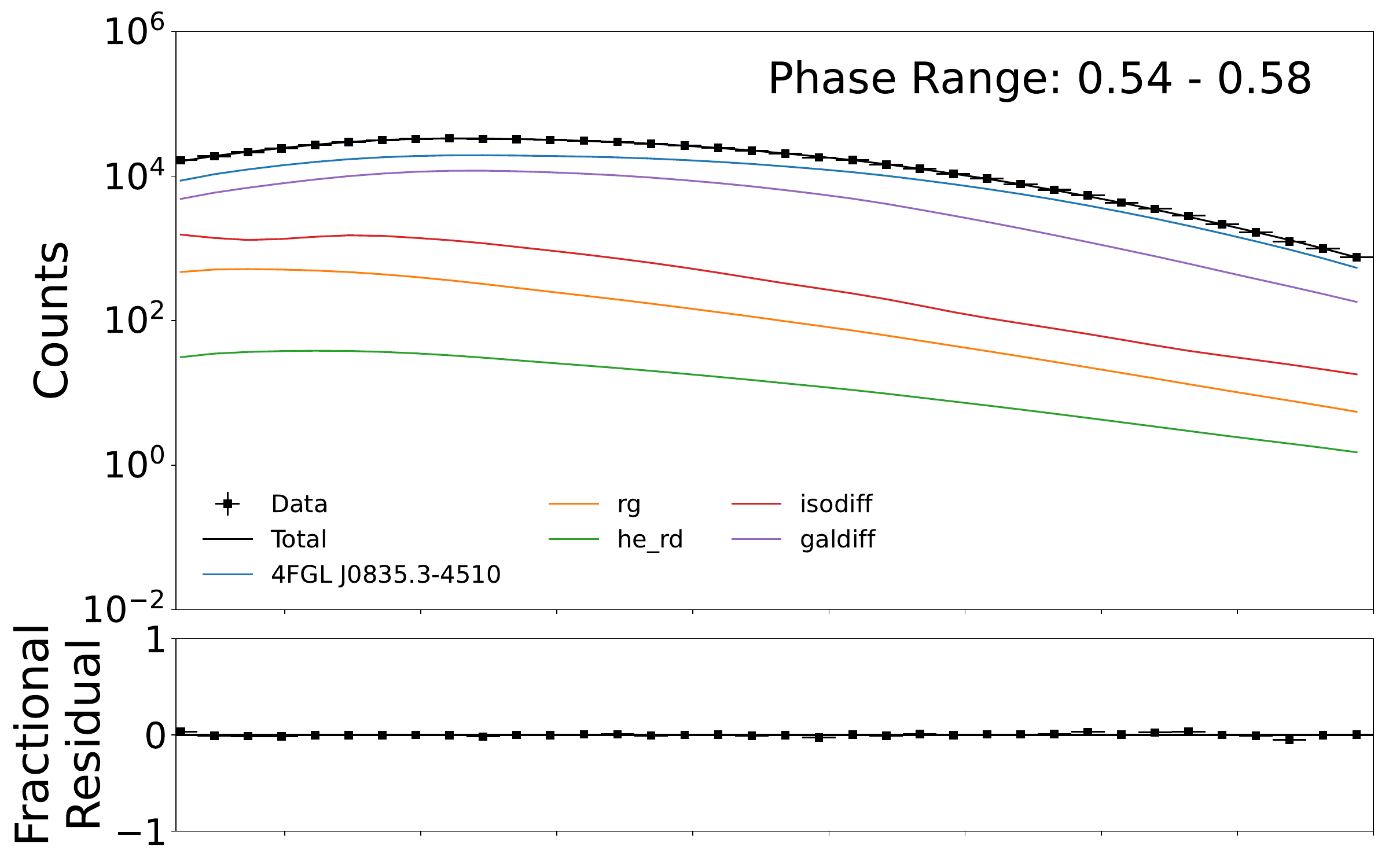}
    \end{minipage}%
    \begin{minipage}{0.32\textwidth}
        \includegraphics[width=\linewidth]{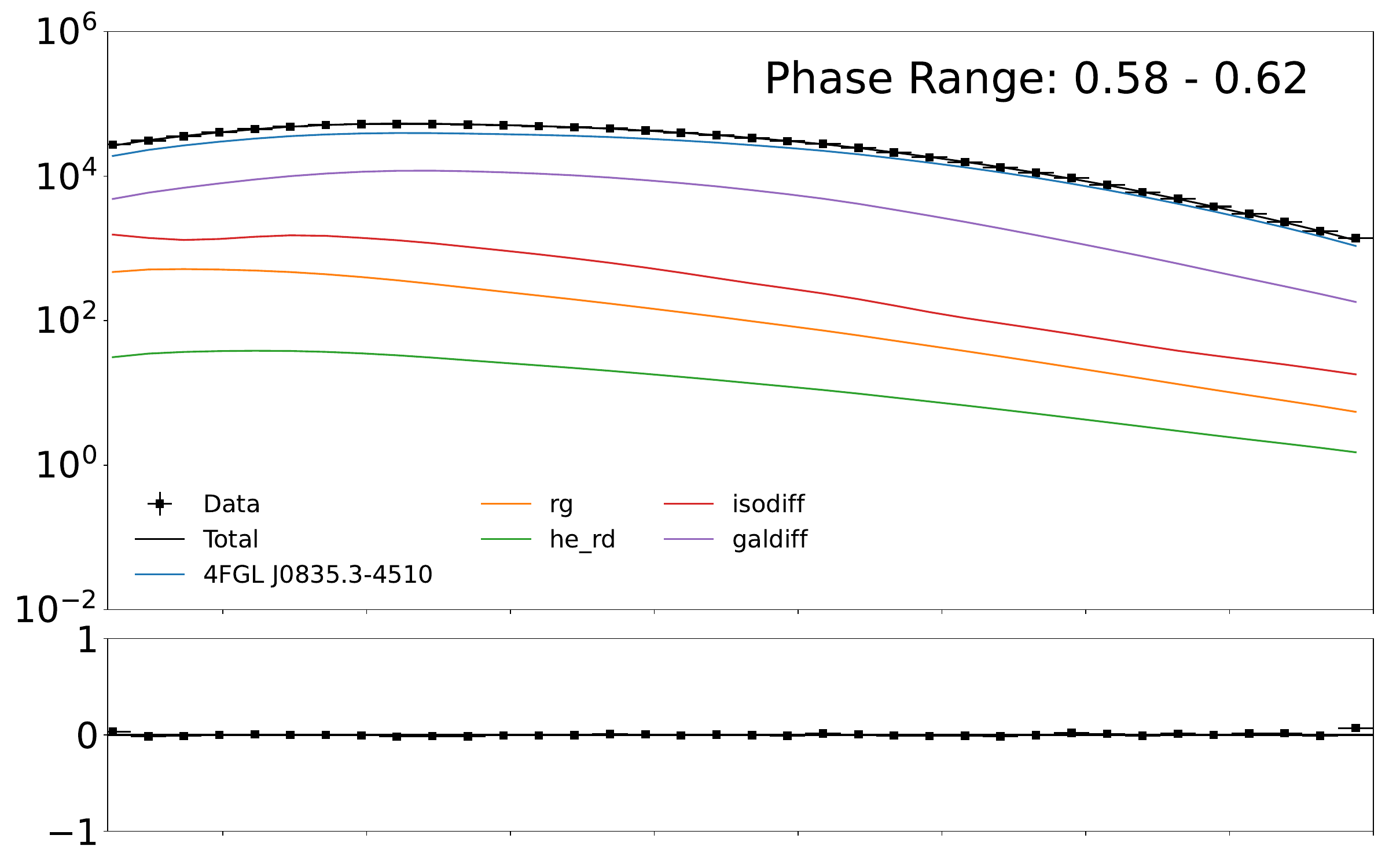}
    \end{minipage}
    \begin{minipage}{0.32\textwidth}
        \includegraphics[width=\linewidth]{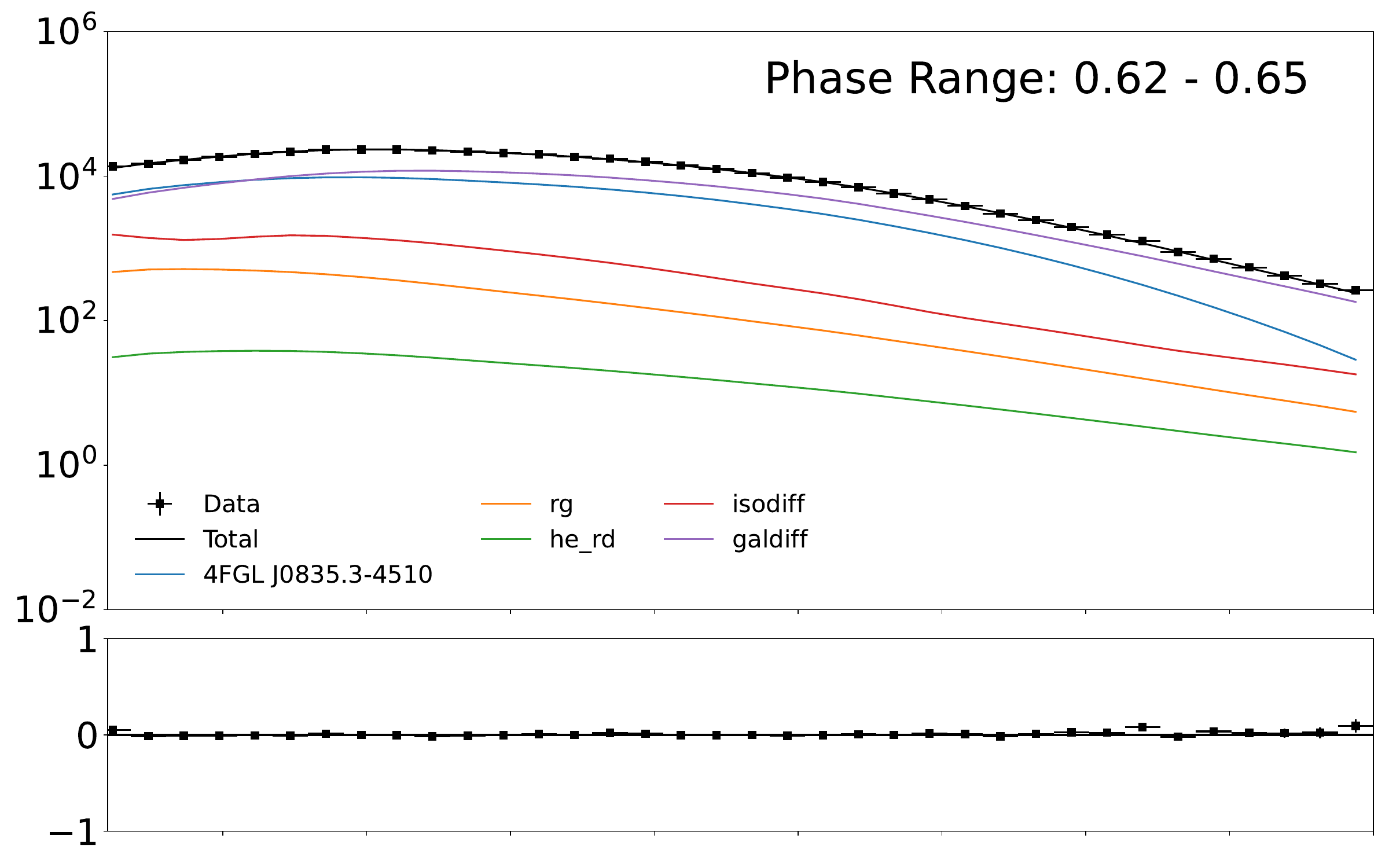}
    \end{minipage}

    \vspace{-2pt}
    
    \begin{minipage}{0.32\textwidth}
        \includegraphics[width=\linewidth]{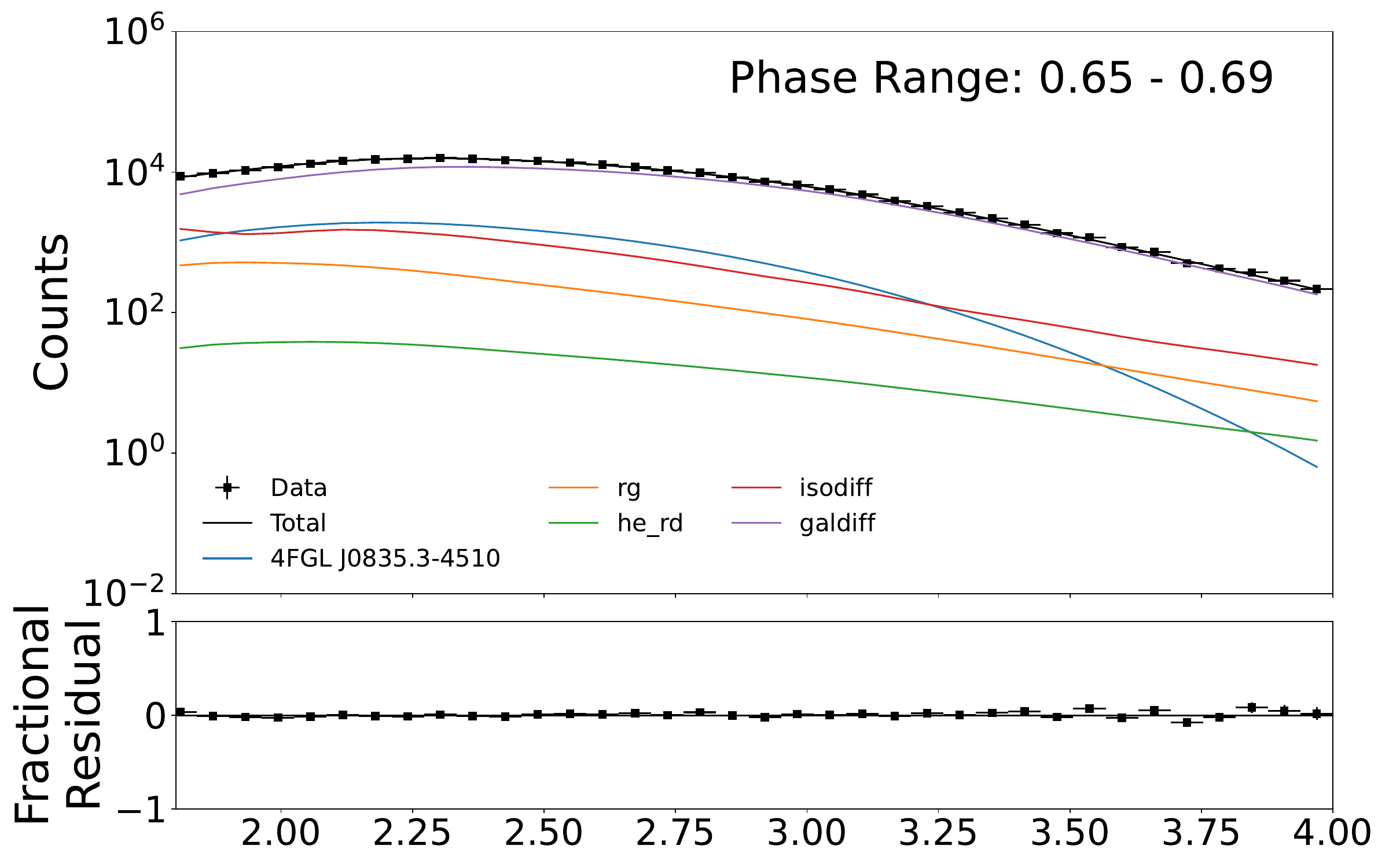}
    \end{minipage}
    \begin{minipage}{0.32\textwidth}
        \includegraphics[width=\linewidth]{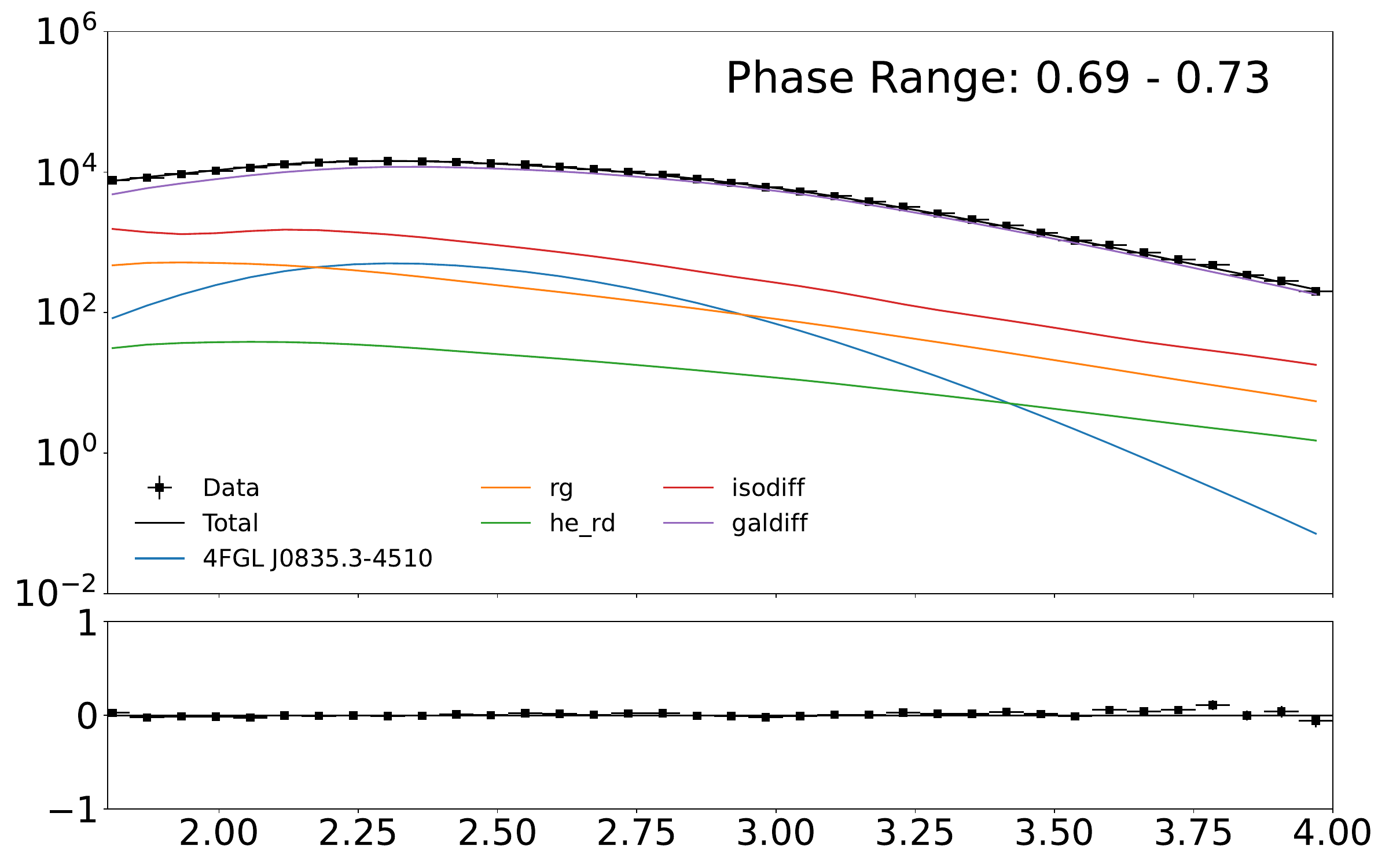}
    \end{minipage}%
    \begin{minipage}{0.32\textwidth}
        \includegraphics[width=\linewidth]{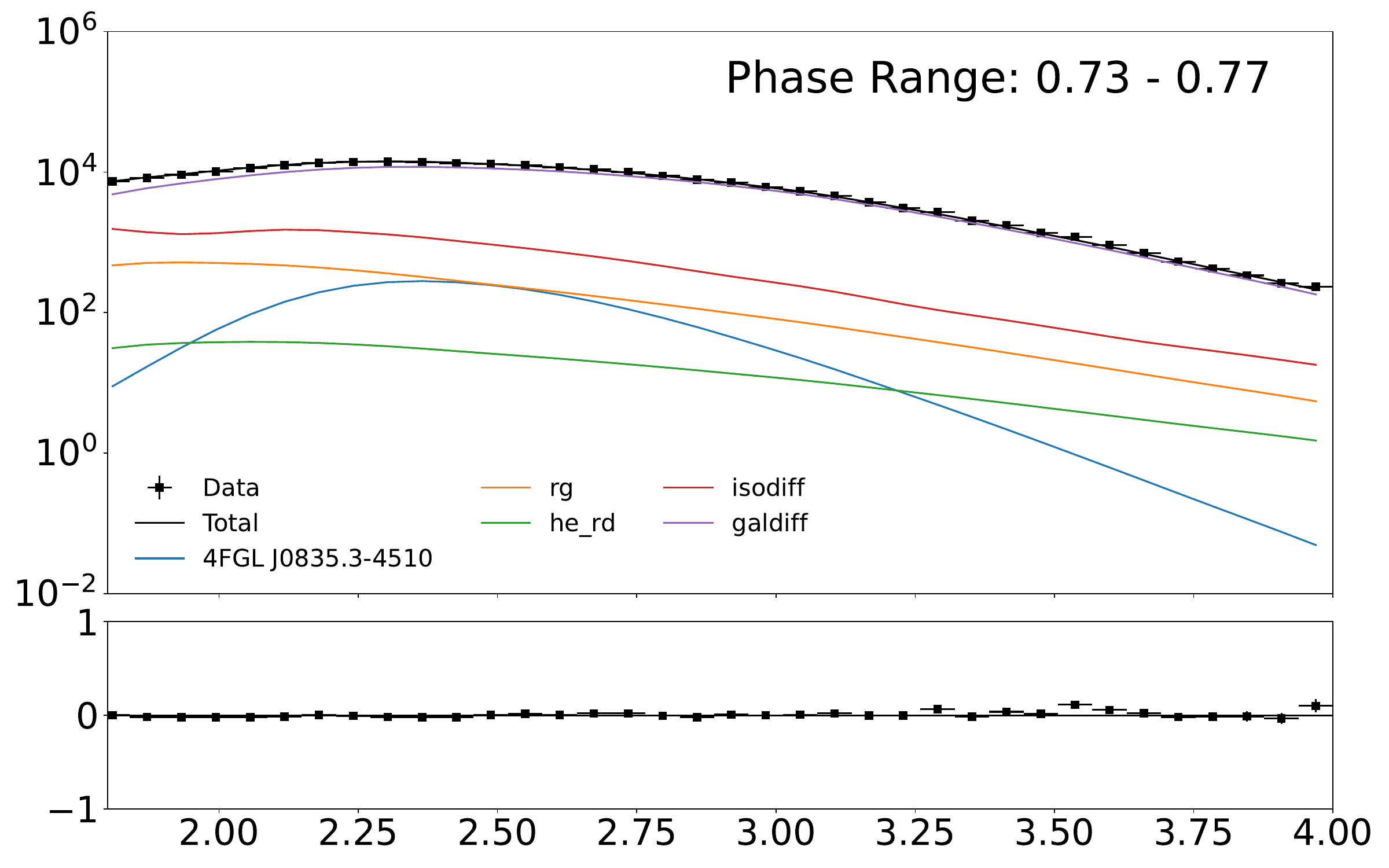}
    \end{minipage}
    
    \vspace{-60pt}
    
    \begin{minipage}{0.32\textwidth}
        \includegraphics[width=\linewidth]{B_label.pdf}
    \end{minipage}
    \begin{minipage}{0.32\textwidth}
        \includegraphics[width=\linewidth]{B_label.pdf}
    \end{minipage}
    \begin{minipage}{0.32\textwidth}
        \includegraphics[width=\linewidth]{B_label.pdf}
    \end{minipage}
    
    \vspace{-40pt} 
    
    \caption{Count spectra of the Vela pulsar for the phase range [0.08, 0.77] when $b$ is free to vary. Spectral data is shown in black. The five most significant sources are shown (including the isotropic and Galactic diffuse backgrounds as well as our radial Gaussian and radial disk PWN components from Section~\ref{sec:pwn}). The bottom panel of each count spectrum show the fractional residuals. The phase bin range is indicated in each plot.
    }
\label{app:count_freeb}
\end{figure*}

\begin{figure*}[!h]

    \centering
    \begin{minipage}{0.32\textwidth}
        \includegraphics[width=\linewidth]{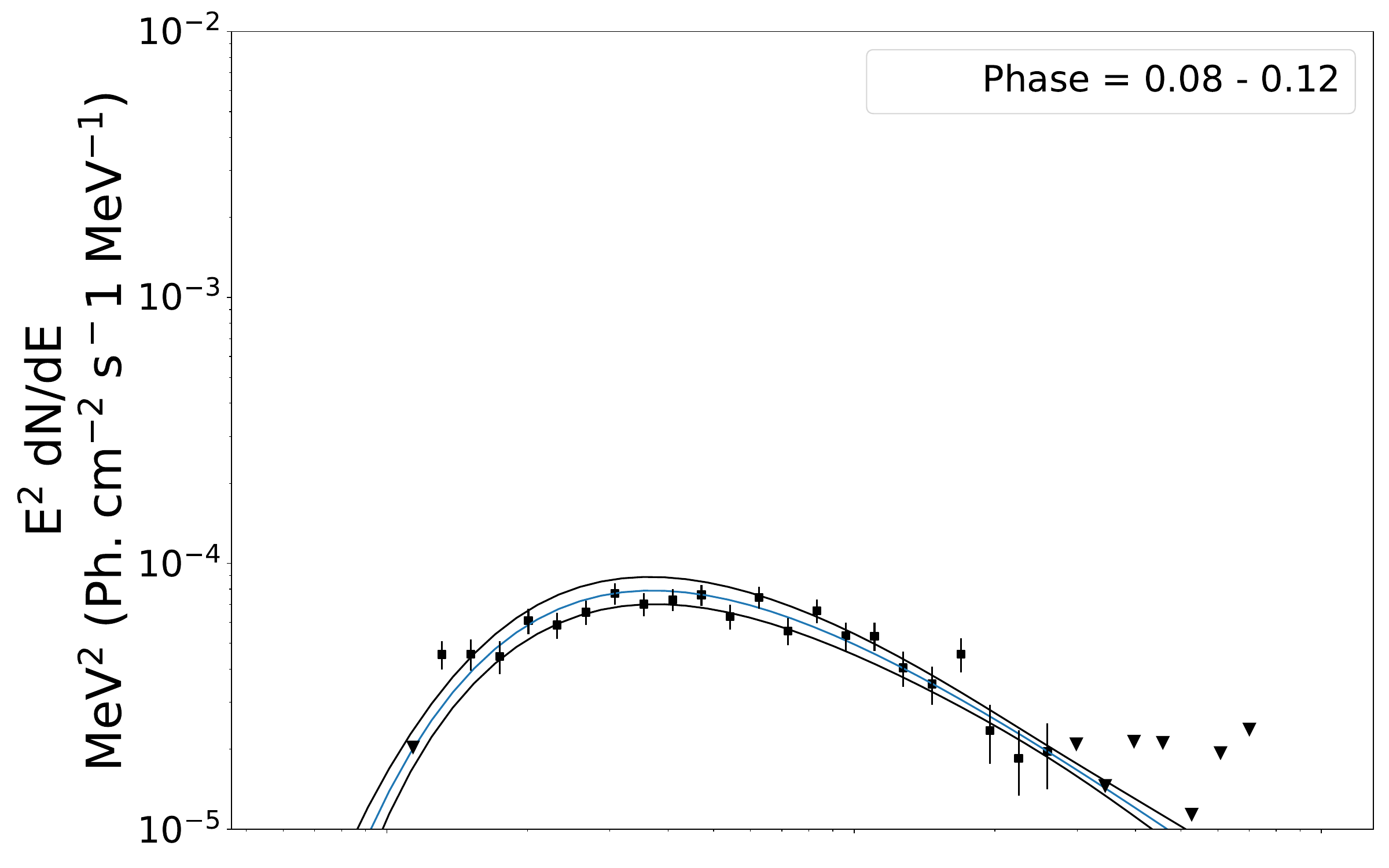}
    \end{minipage}%
    \begin{minipage}{0.32\textwidth}
        \includegraphics[width=\linewidth]{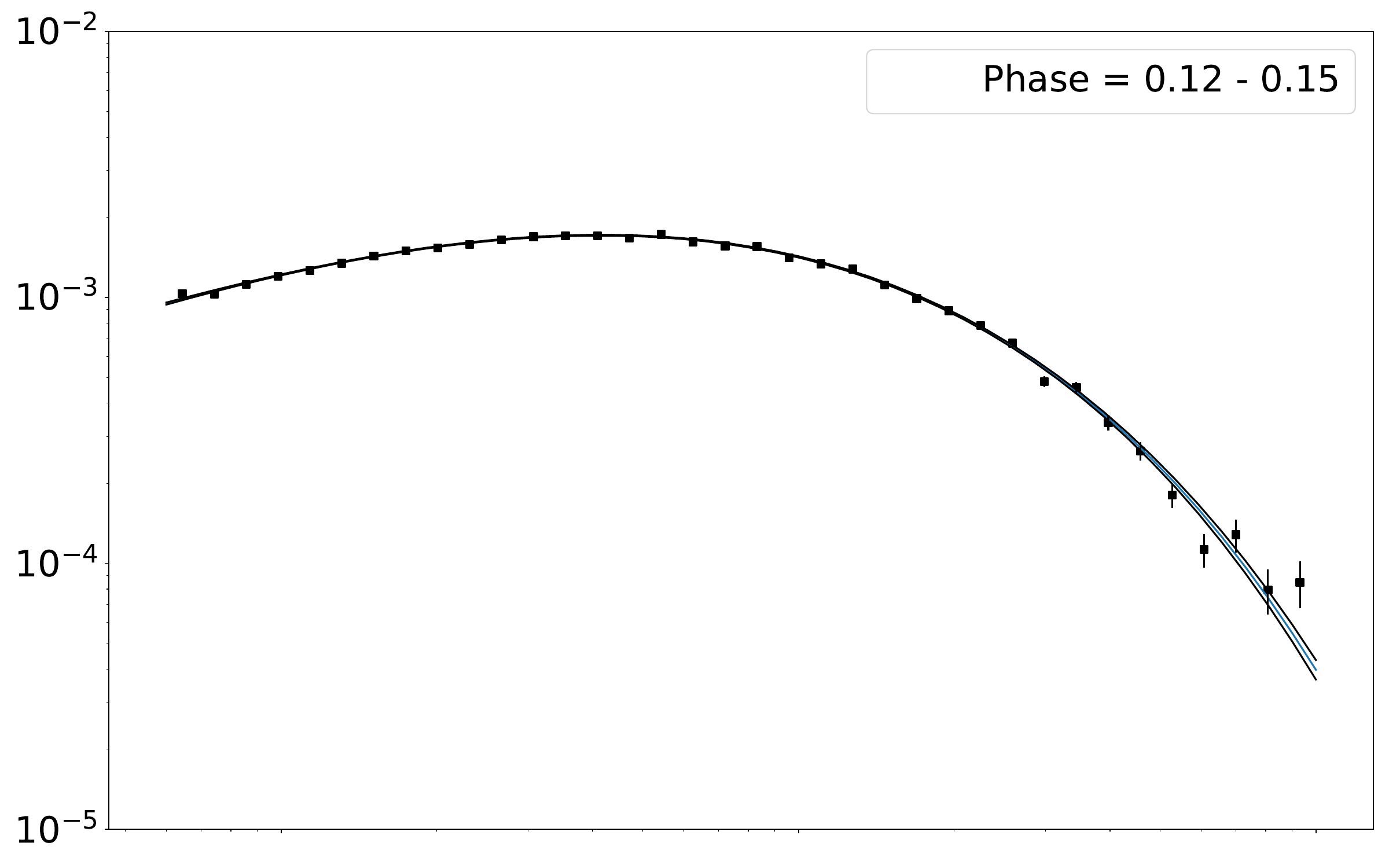}
    \end{minipage}%
    \begin{minipage}{0.32\textwidth}
        \includegraphics[width=\linewidth]{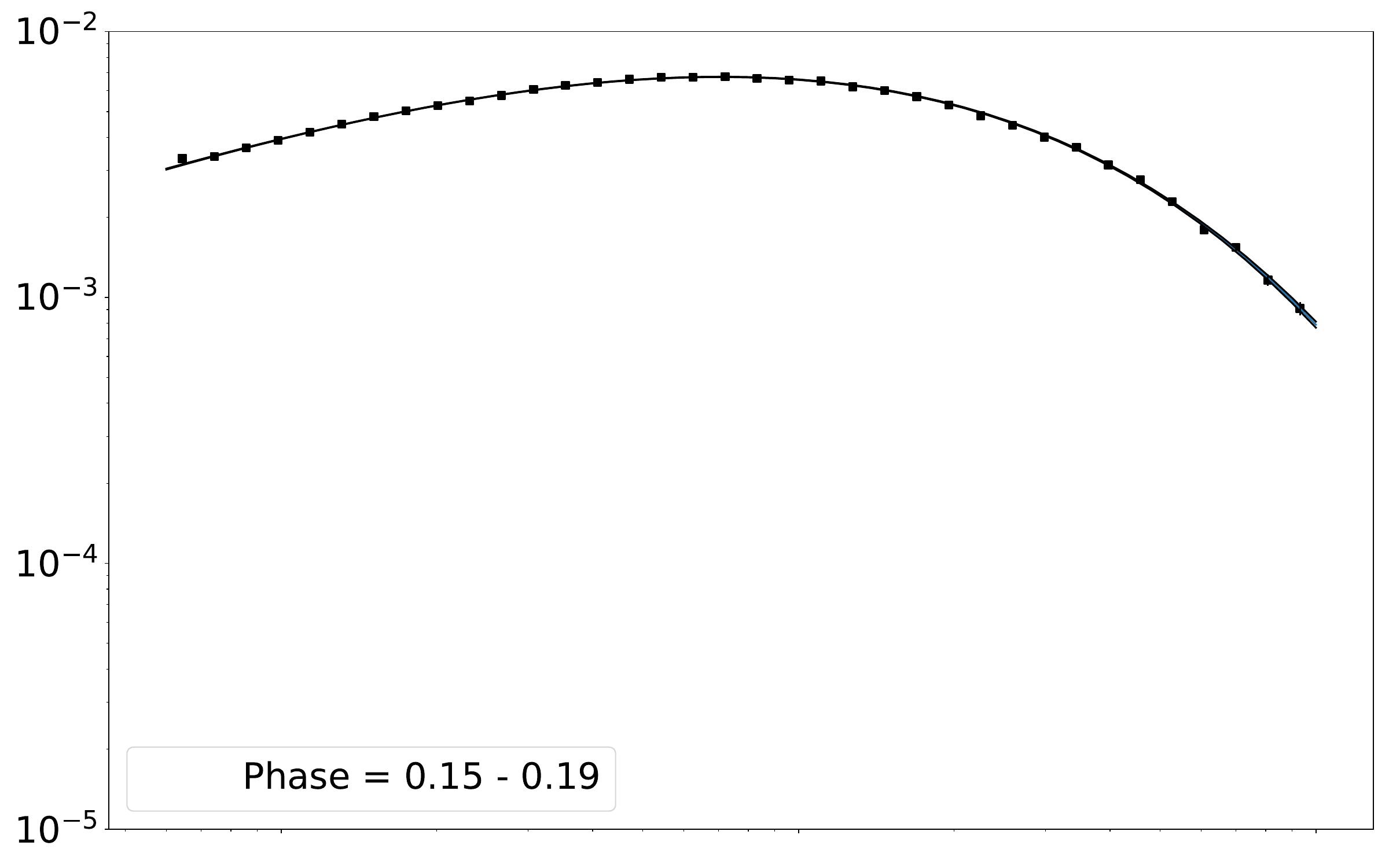}
    \end{minipage}

    \vspace{-2pt} 

    \begin{minipage}{0.32\textwidth}
        \includegraphics[width=\linewidth]{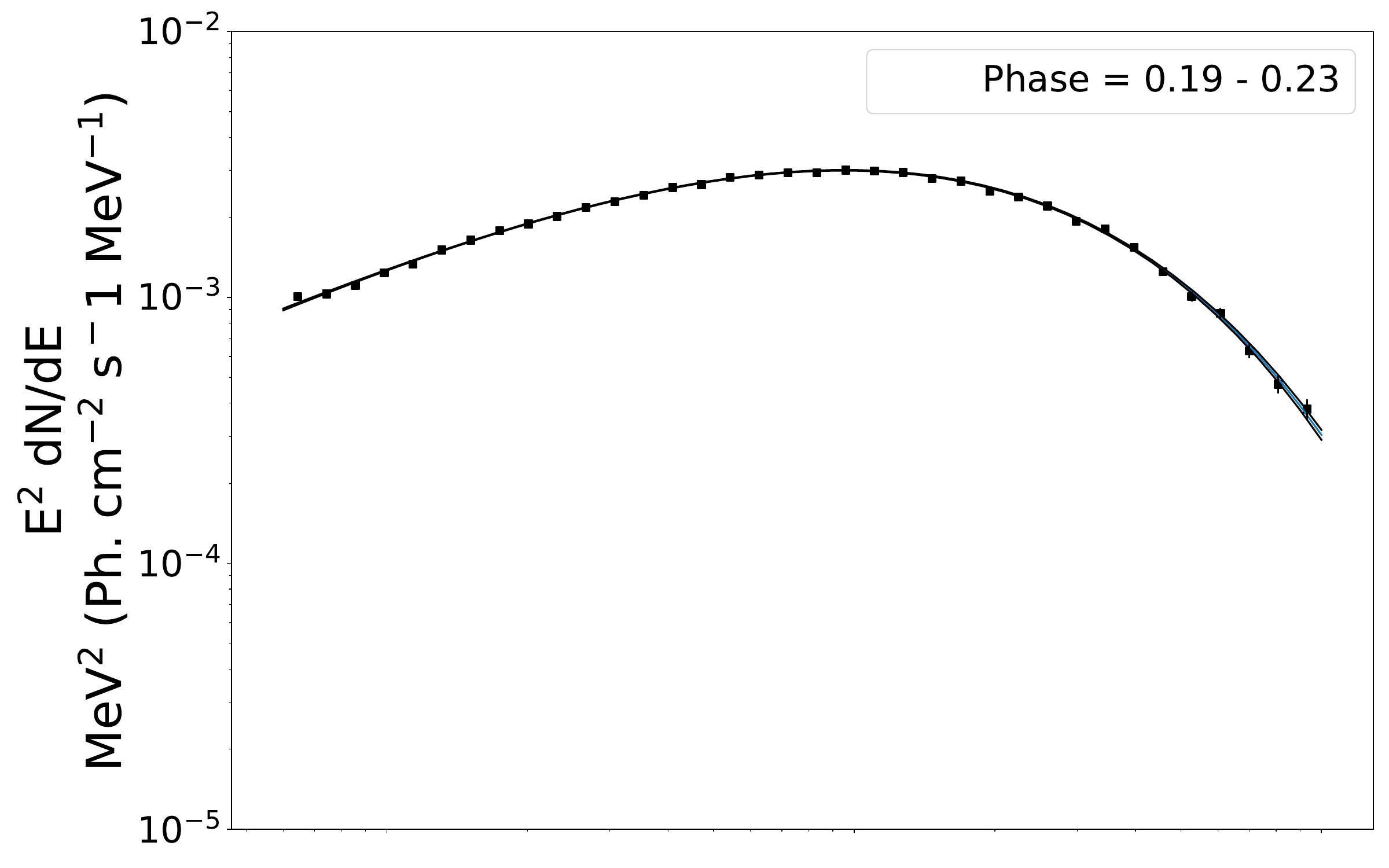}
    \end{minipage}%
    \begin{minipage}{0.32\textwidth}
        \includegraphics[width=\linewidth]{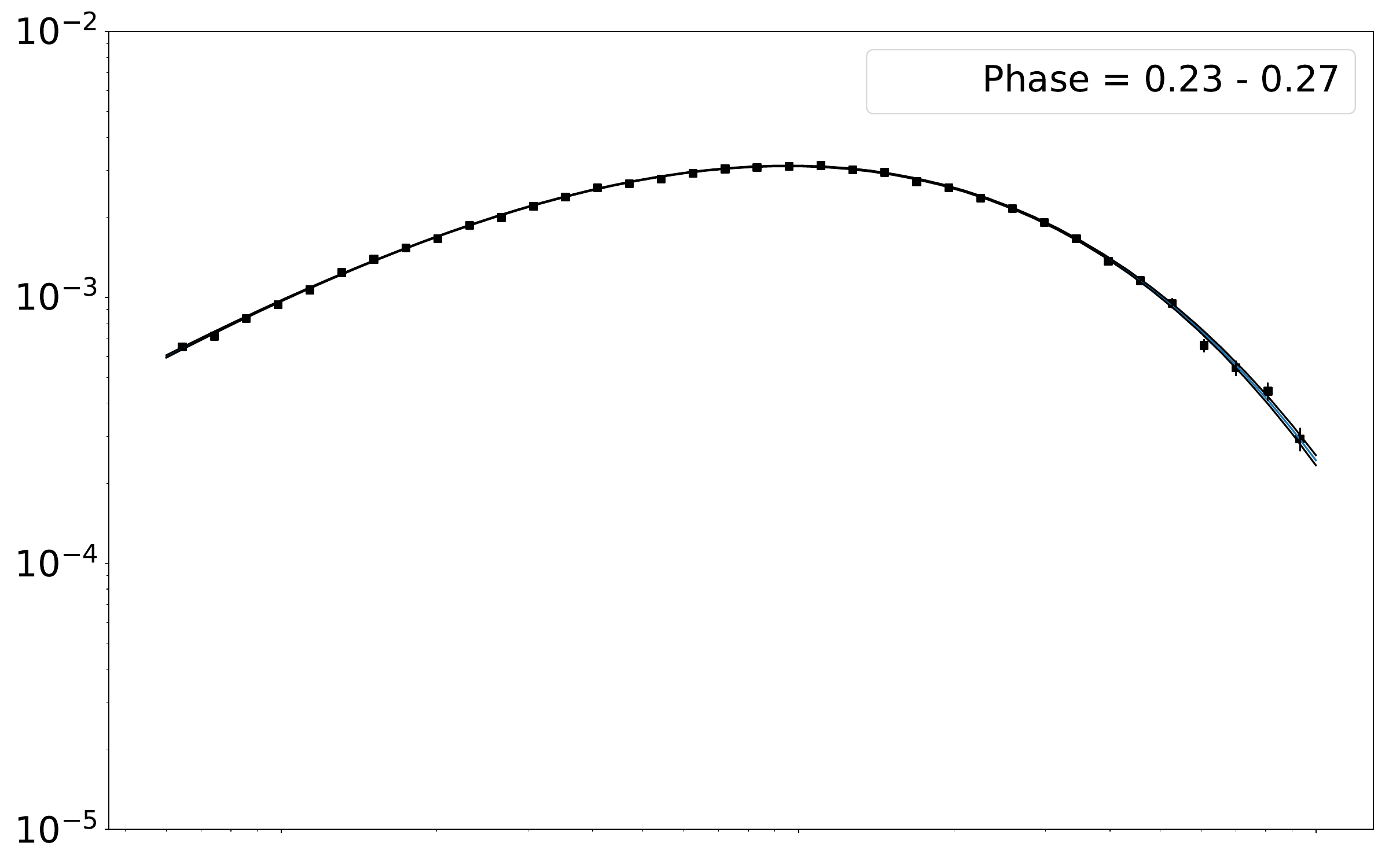}
    \end{minipage}%
    \begin{minipage}{0.32\textwidth}
        \includegraphics[width=\linewidth]{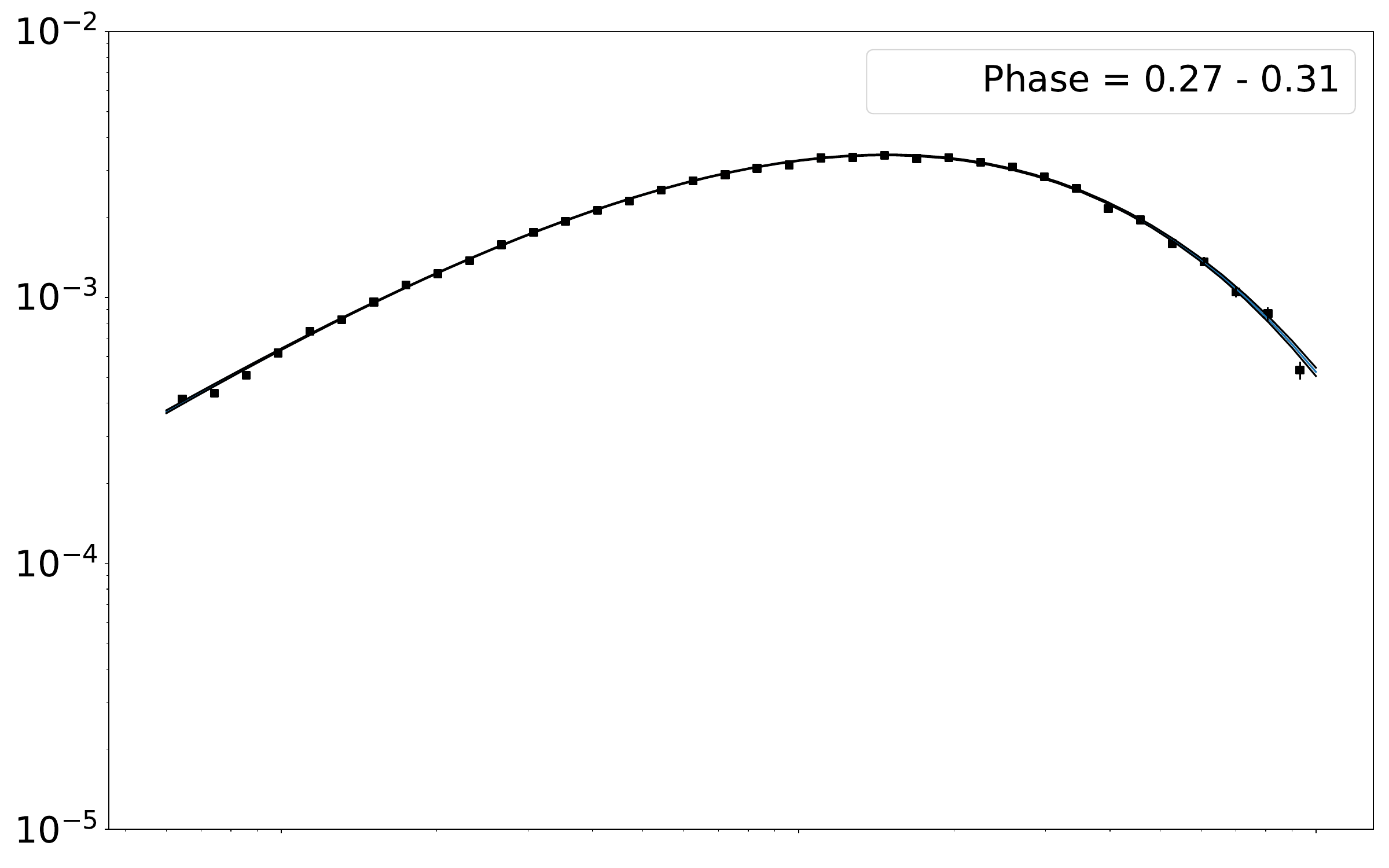}
    \end{minipage}

    \vspace{-2pt}

    \begin{minipage}{0.32\textwidth}
        \includegraphics[width=\linewidth]{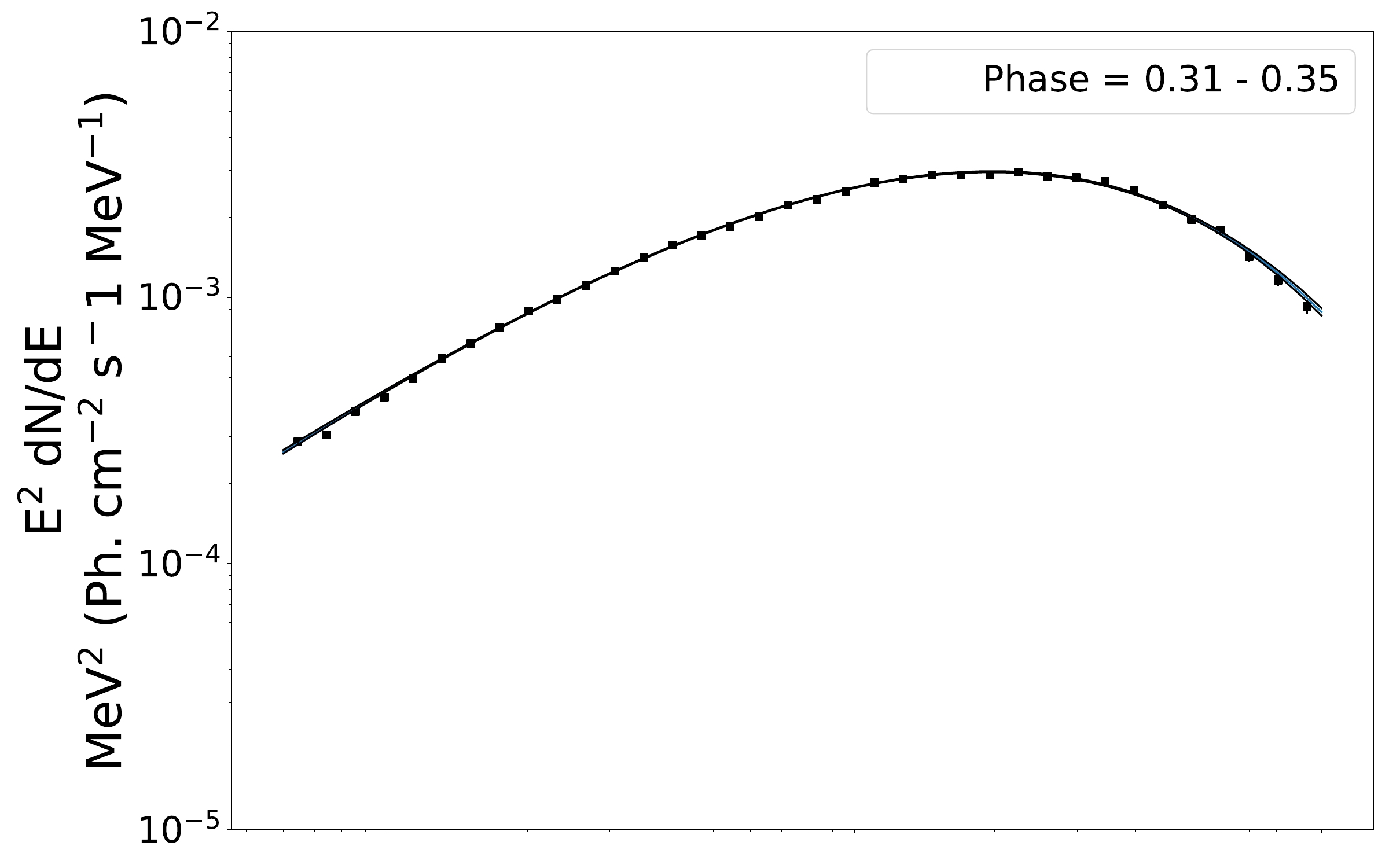}
    \end{minipage}%
    \begin{minipage}{0.32\textwidth}
        \includegraphics[width=\linewidth]{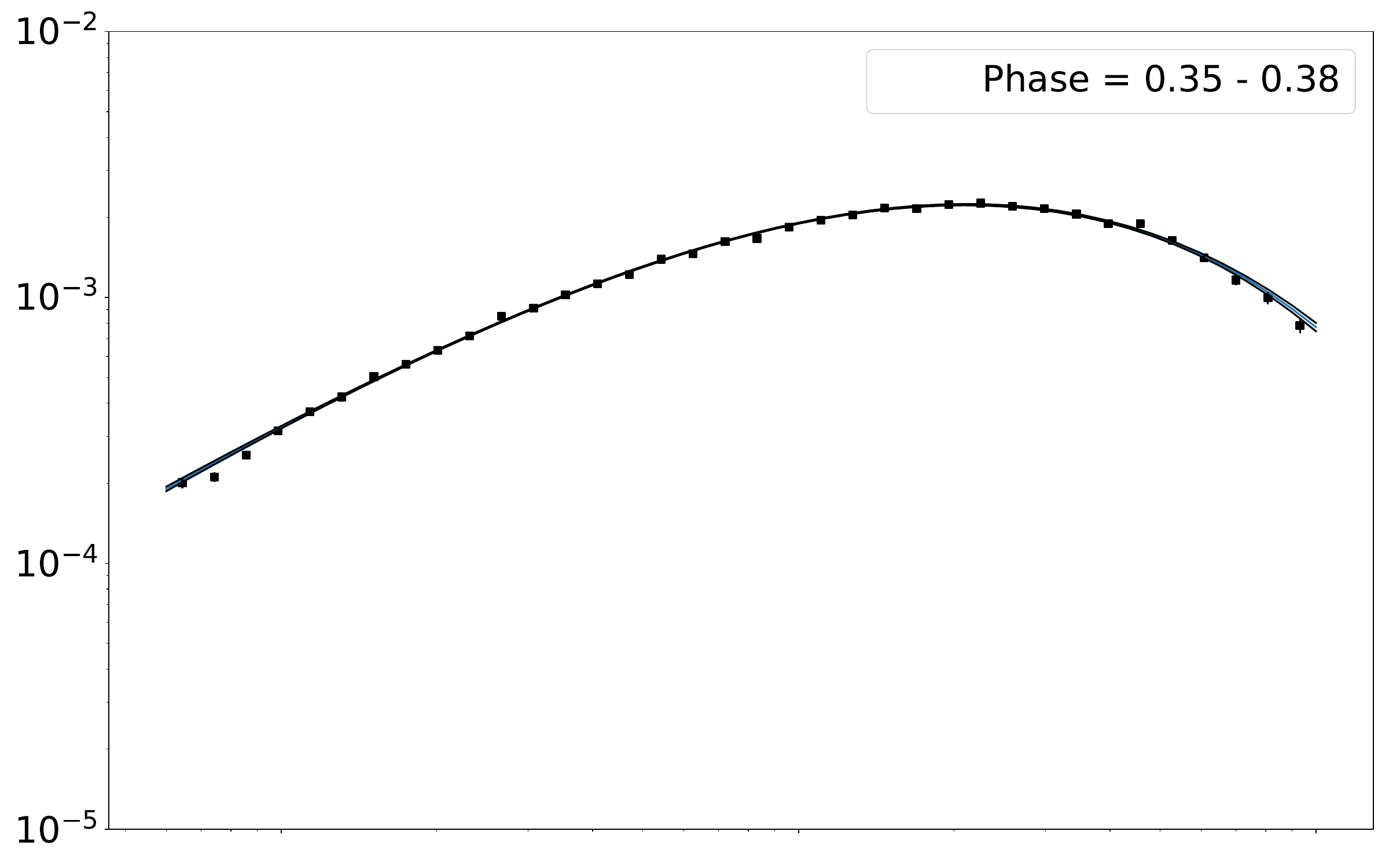}
    \end{minipage}%
    \begin{minipage}{0.32\textwidth}
        \includegraphics[width=\linewidth]{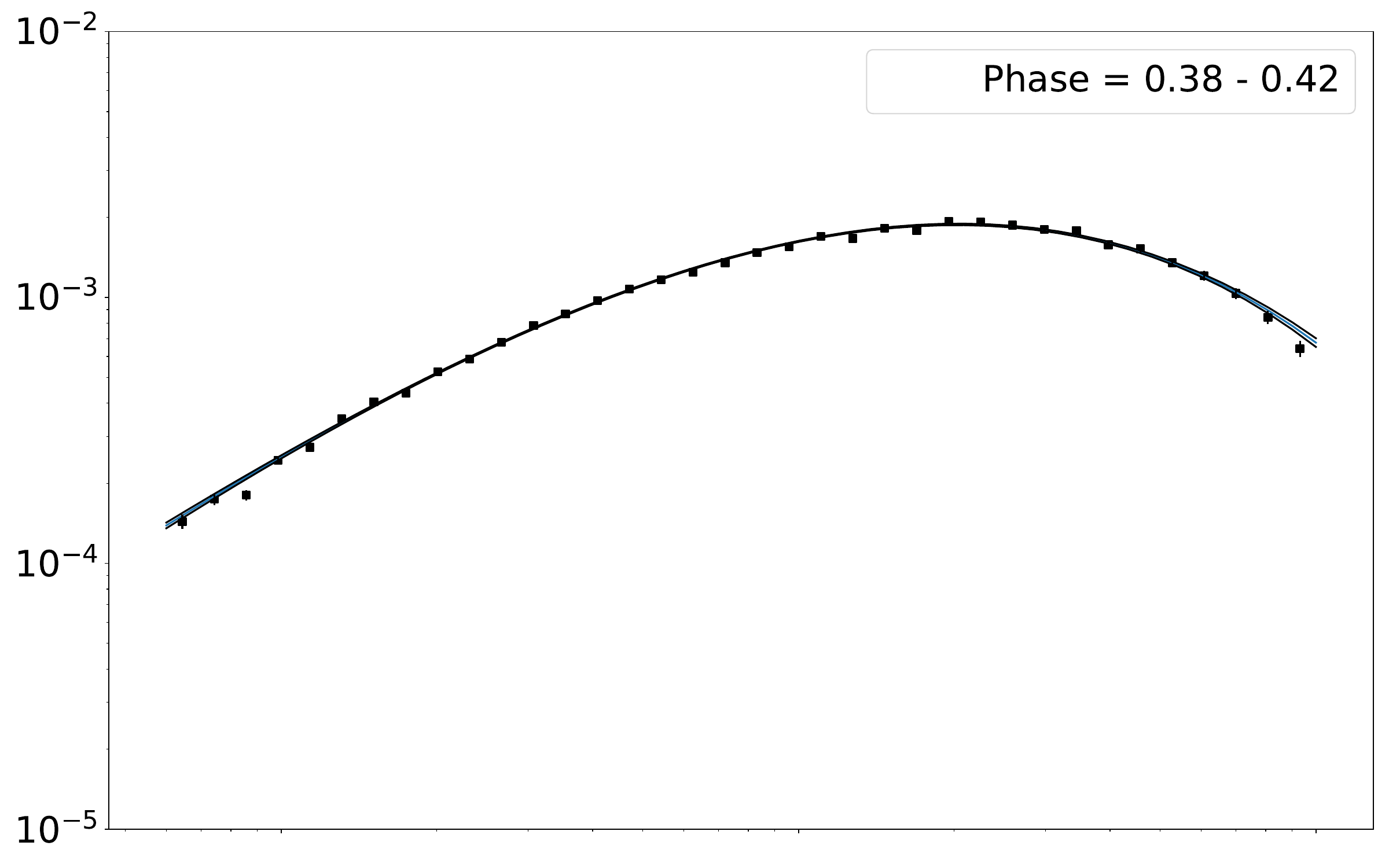}
    \end{minipage}

    \vspace{-2pt} 

    \begin{minipage}{0.32\textwidth}
        \includegraphics[width=\linewidth]{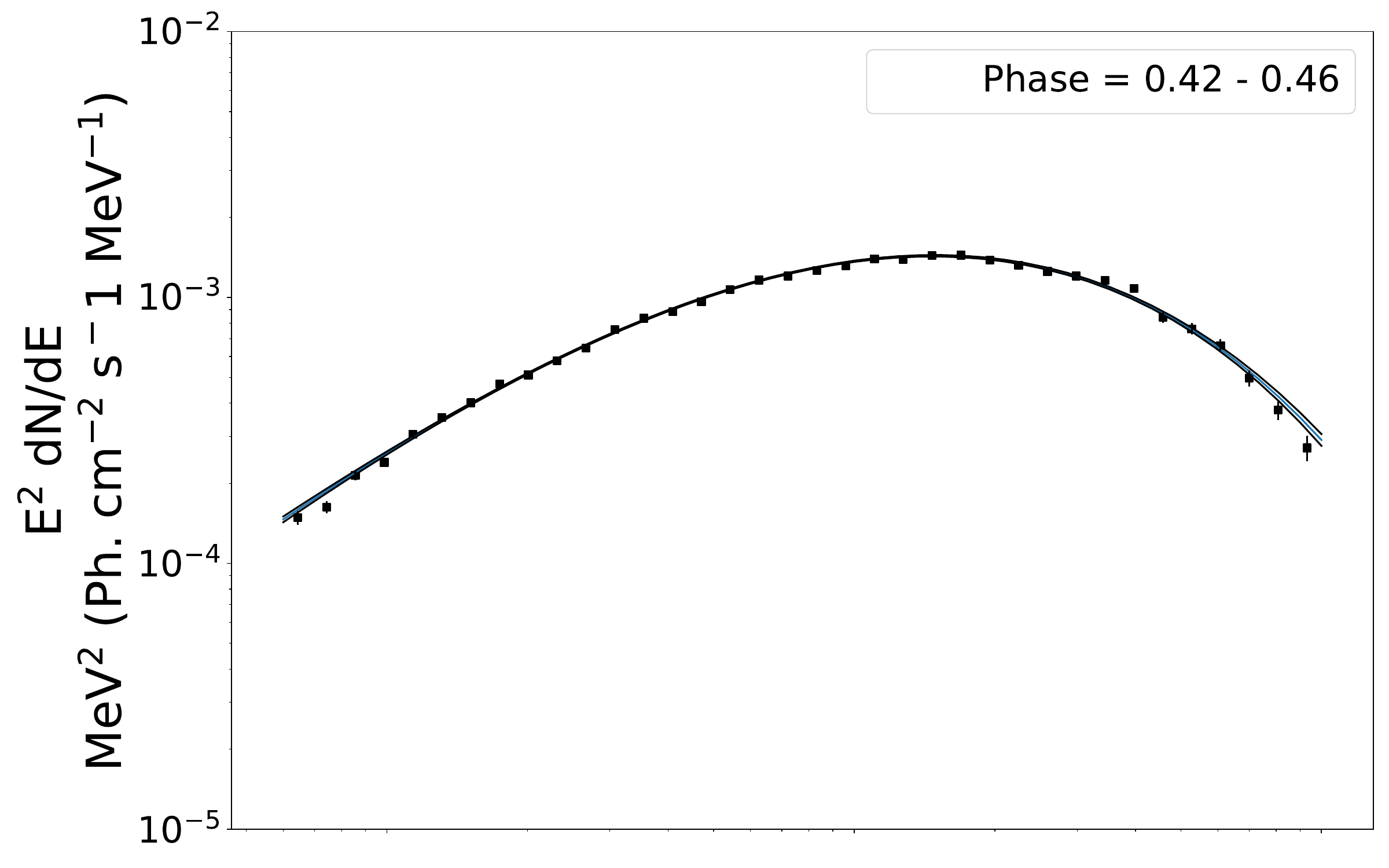}
    \end{minipage}%
    \begin{minipage}{0.32\textwidth}
        \includegraphics[width=\linewidth]{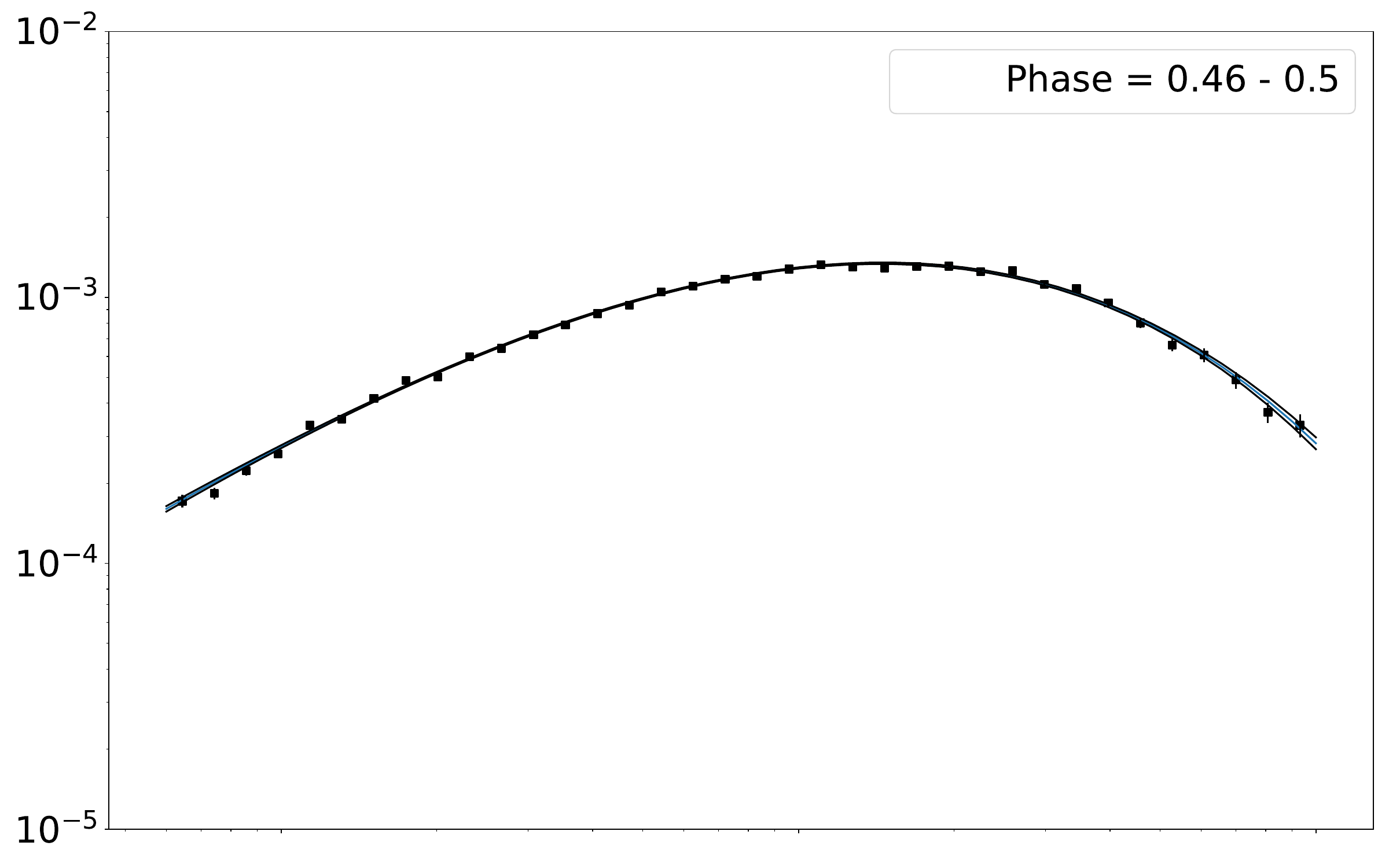}
    \end{minipage}%
    \begin{minipage}{0.32\textwidth}
        \includegraphics[width=\linewidth]{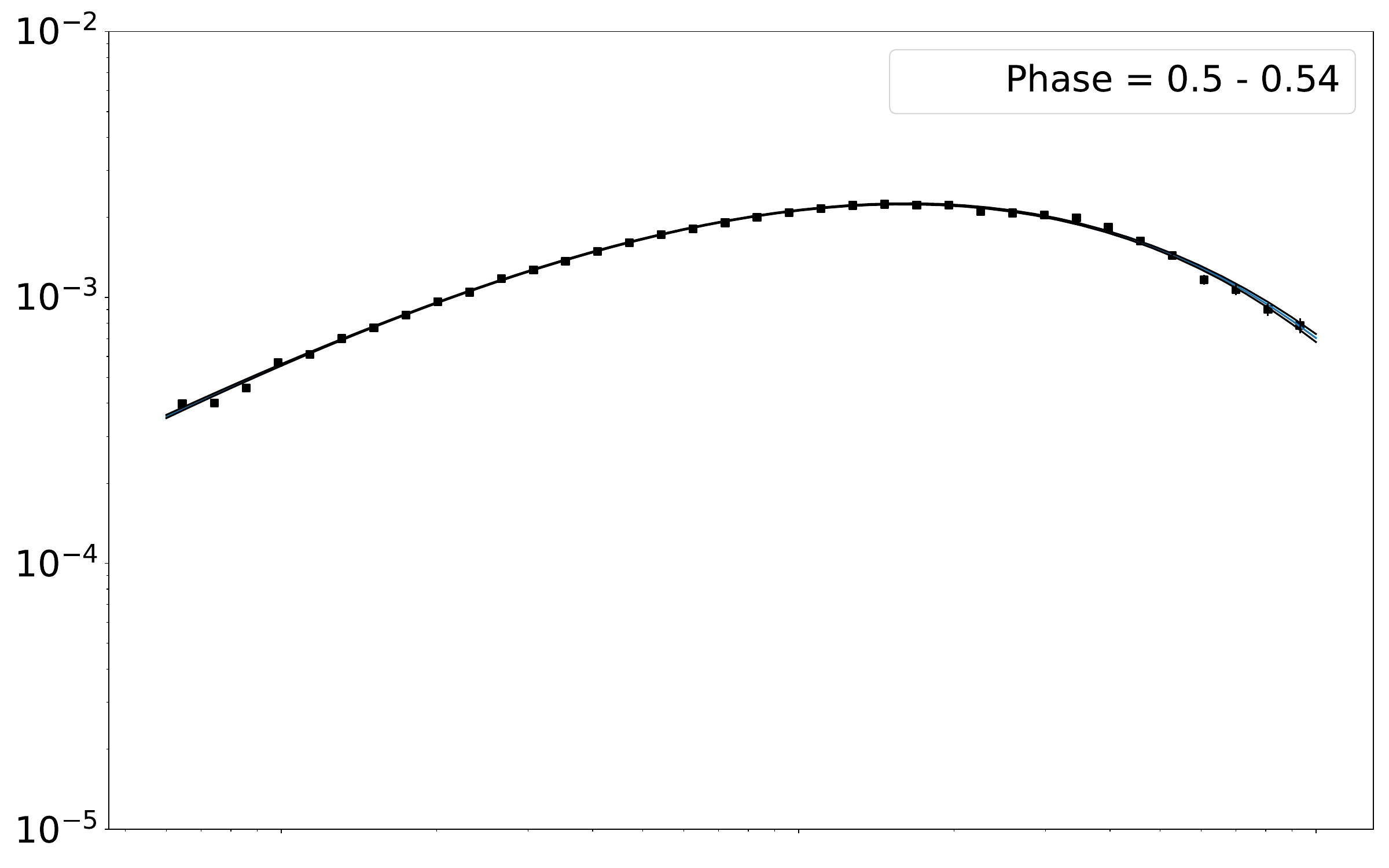}
    \end{minipage}

    \vspace{-2pt} 

    \begin{minipage}{0.32\textwidth}
        \includegraphics[width=\linewidth]{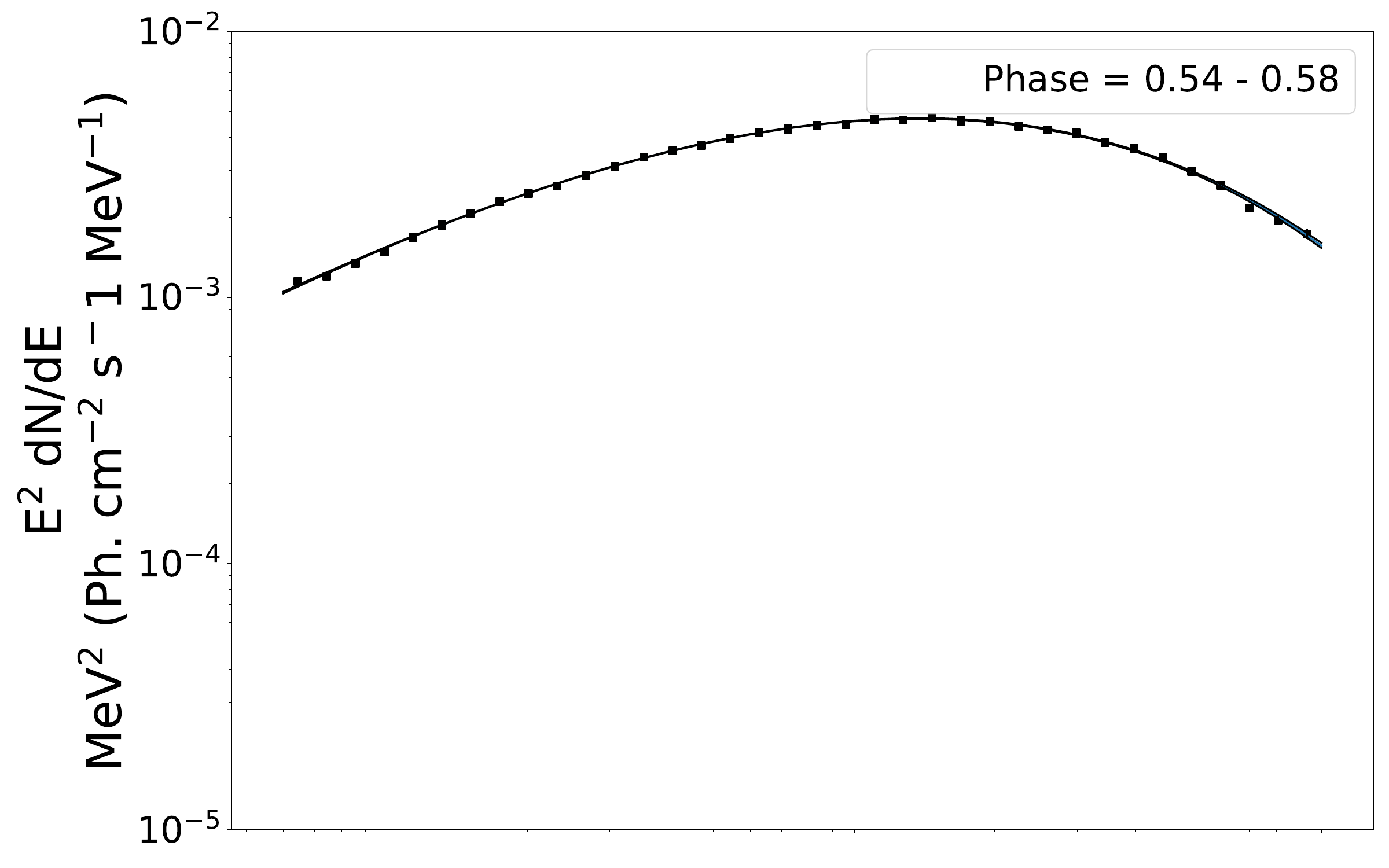}
    \end{minipage}%
    \begin{minipage}{0.32\textwidth}
        \includegraphics[width=\linewidth]{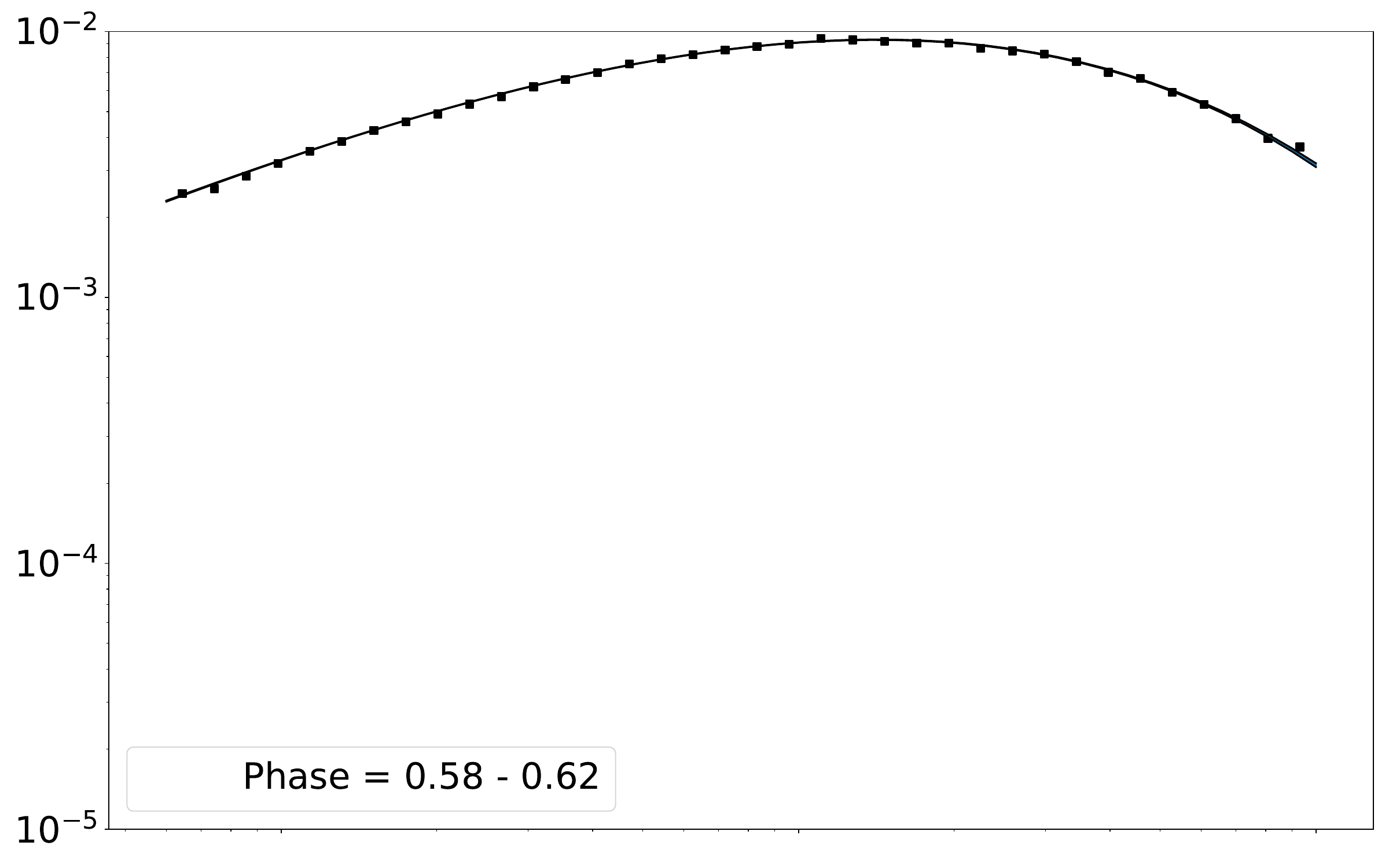}
    \end{minipage}%
    \begin{minipage}{0.32\textwidth}
        \includegraphics[width=\linewidth]{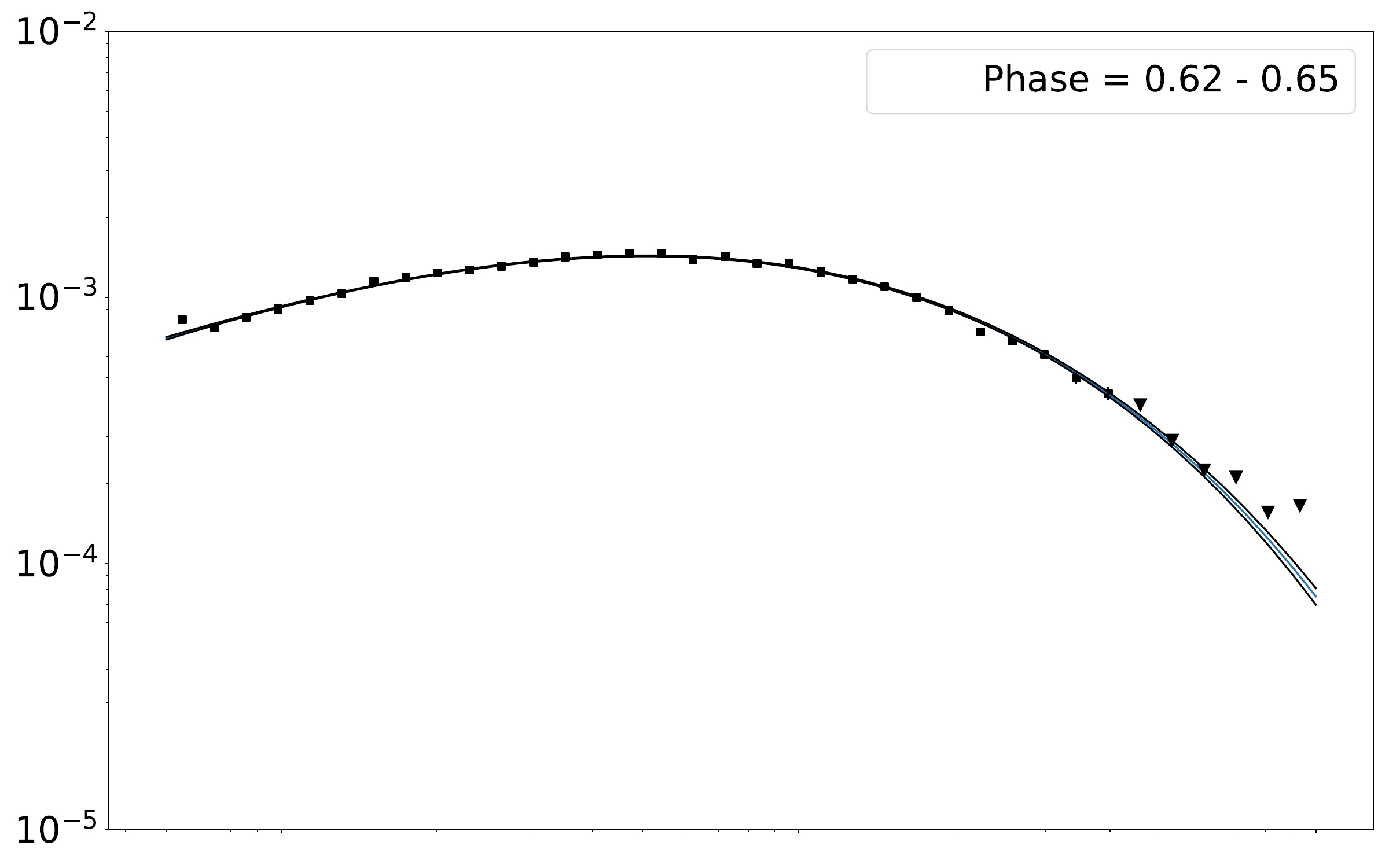}
    \end{minipage}
    
    \vspace{-2pt}
    
    \begin{minipage}{0.32\textwidth}
        \includegraphics[width=\linewidth]{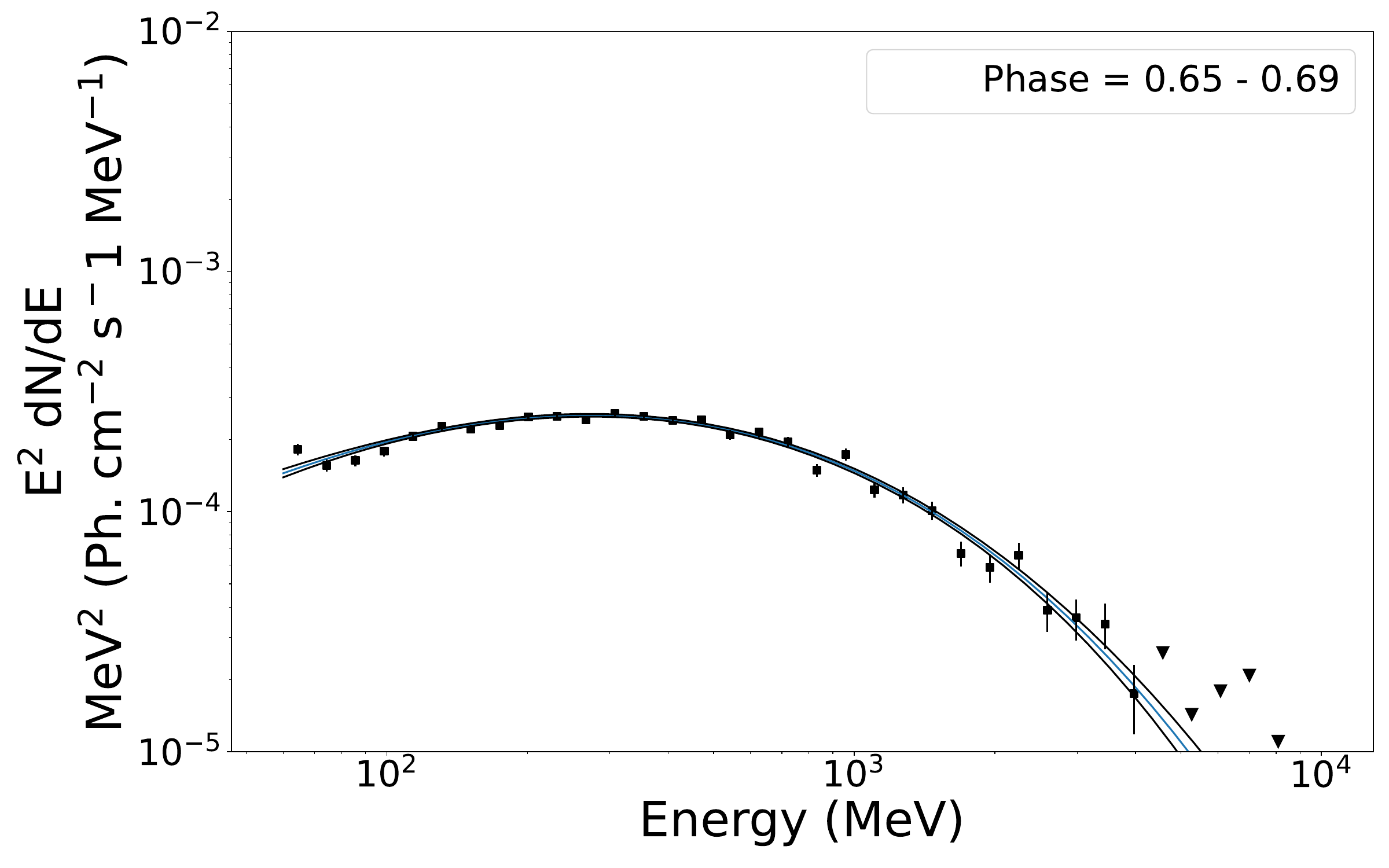}
    \end{minipage}%
    \begin{minipage}{0.32\textwidth}
        \includegraphics[width=\linewidth]{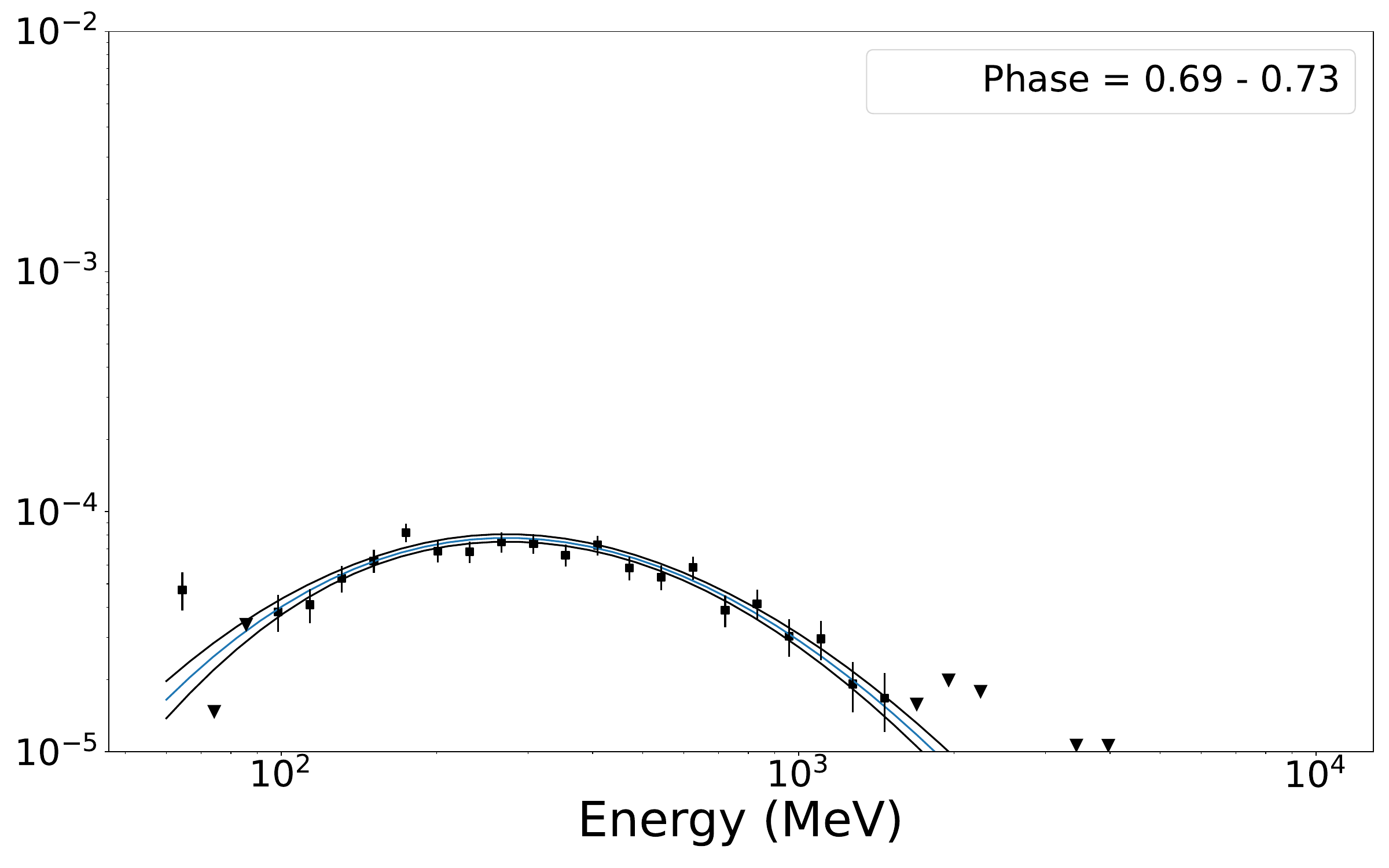}
    \end{minipage}%
    \begin{minipage}{0.32\textwidth}
        \includegraphics[width=\linewidth]{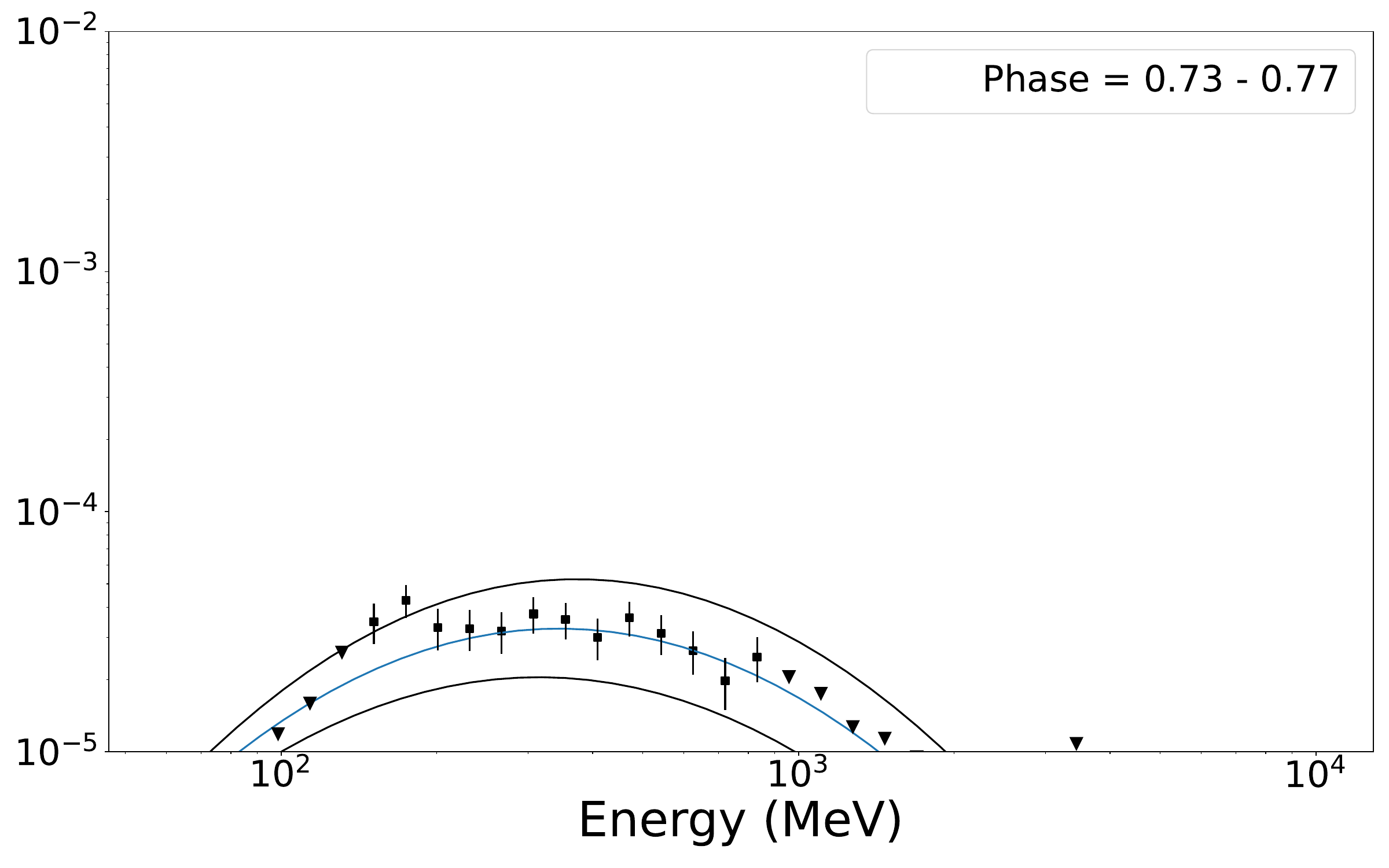}
    \end{minipage}
    
    \caption{The $E^2 dN/dE$ spectrum for the Vela pulsar in the phase range [0.08, 0.77] when $b$ is free to vary. The phase bin is indicated in each plot.  Spectral data are indicated by black crosses while upper-limits are indicated by black triangles. The best-fit model is represented by a solid blue line and the $95\%$ confidence interval is represented by black lines.
   }
\label{app:e2_freeb}
\end{figure*}

We provide the count spectra with fractional residuals along with the best-fit spectra for each phase bin using the PLEC4 spectral shape from the pulsar analysis in Section~\ref{sec:pulsar} 
for $b = 0.394$ (Figures~\ref{app:count_fixedb} and \ref{app:e2_fixedb}) and when $b$ is allowed to vary (Figures~\ref{app:count_freeb} and \ref{app:e2_freeb}).

\pagebreak
\bibliographystyle{aa}
\bibliography{main.bib}
\end{document}